\documentclass[onecolumn]{IEEEtran}
\usepackage{cite}
\ifCLASSINFOpdf
  \usepackage[pdftex]{graphicx}
\else
  \usepackage[dvips]{graphicx}
  \usepackage{epsfig}
\fi

\usepackage{amsfonts}
\usepackage{amsmath}
\usepackage{amssymb}
\usepackage{amscd}
\usepackage{mathrsfs}
\usepackage{color}
\usepackage{amsthm}
\usepackage{mathtools}
\allowdisplaybreaks
\usepackage{algorithmic}
\usepackage{array}
\usepackage{stfloats}
\usepackage{url}
\usepackage{eqparbox}
\usepackage{fancyhdr}
\newtheorem{thrm}{Theorem}
\newtheorem{prop}{Proposition}
\newtheorem{lemma}{Lemma}

\newtheorem{remark}{Remark}
\newtheorem{assumption}{Assumption}
\renewcommand{\baselinestretch}{1.5}

\usepackage[bookmarks=false]{hyperref}

\begin{document}
\title{Learning to Detect an Odd Restless Markov \\Arm with a Trembling Hand}

\author{P. N. Karthik and
        Rajesh Sundaresan

         \thanks{P. N. Karthik is with the Department of Electrical Communication Engineering at the Indian Institute of Science, Bangalore 560012, Karnataka, India. Rajesh Sundaresan is with the Department of Electrical Communication Engineering and the Robert Bosch Centre for Cyber Physical Systems at the Indian Institute of Science, Bangalore 560012, Karnataka, India. Email: (periyapatna, rajeshs)@iisc.ac.in.}

        \thanks{This work was supported by the Science and Engineering Research Board, Department of Science and Technology (grant no. EMR/2016/002503), by the Robert Bosch Centre for Cyber Physical Systems, and the Centre for Networked Intelligence at the Indian Institute of Science.}
       \thanks{A shorter version of this paper has been accepted for presentation at the 2021 IEEE International Symposium on Information Theory (ISIT).}
}

\maketitle

\begin{abstract}
\label{sec:abstract}
This paper studies the problem of finding an anomalous arm in a multi-armed bandit when (a) each arm is a finite-state Markov process, and (b) the arms are restless. Here, anomaly means that the transition probability matrix (TPM) of one of the arms (the odd arm) is different from the common TPM of each of the non-odd arms. The TPMs are unknown to a decision entity that wishes to find the index of the odd arm as quickly as possible, subject to an upper bound on the error probability. We derive a problem instance-specific asymptotic lower bound on the expected time required to find the odd arm index, where the asymptotics is as the error probability vanishes. Further, we devise a policy based on the principle of certainty equivalence, and demonstrate that under a continuous selection assumption and a certain regularity assumption on the TPMs, the policy achieves the lower bound arbitrarily closely. Thus, while the lower bound is shown for all problem instances, the upper bound is shown only for those problem instances satisfying the continuous selection and the regularity assumptions. Our achievability analysis is based on resolving the identifiability problem in the context of a certain lifted countable-state controlled Markov process.
\end{abstract}


\section{Introduction}
\label{sec:introduction}
Consider a multi-armed bandit in which each arm is a time-homogeneous and ergodic discrete-time Markov process taking values in a common finite state space. Assume that the Markov process of each arm is independent of those of the other arms. Suppose that the state evolution on one of the arms (the {\em  odd} arm) is governed by a transition probability matrix (TPM) $P_1$ and that on each of the non-odd arms is governed by $P_2\neq P_1$. A decision entity that knows neither $P_1$ nor $P_2$ wishes to find the index of the odd arm as quickly as possible subject to an upper bound on the error probability. Clearly, the smaller the error probability, the longer the decision entity will have to wait before declaring the odd arm index. Our goal is to characterise the asymptotic growth and growth rate of the expected time required to find the odd arm index as the error probability vanishes.

To arrive at a decision of the odd arm index, the decision entity samples the arms sequentially, one at each discrete time instant $t\in \{0, 1, 2, \ldots\}$. Following \cite{karthik2021detecting}, we assume that the decision entity has a {\em trembling hand}. This means that for some fixed $\eta>0$, the decision entity samples the intended arm with probability $1-\eta$, but with probability $\eta$, the decision entity samples a uniformly randomly chosen arm. Here, $\eta$ is known as the trembling hand parameter. {\color{black} See \cite{karthik2021detecting} for an example where human subjects exhibit a trembling hand in a visual search experiment that involves searching for an oddball image in a sea of distracter images. It is likely that in such visual search experiments, the human subject scans multiple images at once before narrowing down the search to the oddball image \cite{naghshvar2013two}. While \cite{karthik2021detecting} does not model such nuances, it broadly captures the search dynamics in a way that makes the problem amenable to analysis. We therefore follow the model in \cite{karthik2021detecting}.
%It is likely that in such visual search experiments, the human subject scans multiple images at once before narrowing down the search to the oddball image \cite{naghshvar2013two}, in a way different from that modelled in \cite{karthik2021detecting}. Nevertheless, the model in \cite{karthik2021detecting} captures the search dynamics in a way that makes the problem amenable to analysis. We therefore follow the model in \cite{karthik2021detecting}.
}

At each time $t$, the decision entity observes the (noiseless) state of the pulled arm post-trembling. The Markov processes of the unobserved arms continue to evolve, thus making the arms {\em restless} \cite{Whittle1988}. The decision entity continues to sample the arms sequentially until it is sufficiently confident\footnote{This confidence is to be viewed as the decision having a sufficiently small error probability.} of its decision on the odd arm index, at which time it stops further sampling and declares its decision on the odd arm index.

For the above described problem of finding the odd arm in a restless multi-armed bandit, the paper \cite{karthik2021detecting} studies the case when both $P_1$ and $P_2$ are known to the decision entity beforehand. In this paper, we study the more difficult case when neither $P_1$ nor $P_2$ is known beforehand.

\subsection{Observation `Delays' and a Markov Decision Problem}
The restless nature of the arms makes it necessary for the decision entity to maintain a record of (a) the time elapsed since each arm was previously sampled (called the arm’s {\em delay}), and (b) the state of each arm as observed at its previous sampling instant (called the arm's {\em last observed state}). As noted in \cite{karthik2021detecting}, the notion of arm delays is meaningful when the arms are restless, and superfluous when either each arm yields independent and identically distributed (iid) observations or when each arm yields Markov observations and the arms are {\em rested}. The arm delays, being positive and integer-valued, introduce a countably infinite dimension to the problem. As demonstrated in \cite[Section II.C]{karthik2021detecting}, the delays and the last observed states of the arms collectively form a {\em  controlled Markov process} and, in turn, lead to a Markov decision problem (MDP) whose state space is countably infinite and action space is the set of arms. Further, it is evident from \cite[Section II.C]{karthik2021detecting} that the transition probabilities of the MDP are stationary across time and are functions of the odd arm index, $P_1$ and $P_2$. However, the objective is not to maximise rewards (or minimise regret), as is typical in MDPs, but to find the odd arm index quickly and accurately.

\subsection{Certainty Equivalence and  Identifiability}
When neither $P_1$ nor $P_2$ is known beforehand, as is the case in this paper, the transition probabilities of the MDP may be regarded as being parameterised by a triplet of unknowns consisting of (a) the odd arm index, (b) the TPM $P_1$ of the odd arm, and (c) the common TPM $P_2$ of each non-odd arm. Call this triplet an {\em arms configuration}. Because the true (underlying) arms configuration is not known beforehand, it must  at least partially be learnt along the way. A commonly used approach to learn the true parameter governing the transition probabilities of an MDP is {\em certainty equivalence}. The idea behind this approach is to (a) maintain an estimate of the parameter at each time $t$, and (b) take an action at time $t$ supposing that the estimated value is indeed the true parameter value. The key challenge in this approach is to show that the parameter estimates converge to the true parameter value as $t\to \infty$, i.e., the system is {\em identifiable}.

Sufficient conditions that lead to system identifiability have been proposed in the literature. In an important paper \cite{mandl1974estimation}, Mandl demonstrated that for a finite-state MDP, when the parameter estimate at each time $t$ is chosen so as to minimise a ``contrast'' function computed using all the observations and actions up to time $t$, the parameter estimates converge to the true parameter  value as $t\to \infty$ \cite[Theorem 6]{mandl1974estimation}. In particular, when the contrast function is the negative log-likelihood, the resulting parameter estimates are the maximum likelihood (ML) estimates. Mandl demonstrated the convergence of the ML estimates to the true parameter value under an additional condition (known as {\em Mandl's identifiability condition} \cite[Eq. (35)]{mandl1974estimation}) on the MDP transition probabilities.  However, it is not clear if Mandl's identifiability condition is sufficient for identifiability in countable-state MDPs. In \cite{borkar1982identification}, the authors consider the same problem as Mandl's, but for countable-state MDPs when Mandl's identifiability condition is relaxed. The authors of \cite{borkar1982identification} show that under some regularity assumptions on the MDP transition probabilities, certainty equivalence based on ML estimation renders the system identifiable.

In this paper, we use certainty equivalence with ML estimation as in \cite{borkar1982identification} to learn the true arms configuration. Due to the presence of arm delays in the likelihood function, closed-form expressions for the ML estimates of the TPMs are not available. Nevertheless, we show that the system is identifiable under a mild regularity assumption on the TPMs.

\subsection{Prior Works on Odd Arm Identification, {\color{black} Restless Arms, and Other Related Works}}
As mentioned earlier, the paper \cite{karthik2021detecting} studies the problem of odd arm identification in a restless multi-armed bandit when the TPMs of the odd arm and the non-odd arms are known beforehand. In addition to \cite{karthik2021detecting}, prior works on odd arm identification have studied the cases when either each arm yields iid observations or each arm yields Markov observations and the arms are rested. The specifics of such works are as follows. Paper \cite{Vaidhiyan2017} studies the case when each arm yields iid Poisson observations and the Poisson rates of the odd arm and the non-odd arms are known beforehand. A study of the case of unknown Poisson rates appears in \cite{vaidhiyan2017learning}. Extension of the results in \cite{vaidhiyan2017learning} to the case when each arm yields iid observations coming from a generic exponential family appears in \cite{prabhu2017optimal}. The setting when each arm yields Markov observations and the arms are rested has been studied in \cite{pnkarthik2019learning}. The recent papers \cite{deshmukh2018controlled, deshmukh2021sequential, prabhu2020sequential} study more general sequential hypothesis testing problems in multi-armed bandits of which the problem of odd arm identification is a special case. For a related problem of best arm identification instead of odd arm identification, see \cite{Kaufmann2016, moulos2019optimal}.

{\color{black} Restless multi-armed bandits have been studied in the works of \cite{Whittle1988,ortner2012regret, grunewalder2019approximations}. In contrast to these works which are based on the theme of minimising regret, our work is based on the theme of optimal stopping. In \cite{liu2012learning}, a policy based on repeated sampling of the arms has been proposed to achieve a logarithmic order regret for a restless multi-armed bandit with unknown dynamics. The setting of our paper is similar to that of \cite{liu2012learning} in that the TPMs of the odd arm and the non-odd arms are not known beforehand. However, it is not clear if the policy of \cite{liu2012learning}, or more generally, a policy that learns the TPMs by repeatedly sampling the same arm before moving to the next, performs well for the optimal stopping problem studied in this paper.

We note here that the problem of finding the odd (anomalous) arm is closely related to the problem of \emph{anomaly detection} that has been explored sufficiently well in the literature. In works on anomaly detection, the time at which the anomaly occurs is a random variable, and the goal is to detect the anomaly in the shortest possible time after it occurs. However, in our work, it is known to the decision entity beforehand that one of the arms is anomalous.}

\subsection{Prior Works on Certainty Equivalence, Identification, and Adaptive Control of Markov Processes}
The principle of certainty equivalence seems to have been rigorously explored first in the context of linear systems by \r{A}str\"{o}m and Wittenmark \cite{aastrom1973self} where it has been referred to as a {\em self-tuning regulator}. The paper by Mandl \cite{mandl1974estimation} applies the principle of certainty equivalence to the problem of adaptively controlling a finite-state controlled Markov process (equivalently, a finite-state MDP). Mandl demonstrated that under a regularity condition (known as Mandl's identifiability condition \cite[Eq. (35)]{mandl1974estimation}) on the MDP transition probabilities, certainty equivalence based on ML estimation renders the system identifiable. Doshi and Shreve \cite{doshi1980strong} consider a problem similar to Mandl's and show that under a slightly weaker condition than Mandl's, identifiability holds for a scheme based on certainty equivalence with modified maximum likelihood estimates (estimates that nearly maximise the log-likelihood).

It is not clear if either Mandl's identifiability condition or its weaker version in \cite{doshi1980strong} is sufficient for identifiability in infinite-state MDPs. Also, as remarked in \cite{borkar1979adaptive}, Mandl's condition may be too restrictive in practical applications. The authors of \cite{borkar1979adaptive} provide an example of a linear, real-valued Markovian system for which Mandl's identifiability condition fails to hold. For this example, the authors of \cite{borkar1979adaptive}  demonstrate that the parameter estimates may not converge, and even if they do, the convergence is not necessarily to the true parameter value. This suggests that in order to show system identifiability in infinite-state MDPs, it may be necessary to impose additional regularity conditions on the MDP transition probabilities.

 One such set of such regularity conditions that ensures identifiability in countable-state MDPs when Mandl's condition is relaxed may be found in  \cite{borkar1982identification}. We show that the regularity conditions of \cite{borkar1982identification} are satisfied for the problem studied in this paper under a mild assumption on the unknown TPMs of the odd arm and the non-odd arm Markov processes. For a detailed survey of results in stochastic adaptive control of Markov chains, see \cite{kumar1985survey}.
%  which enables us to infer identifiability in the context of the countable-state MDP arising from the arm delays and the last observed states.
%, the authors consider the same problem as Mandl's but for countable-state MDPs and with Mandl's identifiability condition relaxed. The authors of \cite{borkar1982identification} show that under some regularity conditions on the MDP transition probabilities, certainty equivalence based on ML estimation renders the system identifiable.

\subsection{Contributions and Key Challenges to Overcome}
In this section, we bring out the contributions of this paper and highlight the key challenges that we must overcome in our work.
\begin{enumerate}
	\item  We derive an asymptotic lower bound on the expected time required to find the odd arm index subject to an upper bound on the error probability, where the asymptotics is as the error probability vanishes. Specifically, given an arms configuration $C=(h, P_1, P_2)$ in which $h$ is the odd arm index, $P_1$ is the TPM of arm $h$, and $P_2\neq P_1$ is the common TPM of each non-odd arm, we show that the lower bound is $1/R^*(h, P_1, P_2)$, where $R^*(h, P_1, P_2)$ is an arms configuration dependent (or problem instance-dependent) constant and is the value of a max-min optimisation problem. See Section \ref{sec:lower_bound}, equation \eqref{eq:R_delta^*(h,P_1,P_2)} for the exact mathematical expression for $R^*(h, P_1, P_2)$. The max-min expression in  \eqref{eq:R_delta^*(h,P_1,P_2)} consists of an outer supremum over all conditional probability distributions on the arms conditioned on the arm delays and the last observed states, and an inner infimum over all arms configurations in which the odd arm index is any $h'\neq h$.
	
	\item Given an arms configuration $C=(h, P_1, P_2)$, the question of whether there exists an optimal conditional distribution that attains the outer supremum in the expression for $R^*(h, P_1, P_2)$ is still under study. Notwithstanding this, we look at conditional distributions that attain the supremum to within a factor of $1/(1+\delta)$ for any $\delta>0$. We refer to such conditional distributions as $\delta$-optimal solutions for the arms configuration $C=(h, P_1, P_2)$, and note that there may be multiple such solutions in general for each arms configuration.
%Additionally, the presence of arm delays poses difficulty in simplifying the inner infimum in the expression for $R^*(h, P_1, P_2)$.
	
	It is worth noting here that in the prior works \cite{vaidhiyan2017learning, prabhu2017optimal, pnkarthik2019learning}, the outer supremum in the expression for the constant appearing in the lower bound is over all {\em unconditional} probability distributions on the arms which are simpler objects to deal with than conditional probability distributions. Further, in these works, this outer supremum is attained by a unique (unconditional) probability distribution on the arms.
%	Because the arm delays are superfluous in these works, the task of simplifying the inner infimum in the expression for the constant in the lower bound is considerably easier.

	\item An important property of the optimal solutions to the lower bounds in \cite{vaidhiyan2017learning, prabhu2017optimal, pnkarthik2019learning} is that they are continuous functions of the arms configuration. It is this continuity property that is used to show that the certainty equivalence-based arm sampling policies of these works (using ML estimation) render the system identifiable. This, in turn, is used to show that these policies achieve the respective lower bounds asymptotically. Furthermore, a close look at \cite{vaidhiyan2017learning, prabhu2017optimal, pnkarthik2019learning} reveals that the ML estimates in these works have closed-form expressions. This readily helps in showing the convergence of the ML estimates to their true values by using either the law of large numbers (for the case of iid observations from the arms in \cite{vaidhiyan2017learning, prabhu2017optimal}) or the ergodic theorem (for the case of Markov observations from the arms in  \cite{pnkarthik2019learning}).
	
	In the context of this paper, we note that neither the existence of unique optimal solutions to $R^*(h, P_1, P_2)$ nor the above mentioned continuity property is available. Furthermore, the presence of arm delays in the likelihood function poses difficulty in obtaining closed-form expressions for the ML estimates of the TPMs of the odd arm and the non-odd arm Markov processes As a result, showing that the ML estimates converge to their true values is a challenging task.
%	We stick to ML estimation for estimating the TPMs because most of the papers in the literature talk about identifiability with ML estimation, and there is very less work on identifiability with other estimation techniques such as mmse.
	
	\item Notwithstanding the difficulties highlighted in the previous point, we demonstrate that under two key assumptions, certainty equivalence based on ML estimation renders the system identifiable. The first of these assumptions is on the existence of a continuous selection of $\delta$-optimal solutions (the analogue of continuity of optimal solutions in the prior works). The second assumption, analogous to that appearing in \cite{borkar1979adaptive}, is a mild regularity on the TPMs of the odd arm and the non-odd arm Markov processes that requires the non-zero entries of the powers of the TPMs to be larger than a certain constant $\bar{\varepsilon}^*\in (0,1)$. 
	
	\item Given an error probability $\epsilon>0$ and $\delta>0$, we construct a policy based on the principle of certainty equivalence with ML estimation and using the $\delta$-optimal solutions. Under the assumptions mentioned in the previous point, we demonstrate that the policy stops in finite time almost surely and that the error probability of the policy at stoppage is upper bounded by $\epsilon$. Further, we show that as $\epsilon \downarrow 0$, the policy's expected time to find the odd arm index satisfies an upper bound that is away from the lower bound only by a multiplicative factor of $(1+\delta)^2$. Our achievability analysis relies crucially on (a) resolving the identifiability problem for the countable-state MDP arising from the arm delays and the last observed states, and (b) showing that the policy eventually samples the arms according to the $\delta$-optimal solution for the underlying arms configuration. Finally, by letting $\delta \downarrow 0$, we show that our policy meets the lower bound.
	
	In summary, we prove the lower bound for all arms configurations, and prove the upper bound only for those arms configurations satisfying two regularity assumptions: (a) the existence of continuous selection of $\delta$-optimal solutions, and (b) the regularity of the TPMs.
	\item {\color{black} As highlighted in point 3 above, because of the presence of arm delays in the expression for the likelihood function, it is difficult to obtain closed-form expressions for the ML estimates of the TPMs (and hence for the maximum likelihood) at any given time. This difficulty can be circumvented by repeatedly sampling the Markov process of each arm and using the successive observations from each of the arms to estimate the TPMs. Because the Markov process of each arm is ergodic, these TPM estimates will converge to their true values asymptotically.

 However, unlike the policy we propose in Section \ref{sec:achievability}, repeated sampling of the arms may not ensure that given $\delta>0$ and an underlying arms configuration $C$, the arms are eventually sampled according to the $\delta$-optimal solution for $C$, a condition that is crucial in order to meet the constant in the lower bound. Moreover, it is not clear if the desired error probability criterion can be met. While policies that sample the arms repeatedly have been proposed and shown to perform well for the problem of minimising regret \cite{liu2012learning}, it is not clear if such policies perform well for optimal stopping problems such as that studied in this paper.
	
%	Because an arm's delay is equal to $1$ for as long as it is sampled repeatedly, the TPM estimates can be expressed in closed-form. Also, because the Markov process of each arm is ergodic, these estimates will clearly converge to their true values.
%	
%	Using the above TPM estimates instead of the ML estimates leads to a ``new likelihood'' that is (a) different from the maximum likelihood, and (b) can be computed in closed-form. However, it is not clear if a policy that uses the above mentioned ``new likelihood'' instead of the maximum likelihood meets the stoppage error probability criterion. Whereas, in spite of the computational intractability of the ML estimates, our policy achieves the desired error probability and is hence asymptotically optimal. A key step in showing that our policy meets the desired error probability criterion is to upper bound the likelihood function under a given arms configuration $C=(h, P_1, P_2)$ by the maximum likelihood under the hypothesis that arm $h$ is the odd arm. Clearly, such an upper bound does not hold for the new likelihood.
}
\end{enumerate}

\subsection{Organisation of the Paper}
The rest of this paper is organised as follows. In Section \ref{sec:notations}, we set up the notations and provide some preliminaries on MDPs that will be used in the remainder of the paper. In Section \ref{sec:lower_bound},  we present the asymptotic lower bound on the expected time to identify the odd arm index. In Section \ref{sec:achievability}, we state the two assumptions, present a policy based on certainty equivalence with ML estimation, and demonstrate that it achieves the lower bound asymptotically. We state the main result of this paper in Section \ref{sec:main_result} and provide some concluding remarks in Section \ref{sec:conclusions}. The proofs are contained in Appendices \ref{appndx:proof_of_prop_lower_bound}-\ref{appndx:proof_of_prop_upper_bound}.

\section{Notations and Preliminaries}\label{sec:notations}
Consider a multi-armed bandit with $K\geq 3$ arms, and let $\mathcal{A}=\{1, \ldots, K\}$ denote the set of arms. For each $a\in \mathcal{A}$, consider a discrete-time Markov process $\{X_t^a:t\geq 0\}$ associated with arm $a$ that is time-homogeneous, ergodic and takes values in a common finite set $\mathcal{S}$. Assume that the Markov process of each arm is independent of those of the other arms. Let the TPM of one of the arms (the {\em odd} arm) be $P_1$, and that of each non-odd arm be $P_2$, where $P_2\neq P_1$. Write $C=(h, P_1, P_2)$ to denote an arms configuration in which $h$ is the odd arm index, $P_1$ is the TPM of the odd arm $h$, and $P_2\neq P_1$ is the TPM of each non-odd arm. Let $\mathcal{H}_h$ denote the composite hypothesis that $h$ is the odd arm index.

%A decision maker, who knows neither $P_1$ nor $P_2$, intends to identify the true hypothesis as quickly as possible.  Given a pre-specified error probability $\epsilon>0$, the decision maker also has to ensure that the probability of his decision error is within $\epsilon$. In order to formulate his decision about the true odd arm location, the decision maker samples the arms sequentially, one at each discrete time instant $t\in \{0,1,2,\ldots\}$, and observes the state of the arm sampled. This process repeats until the decision maker is sufficiently confident about his estimate of the odd arm location. Although the decision maker observes the state of only one of the arms at any given time instant, the Markov process of each unobserved arm continues to evolve according to its respective transition probability matrix, thereby making the arms \emph{restless}.  Given an arms configuration $C=(h, P_1, P_2)$, our goal is to characterise the limiting growth rate of the expected time taken by the decision maker to identify the location of the odd arm as a function of the error probability $\epsilon$, in the limit as $\epsilon \downarrow 0$.
A decision entity that knows neither $P_1$ nor $P_2$ wishes to find the odd arm index as quickly as possible while keeping the probability of its decision error small. Although the unknowns in the problem are (a) the odd arm index, (b) $P_1$, and (c) $P_2$, the objective of the decision entity is to only find the odd arm index correctly with high probability. The decision entity samples the arms sequentially, one at each time $t\geq 0$. Let $B_t$ denote the arm that the decision entity intends to sample at time $t$. The decision entity has a trembling hand, and the arm $A_t$ is instead sampled at time $t$, where $A_t$ and $B_t$ satisfy the probabilistic relation
\begin{equation}
P(A_t=a|B_t=b)=\frac{\eta}{K} + (1-\eta)\,\mathbb{I}_{\{a=b\}}
\label{eq:trembling_hand_relation}
\end{equation}
for some fixed $\eta>0$; here, $\eta$ is called the {\em trembling hand parameter}. Notice that $A_t$ is chosen uniformly at random with probability $\eta$, and $A_t=B_t$ with probability $1-\eta$. The decision entity observes $A_t$ and therefore knows whether its hand trembled at time $t$. Further, the decision entity observes the (noiseless) state of the sampled arm $A_t$, which we denote by $\bar{X}_t$. Therefore, at any given time $t$, the decision entity has knowledge of the history $(B_0, A_0, \bar{X}_0, \ldots, B_t, A_t, \bar{X}_t)$ of all the intended arm samples, the actual arm samples and the states of the actually sampled arms up to time $t$. We write $(B^t, A^t,\bar{X}^t)$ to compactly represent the history $(B_0, A_0, \bar{X}_0, \ldots, B_t, A_t, \bar{X}_t)$. While the decision entity observes the state of only one arm at each time instant, the Markov processes of the other arms continue to evolve ({\em restless} arms).

Define a {\em policy} $\pi$ of the decision entity as a collection of functions $\{\pi_t:t\geq 0\}$ such that $\forall$ $t\geq 0$, $\pi_t$ takes as input the history $(B_0, A_0, \bar{X}_0, \ldots, B_t, A_t, \bar{X}_t)$ and outputs one of the following:
\begin{itemize}
	\item sample arm $B_{t+1}$ according to a deterministic or a randomised rule.
	\item stop sampling and declare the odd arm index.
\end{itemize}
Let $\tau(\pi)$ and $\theta(\tau(\pi))$ denote respectively the stopping time of the policy $\pi$ and the odd arm index put out by the policy $\pi$ at stoppage. Because the decision entity is oblivious to the underlying arms configuration, any sequential arm sampling policy of the decision entity must meet the error probability constraint for all arms configurations. Given an error probability $\epsilon>0$, let $\Pi(\epsilon)$ denote the set of all policies whose error probability at stoppage is $\leq \epsilon$ for all arms configurations, i.e.,
\begin{align}
\Pi(\epsilon) & \coloneqq \bigg\lbrace\pi: P^\pi\bigg(\theta(\tau(\pi)) \neq h ~\bigg \vert~ C=(h, P_1, P_2)\bigg) \leq \epsilon \quad \forall ~C=(h, P_1, P_2),~h\in \mathcal{A},~P_2\neq P_1 \bigg\rbrace.
\label{eq:Pi(epsilon)}
\end{align}
In \eqref{eq:Pi(epsilon)} and throughout the paper, $P^\pi(\cdot|C)$ denotes probabilities computed under the policy $\pi$ and the arms configuration $C$. Similarly, $E^\pi[\cdot|C]$ will be used to denote expectations.

%For an error probability $\epsilon>0$, our interest is in the set
%\begin{equation}
%\Pi(\epsilon)\coloneqq \bigg\lbrace\pi: P^\pi\bigg(\theta(\tau(\pi)) \neq h ~\bigg \vert~ C=(h, P_1, P_2)\bigg) \leq \epsilon\quad \forall ~C=(h, P_1, P_2),~P_1\neq P_2 \bigg\rbrace
%\label{eq:Pi(epsilon)}
%\end{equation}
%of policies whose probability of error at stoppage is within $\epsilon$ for all possible arm configurations.
%Note that the policies in \eqref{eq:Pi(epsilon)} must work for all possible arm configurations, since no policy $\pi$ has prior knowledge of the underlying arms configuration. Since there are uncountably infinitely many possible choices for the transition probability matrices $P_1$ and $P_2$, $P_1\neq P_2$, it is not clear {\em a priori} that the set in \eqref{eq:Pi(epsilon)} is non-empty for each $\epsilon>0$. Indeed, the reader may be tempted to think that $P_1$ and $P_2$ may be chosen in a way that their entries are arbitrarily close to one another (with some entries of $P_1$ possibly matching with those of $P_2$), so that for any policy $\pi$, the condition of its error probability being within $\epsilon$ is violated. However, as we show later, for such a choice of the transition probability matrices $P_1$ and $P_2$, each policy $\pi$ might have to wait longer before it stops and declares the location of the odd arm with an error probability $\leq \epsilon$.
Note the requirement of policies in \eqref{eq:Pi(epsilon)} to work for all arms configurations. A careful reader may raise the question that the error probability criterion may not hold for arms configurations $C=(h, P_1, P_2)$ in which the entries of $P_1$ and $P_2$ are arbitrarily close to each another, with some entries of $P_1$ possibly matching with those of $P_2$. However, the key here is to note that, for such arms configurations, a policy $\pi\in \Pi(\epsilon)$ might have to observe longer before it stops and declares the odd arm index with an error probability $\leq \epsilon$. Therefore, the  ``closer'' the TPMs $P_1$ and $P_2$ are, the harder they are to distinguish, which results in larger stopping times of the policies.

%In other words, given an arms configuration $C=(h, P_1, P_2)$, the value of
%\begin{equation}
%\lim\limits_{\epsilon\downarrow 0} \inf\limits_{\pi\in \Pi(\epsilon)}\frac{E^\pi[\tau(\pi)|C]}{\log (1/\epsilon)}
%\label{eq:what_information_theorists_study}
%\end{equation}
%is, in general, a function of the transition probability matrices $P_1$ and $P_2$, and `closer' the transition probability matrices are, larger is the amount of time required to identify the true location of the odd arm. In this paper, we (a) make precise this notion of closeness between the transition probability matrices of the odd arm and the non-odd arm Markov processes, and (b) obtain a precise characterisation of the quantity in \eqref{eq:what_information_theorists_study} in terms of the above defined notion of closeness between the transition probability matrices.
We anticipate from the prior works that for every arms configuration $C=(h, P_1, P_2)$,
\begin{equation}
	\inf\limits_{\pi\in \Pi(\epsilon)}~E^\pi[\tau(\pi)|C] = \Theta(\log (1/\epsilon)).
	\label{eq:objective}
\end{equation}
The constant multiplying $\log(1/\epsilon)$ is, in general, a function of $C$. Our interest is in characterising the best (smallest) constant factor multiplying $\log(1/\epsilon)$ in the limit as $\epsilon\downarrow 0$. For simplicity, we assume that under every policy, $A_0=1$, $A_1=2$ and so on until $A_{K-1}=K$. If this is not the case, the arms may be sampled uniformly at random until the above sequence of arm samples is observed. Such an exercise of sampling the arms to first see the above sequence will only result in a finite delay (independent of $\epsilon$) almost surely when $\eta>0$, and does not affect the asymptotic analysis as $\epsilon \downarrow 0$.

\subsection{Arm Delays and Last Observed States}
Due to the restless nature of the arms, the decision entity has to keep a record of (a) the time elapsed since an arm was previously selected (the arm's \emph{delay}), and (b) the state of the arm at its previous selection time instant (the arm's \emph{last observed state}). As noted in \cite{karthik2021detecting}, arm delays and the last observed states are striking features of the setting of restless arms. When each arm yields independent and identically distributed (iid) observations, or when each arm yields Markov observations and the arms are rested, the notion of arm delays is superfluous.

Following the notations in \cite{karthik2021detecting}, let $d_a(t)$ and $i_a(t)$ respectively denote the delay and the last observed state of arm $a\in \mathcal{A}$ at time $t$. Let $\underline{d}(t)=(d_a(t):a\in \mathcal{A})$ and $\underline{i}(t)=(i_a(t):a\in \mathcal{A})$ denote the vectors of arm delays and last observed states at time $t$. The arm delays and the last observed states make sense when each arm is sampled at least once, and shall therefore be defined for $t\geq K$ keeping in mind that $A_0=1, \ldots, A_{K-1}=K$ under every policy. For $t=K$, we set $\underline{d}(K)=(K, K-1, \ldots, 1)$. This means that with reference to time $t=K$, arm $1$ was sampled $K$ time instants earlier (i.e., at $t=0$), arm $2$ was sampled $K-1$ time instants earlier (i.e., at $t=1$) and so on. The rule for updating $(\underline{d}(t), \underline{i}(t))$, based on the value of $A_t$, is straightforward and is as follows: if $A_{t}=a$, then
\begingroup \allowdisplaybreaks\begin{align}
	{d}_{\tilde{a}}(t+1)=\begin{cases}
		d_{\tilde{a}}(t)+1, &\tilde{a}\neq a,\\
		1,& \tilde{a}=a',
	\end{cases} \qquad \qquad
	i_{\tilde{a}}(t+1)=\begin{cases}
		i_{\tilde{a}}(t),& \tilde{a}\neq a,\\
		\bar{X}_{t},& \tilde{a}=a',
	\end{cases}
	\label{eq:delays_and_last_observed_states_update_rule}
\end{align}\endgroup
where, to recall, $\bar{X}_{t}$ is the state of the arm $A_t=a$ at time $t$. Therefore, the process $\{(\underline{d}(t), \underline{i}(t)):t\geq K\}$ takes values in a subset, say $\mathbb{S}$, of $\{1, 2, \ldots\}^K\times \mathcal{S}^K $. The set $\mathbb{S}$ is countably infinite and includes among many others the constraint that, for each $t\geq K$, exactly one component of the vector $\underline{d}(t)$ is equal to $1$ and all other components are strictly greater than $1$.

Therefore, it follows that under any policy $\pi$, the decision entity first samples each of the $K$ arms in such a way that $A_0=1$, $A_1=2$ and so on until $A_{K-1}=K$. Then, for all $K\leq t<\tau(\pi)$, based on (a) the history $\{(\underline{d}(s), \underline{i}(s)): K\leq s\leq t\}$ of arm delays and last observed states, and (b) the history $\{B_s:K\leq s<t\}$ of all the intended arm samples, the decision maker chooses to sample the arm $B_{t}$. Notice that this is equivalent to choosing $B_{t}$ based the history $(B^{t-1}, A^{t-1}, \bar{X}^{t-1})$. Due to the trembling hand, the decision entity observes that arm $A_{t}$ is instead pulled at time $t$. Subsequently, the decision entity observes the state $\bar{X}_{t}$ of arm $A_t$, and updates $(\underline{d}(t), \underline{i}(t))$ to $(\underline{d}(t+1), \underline{i}(t+1)) $ based on the update rule in \eqref{eq:delays_and_last_observed_states_update_rule}. At time $t=\tau(\pi)$, the decision entity announces its estimate $\theta(\tau(\pi))$ of the odd arm index.

\subsection{Controlled Markov Process and the Associated Markov Decision Problem}
From \cite[Section II.C]{karthik2021detecting}, we know that $\{(\underline{d}(t), \underline{i}(t)):t\geq K\}$ is a {\em controlled Markov process} with $(\underline{d}(t), \underline{i}(t))$ regarded as the state at time $t$ and $B_t$ regarded as the control at time $t$. That is, we are in the setting of a Markov decision problem (MDP) whose state space is $\mathbb{S}$, action space is $\mathcal{A}$, and the transition probabilities under an arms configuration $C$ are as follows: for all $(\underline{d}', \underline{i}'), (\underline{d}, \underline{i})\in \mathbb{S}$ and $b\in \mathcal{A}$,
\begin{align}
	&P^\pi(\underline{d}(t+1)=\underline{d}',\underline{i}(t+1)=\underline{i}'\mid \underline{d}(t)=\underline{d},\underline{i}(t)=\underline{i}, B_t=b, C)\nonumber\\
	&\hspace{4cm}=\begin{cases}
		\left(\frac{\eta}{K}+(1-\eta)\,\mathbb{I}_{\{a=b\}}\right)\,(P_C^a)^{d_a}(i_a'|i_a),&\text{if }d_a'=1\text{ and }d'_{\tilde{a}}=d_{\tilde{a}}+1\text{ for all }\tilde{a}\neq a,\\
		&i_{\tilde{a}}'=i_{\tilde{a}}\text{ for all }\tilde{a}\neq a,\\
		0,&\text{otherwise}.
	\end{cases}\label{eq:MDP_transition_probabilities}
\end{align}
In \eqref{eq:MDP_transition_probabilities}, $P_C^a$ is the TPM of arm $a$ under the arms configuration $C$. For instance, if $C=(h, P_1, P_2)$, then
\begin{equation}
P_C^a=\begin{cases}
P_1, & a=h, \\
P_2, & a\neq h.
\end{cases}
\label{eq:P_C^a}
\end{equation}
Also, $d_a'$ and $i_a'$ denote the component corresponding to arm $a$ in the vectors $\underline{d}'$ and $\underline{i}'$ respectively.  Similarly, $d_{\tilde{a}}$ and $i_{\tilde{a}}$ denote the component corresponding to arm $\tilde{a}$ in the vectors $\underline{d}$ and $\underline{i}$ respectively. Notice that the transition probabilities in \eqref{eq:MDP_transition_probabilities} are stationary across time and are parameterised by the arms configuration. We write $Q_C(\underline{d}', \underline{i}'\mid \underline{d}, \underline{i}, b)$ as a short hand representation of the transition probabilities in \eqref{eq:MDP_transition_probabilities} to be in line with the standard notation for controlled state transitions in MDPs.

Our objective, however, is nonstandard in the context of MDPs and more in line with what information theorists study. Given an arms configuration $C=(h, P_1, P_2)$, we are interested in determining the following:
\begin{equation}
  \lim_{\epsilon \downarrow 0} ~ \inf_{\pi \in \Pi(\epsilon)} ~ \frac{E^\pi [\tau(\pi)|C]}{\log (1/\epsilon)}.
  \label{eq:main_quantity}
\end{equation}

\subsection{SRS Policies and State-Action Occupancy Measures}
Call a policy $\pi$ a {\em stationary randomised strategy (SRS)} if there exists a Cartesian product $\lambda$ of the form\footnote{Writing $\mathcal{P}(
	\mathcal{A})$ to denote the space of all probability distributions on $\mathcal{A}$, it follows that  \eqref{eq:SRS_Phi_defn} is an element of the product space $\bigotimes \limits_{(\underline{d},\underline{i})\in\mathbb{S}} \mathcal{P}(\mathcal{A})$.}
\begin{equation}
	\lambda=\bigotimes \limits_{(\underline{d},\underline{i})\in\mathbb{S}} \lambda(\cdot|\underline{d}, \underline{i}),\label{eq:SRS_Phi_defn}
\end{equation}
with $\lambda(\cdot|\underline{d}, \underline{i})$ being a probability measure on $\mathcal{A}$ indexed by $(\underline{d}, \underline{i})\in \mathbb{S}$, with the interpretation that $B_{t}$ is sampled according to $\lambda(\cdot|\underline{d}(t), \underline{i}(t))$ for all $t\geq K$. Let such an SRS policy be denoted more explicitly as $\pi^\lambda$, and let $\Pi_{\textsf{SRS}}$ be the space of all SRS policies. Clearly, $\{(\underline{d}(t), \underline{i}(t)):t\geq K\}$ is a Markov process under every SRS policy. Further, \cite[Lemma 1]{karthik2021detecting} shows that this Markov process is ergodic when the trembling hand parameter $\eta>0$, and therefore possesses a unique stationary distribution. Let $\mu^\lambda=\{\mu^\lambda(\underline{d}, \underline{i}): (\underline{d}, \underline{i})\in \mathbb{S}\}$ be the stationary distribution under $\pi^\lambda$. Also, for $(\underline{d},\underline{i})\in \mathbb{S}$ and $a\in \mathcal{A}$, let
\begin{eqnarray}
		\nu^\lambda(\underline{d},\underline{i},a)\coloneqq \mu^\lambda(\underline{d},\underline{i})\left(\frac{\eta}{K}+(1-\eta)\,\lambda(a|\underline{d},\underline{i})\right)\label{eq:ergodic_state_action_occupancy_measure}
	\end{eqnarray}
denote the {\em ergodic state-action occupancy measure} under $\pi^\lambda$.

\section{Converse: Lower Bound}\label{sec:lower_bound}
In this section, we present a lower bound for \eqref{eq:main_quantity}. Given two probability distributions $\mu$ and $\nu$ on $\mathcal{S}$, the Kullback-Leibler (KL) divergence (also called the {\em relative entropy}) between $\mu$ and $\nu$ is defined as
\begin{equation}
	D(\mu\|\nu)\coloneqq \sum\limits_{i\in\mathcal{S}}\mu(i)\log \frac{\mu(i)}{\phi(i)},\label{eq:D(mu||nu)}
\end{equation}
where, by convention, $0\log \frac{0}{0}=0$. Also, given a TPM $P$ on $\mathcal{S}$, an integer $d\geq 1$, and $i, j\in \mathcal{S}$, let $P^d(j|i)$ denote the $(i, j)$th entry of the matrix $P^d$.

\begin{prop}
	\label{prop:lower_bound}
	Under the arms configuration $C=(h, P_1, P_2)$,
	\begin{equation}
		\liminf\limits_{\epsilon\downarrow 0}\inf\limits_{\pi\in\Pi(\epsilon)}\frac{E^\pi[\tau(\pi)| C]}{\log(1/\epsilon)}\geq \frac{1}{R^*(h,P_1,P_2)},\label{eq:lower_bound}
	\end{equation}
	where $R^*(h,P_1,P_2)$ is given by
	\begingroup \allowdisplaybreaks\begin{align}
		R^*(h,P_1,P_2)
	\coloneqq \sup\limits_{\pi^\lambda\in\Pi_{\textsf{SRS}}}~ \inf\limits_{\substack{C'=(h', P_1', P_2'):\\h'\neq h,\\P_1'\neq P_2'}} ~ \sum\limits_{(\underline{d},\underline{i})\in\mathbb{S}}~\sum\limits_{a=1}^{K}  ~\nu^\lambda(\underline{d},\underline{i},a) \,\textcolor{black}~{k_{CC'}(\underline{d}, \underline{i}, a)},\label{eq:R_delta^*(h,P_1,P_2)}
	\end{align}\endgroup
	with
	\begin{align}
		\textcolor{black}{k_{CC'}(\underline{d}, \underline{i}, a)} & \coloneqq D((P_C^a)^{d_a}(\cdot | i_a)\| (P_{C'}^a)^{d_a}(\cdot| i_a))\nonumber\\
		&=
		\begin{cases}
			D(P_1^{d_a}(\cdot|i_a)\|(P_2')^{d_a}(\cdot|i_a)),&a=h,\\
			{D(P_2^{d_a}(\cdot|i_a)\|(P_1')^{d_a}(\cdot|i_a))},&a=h',\\
			D(P_2^{d_a}(\cdot|i_a)\|(P_2')^{d_a}(\cdot|i_a)),& a\neq h,h'.
		\end{cases}\label{eq:k(a,d_a,i_a)}
	\end{align}
The infimum in \eqref{eq:R_delta^*(h,P_1,P_2)} is over all alternative odd arm configurations $C'=(h', P_1', P_2')$ satisfying (a) $h'\neq h$, and (b) $P_1'\neq P_2'$.
\end{prop}
\begin{IEEEproof}
	See Appendix \ref{appndx:proof_of_prop_lower_bound}.
\end{IEEEproof}
Notice that closer the TPMs $P_1$ and $P_2$ are (in terms of relative entropies of the corresponding rows), the smaller is the value of $R^*(h, P_1, P_2)$, resulting in a larger value of the lower bound \eqref{eq:lower_bound}. The key ingredients in the proof of the lower bound are (a) a data processing inequality for the setting of restless arms based on a change of measure argument presented in \cite{Kaufmann2016}, (b) a Wald-type lemma for Markov processes, (c) a recognition of the fact that for any $(\underline{d}, \underline{i})\in \mathbb{S}$, the long-term fraction of exits from $(\underline{d}, \underline{i})$ matches the long-term fraction of entries to  $(\underline{d}, \underline{i})$, and (d) the restriction of the supremum in \eqref{eq:R_delta^*(h,P_1,P_2)} to the class of SRS policies, which is possible thanks to an analogue of \cite[Theorem 8.8.2]{puterman2014markov} for countable-state controlled Markov processes. A formal statement of this theorem as applicable to this paper may be found in  \cite[Appendix H]{karthik2021detecting}.

\subsection{Simplifying $R^*(h, P_1, P_2)$}
Note that
\begin{align}
	& R^*(h,P_1,P_2)\nonumber\\
	&= \sup\limits_{\pi^\lambda\in\Pi_{\textsf{SRS}}}\, \inf\limits_{\substack{C'=(h', P_1', P_2'):\\h'\neq h,\\P_1'\neq P_2'}}\sum\limits_{(\underline{d},\underline{i})\in\mathbb{S}}\bigg[\nu^\lambda(\underline{d},\underline{i},h) \,D(P_1^{d_h}(\cdot|i_h)\|(P_2')^{d_h}(\cdot|i_h)) + \nu^\lambda(\underline{d},\underline{i},h') \,D(P_2^{d_{h'}}(\cdot|i_{h'})\|(P_1')^{d_{h'}}(\cdot|i_{h'})) \nonumber\\
	&\hspace{8cm}+ \sum\limits_{a\neq h, h'}\nu^\lambda(\underline{d},\underline{i},a) \,D(P_2^{d_a}(\cdot|i_a)\|(P_2')^{d_a}(\cdot|i_a))\bigg].
	\label{eq:simplifying_R*_1}
\end{align}
Because $P_1'$ appears only in the second term within the square brackets, it follows that the infimum over all $P_1'$ of this term is equal to zero and may be achieved by setting $P_1'=P_2$. Therefore,
\begin{align}
	& R^*(h,P_1,P_2)\nonumber\\
	&= \sup\limits_{\pi^\lambda\in\Pi_{\textsf{SRS}}}~ \inf\limits_{\substack{h', \, P_2':\\h'\neq h,~P_2'\neq P_2}}~\sum\limits_{(\underline{d},\underline{i})\in\mathbb{S}}\bigg[\nu^\lambda(\underline{d},\underline{i},h) \,D(P_1^{d_h}(\cdot|i_h)\|(P_2')^{d_h}(\cdot|i_h))+ \sum\limits_{a\neq h, h'}\nu^\lambda(\underline{d},\underline{i},a) \,D(P_2^{d_a}(\cdot|i_a)\|(P_2')^{d_a}(\cdot|i_a))\bigg].
	\label{eq:simplifying_R*_2}
\end{align}
It is worth comparing \eqref{eq:simplifying_R*_2} with  \cite[Eq. (19), pp. 4328]{pnkarthik2019learning}, the final expression for the lower bound when the arms are rested. The absence of arm delays in \cite[Eq. (19), pp. 4328]{pnkarthik2019learning} makes it possible to simplify this term further. In fact, a close examination of the steps presented in \cite[pp. 4337-4338]{pnkarthik2019learning} shows that when the arms are rested, the inner infimum over all $h'\neq h$ and $P_2'\neq P_2$ may be simplified by first computing the infimum over all $h'\neq h$ to arrive at the expression  \cite[Eq. (91), pp. 4338]{pnkarthik2019learning}, following which the optimal value of $P_2'$ (the one that attains the infimum over all $P_2'\neq P_2$) may be obtained using the method of Lagrange multipliers. The final expression for the optimal value of $P_2'$ is given by \cite[Eq. (20), pp. 4328]{pnkarthik2019learning}.

The presence of ergodic state-action occupancy measures inside the summation in \eqref{eq:simplifying_R*_2} does not allow the simplification of infimum over $P_2'\neq P_2$, as was possible in the setting of rested arms. Further, because of the presence of arm delays in the relative entropy terms in \eqref{eq:simplifying_R*_2}, it may not be possible to obtain a closed-form expression for the choice of $P_2'$ that attains the inner infimum in \eqref{eq:simplifying_R*_2}.
\subsection{Near-Optimal Solutions to the Supremum in \eqref{eq:simplifying_R*_2}}
It is not clear if there exists an optimal SRS policy $\pi^\lambda$ that attains the supremum in \eqref{eq:R_delta^*(h,P_1,P_2)}. However, for each $\delta>0$, there exists $\lambda=\lambda_{h, P_1, P_2, \delta}(\cdot|\cdot) \in \bigotimes \limits_{(\underline{d},\underline{i})\in\mathbb{S}} \mathcal{P}(\mathcal{A})$ such that
\begin{equation}
\inf\limits_{\substack{h', \, P_2':\\h'\neq h,~P_2'\neq P_2}}~ \sum\limits_{(\underline{d},\underline{i})\in\mathbb{S}}~\sum\limits_{a=1}^{K}  \nu^\lambda(\underline{d},\underline{i},a) ~\textcolor{black}{k_{CC'}(\underline{d}, \underline{i}, a)} \geq \frac{R^*(h, P_1, P_2)}{1+\delta}.
	\label{eq:lambda_h_P1_P2_delta_defining_equation}
\end{equation}
Call $\lambda_{h, P_1, P_2, \delta}$ a {\em $\delta$-optimal solution} for the arms configuration $C=(h, P_1, P_2)$.
More generally, let $\lambda_{h, P, Q, \delta}$ denote a $\delta$-optimal solution for $C=(h, P, Q)$.
Notice that more than one $\lambda$ may satisfy \eqref{eq:lambda_h_P1_P2_delta_defining_equation}.

In the next section, we show that under some regularity on the choice of the $\delta$-optimal solutions for the various possible arms configurations, a time-varying policy based on certainty equivalence and the choice of the $\delta$-optimal solutions achieves the lower bound asymptotically.

\section{Achievability}
\label{sec:achievability}
We begin this section by stating two key assumptions that form the basis for the results to be stated later.
%This section is organised as follows. First, we state two key assumptions that form the basis for the results of this section. The first of these is a regularity assumption on the $\delta$-optimal solutions, and the second a technical condition on the TPMs of any arms configuration. Next, we present a test statistic and use it to construct a policy based on the principle of certainty equivalence with ML estimation. We show that the ML estimates converge to their true values (identification). Lastly, we show that the expected stopping time of the policy satisfies an upper bound that is arbitrarily close to the lower bound. The proof of identification exploits the technical condition on the TPMs, while the proof of the upper bound relies on the regularity assumption.

\subsection{Two Key Assumptions}
Given $\delta>0$, \eqref{eq:lambda_h_P1_P2_delta_defining_equation} suggests that in order to approach the constant $R^*(h, P_1, P_2)$ in the lower bound to within a factor of $1/(1+\delta)$, the arms must eventually be sampled according to $\lambda_{h, P_1, P_2, \delta}$ or one of the $\delta$-optimal solutions for $C=(h, P_1, P_2)$. Because the underlying arms configuration is unknown and may be any one among the uncountably infinite collection $\{C=(h, P, Q):h\in \mathcal{A}, P\neq Q\}$ of all possible arms configurations, at best one can hope for a $(P, Q)$ that approaches $(P_1, P_2)$. A feasible option is to sample the arms according to $\lambda_{h, P, Q, \delta}$ and hope that it is eventually close to $\lambda_{h, P_1, P_2, \delta}$ when $(P, Q)$ is close to $(P_1, P_2)$. For this to work, we need a regularity condition.
\begin{assumption}[Continuous selection]
\label{assmptn:continuous_selection}
For each $\delta>0$, there exists a selection of $\delta$-optimal solutions $\{\lambda_{h, P, Q, \delta}: h\in \mathcal{A}, P\neq Q\}$ such that for each $h\in \mathcal{A}$, the mapping $(P, Q) \mapsto \lambda_{h, P, Q, \delta}$ is continuous under (a) the relative topology arising from the Euclidean metric on the domain set, and (b) the product topology on the range set.
\end{assumption}

The paper \cite{prabhu2020sequential} considers a similar assumption as above for a more general sequential hypothesis testing problem in multi-armed bandits, but with independent observations (see \cite[Assumption A]{prabhu2020sequential}). Also, the analogue of Assumption \ref{assmptn:continuous_selection} for the maximisers, instead of $\delta$-optimal solutions, holds in the settings of the prior works \cite{vaidhiyan2017learning, prabhu2017optimal, pnkarthik2019learning} as a consequence of Berge's maximum theorem \cite{ausubel1993generalized}.
Henceforth, for each $\delta>0$, fix a selection $\{\lambda_{h, P, Q, \delta}: h\in \mathcal{A}, P\neq Q\}$ of $\delta$-optimal solutions satisfying Assumption \ref{assmptn:continuous_selection}.

Let $\mathscr{P}({S})$ denote the space of all TPMs on the finite set $\mathcal{S}$. For $\bar{\varepsilon}^*\in (0,1)$, let
\begin{align}
	\mathscr{P}(\bar{\varepsilon}^*) &\coloneqq \{P\in \mathscr{P}(\mathcal{S}):~P\text{ is ergodic},~\forall~d\geq 1, ~i, j\in \mathcal{S},~ P^d(j|i)>0 \implies P^d(j|i) \geq \bar{\varepsilon}^*\}.
	\label{eq:P(bar_varepsilon)}
\end{align}
Eq. \eqref{eq:P(bar_varepsilon)} defines the class of all ergodic $P\in \mathscr{P}(\mathcal{S})$ such that each non-zero entry of $P^d$ is lower bounded by $\bar{\varepsilon}^*$ uniformly in $d$. Clearly, every ergodic $P$ belongs to $\mathscr{P}(\bar{\varepsilon}^*) $ for some $P$-dependent $\bar{\varepsilon}^*$. To see this, fix an arbitrary ergodic $P$, and let $\mu=(\mu(j):j\in \mathcal{S})$ be the unique stationary distribution for $P$. From \cite[Theorem 4.9]{levin2017markov}, we have $P^d(j|i)\to \mu(j)>0$ as $d\to \infty$ for all $i, j\in \mathcal{S}$. Let $\mu_{\textsf{min}}=\min_j \mu(j)$. Then, there exists $D$ such that $\forall$ $d\geq D$, each non-zero entry of $P^d$ is lower bounded by $\mu_{\textsf{min}}/2$. Further, let $$p_{\textsf{min}}\coloneqq \min\{P^d(j|i)> 0: i, j\in \mathcal{S}, ~d<D\}.$$ Then, we have $P\in \mathscr{P}(\bar{\varepsilon}^*)$, with $\bar{\varepsilon}^*=\min\{p_{\textsf{min}}, \mu_{\textsf{min}}/2\}$. Our next assumption however requires this to hold uniformly across all possible pairs $P, Q$ that can arise  in our problem. Define
\begin{equation}
	\mathcal{C}(\bar{\varepsilon}^*)\coloneqq \{(P, Q):~P(\cdot|i)\text{ is mutually absolutely continuous with }Q(\cdot|i)\text{ for all }i\in \mathcal{S},~P\in \mathscr{P}(\bar{\varepsilon}^*), ~Q\in \mathscr{P}(\bar{\varepsilon}^*)\},
	\label{eq:C(bar{varepsilon}^*)}
\end{equation}
%and let
%\begin{equation}
%	\mathcal{C}\coloneqq \bigcup\limits_{\bar{\varepsilon}^*>0} \mathcal{C}(\bar{\varepsilon}^*).
%	\label{eq:C}
%\end{equation}

% if $P$ has strictly positive entries, then $P\in \mathscr{P}(\bar{\varepsilon}^*)$ with $\bar{\varepsilon}^*=\min\{P(j|i):i, j\in \mathcal{S}\}$. Furthermore, if $P\in \mathscr{P}(\bar{\varepsilon}^*)$, then $P\in \mathscr{P}(\bar{\varepsilon}^*)$ for all $\bar{\varepsilon}<\bar{\varepsilon}^*$. To see how rich the class $\mathscr{P}(\bar{\varepsilon}^*)$ for any given $\bar{\varepsilon}^*$ is, consider the simple case when $|\mathcal{S}|=2$, and note that for all $\alpha, \beta\in (0, 1/2)$ such that $\alpha^2\beta^2>\bar{\varepsilon}^*$, $$P_{\alpha, \beta}=\begin{pmatrix}
%	\alpha & 1-\alpha \\
%	\beta & 1-\beta
%\end{pmatrix} \in \mathscr{P}(\bar{\varepsilon}^*).$$
%Additionally, for all $\alpha\in (0,1/2)$ such that $\alpha^2>\bar{\varepsilon}^*$, $$P_{\alpha}=\begin{pmatrix}
%	\alpha & 1-\alpha \\
%	1 & 0
%\end{pmatrix} \in \mathscr{P}(\bar{\varepsilon}^*).$$

%Consider the following assumption on the TPMs of any arms configuration.
\begin{assumption}
\label{assmptn:technical_assumption_on_TPMs}
	There exists $\bar{\varepsilon}^*\in (0,1)$ such that for every arms configuration $C=(h, P, Q)$, $(P, Q)\in \mathcal{C}(\bar{\varepsilon}^*)$.
\end{assumption}
Some remarks are in order. An arms configuration $C=(h, P, Q)$ satisfying Assumption \ref{assmptn:technical_assumption_on_TPMs}
only increases the difficulty of finding the odd arm index $h$. If $P, Q\in \mathscr{P}(\bar{\varepsilon}^*)$ for some $\bar{\varepsilon}^*>0$, then they are harder to ``distinguish'' from one another. To see this, note that for any ergodic $P, Q$, we know from \cite[Proposition 2.4]{levin2017markov} that that there exists $M=M(P, Q)$ such that all the entries of $P^d$ and $Q^d$ are strictly positive $\forall$ $d\geq M$, thus implying that $\forall$ $d\geq M$ and $i \in \mathcal{S}$,
\begin{equation}
	D(P^d(\cdot|i)\|Q^d(\cdot|i))<\infty,\quad D(Q^d(\cdot|i)\|P^d(\cdot|i))<\infty.
	\label{eq:relative_entropy_terms_finite}
\end{equation}
For $d<M$, it may be the case that one or both of the relative entropy terms in \eqref{eq:relative_entropy_terms_finite} equals $+\infty$ and discrimination becomes easier. However, when $P, Q\in \mathscr{P}(\bar{\varepsilon}^*)$, it follows that $\forall$ $d\geq 1$, each row of $P^d$ is mutually absolutely continuous with the corresponding row of $Q^d$. Furthermore, $\forall$ $d\geq 1$ and $i,j \in \mathcal{S}$ such that $P^d(j|i)>0$, $Q^d(j|i)>0$, the relation
\begin{equation}
	\bar{\varepsilon}^* \leq \frac{P^d(j|i)}{Q^d(j|i)}\leq  \frac{1}{\bar{\varepsilon}^*}
	\label{eq:a_desired_relation}
\end{equation}
holds, thus implying that $\forall d\geq 1$, each of the relative entropy terms in \eqref{eq:relative_entropy_terms_finite} is at most $\log (1/\bar{\varepsilon}^*)$. Therefore, $P, Q\in  \mathscr{P}(\bar{\varepsilon}^*)$ cannot have an arbitrarily large separation (in terms of relative entropy) and are harder to distinguish from one another.

It is worth noting here that Assumption \ref{assmptn:technical_assumption_on_TPMs} is equivalent to \cite[Assumption I]{borkar1979adaptive} which, in the context of this paper, states that there exists $\bar{\epsilon}>0$ such that for all $(\underline{d}, \underline{i}), (\underline{d}', \underline{i}') \in \mathbb{S}$, $b\in \mathcal{A}$ and $C=(h, P_1, P_2)$,
		$$
		Q_{C}(\underline{d}', \underline{i}'\mid \underline{d}, \underline{i}, b)=0 \quad \text{ or }\quad  Q_{C}(\underline{d}', \underline{i}'\mid \underline{d}, \underline{i}, b)>\bar{\epsilon}.
		$$
		Indeed, it is easy to see from \eqref{eq:MDP_transition_probabilities} that
		\begin{equation}
			Q_{C}(\underline{d}', \underline{i}'\mid \underline{d}, \underline{i}, b)=\sum\limits_{a=1}^{K}\left(\frac{\eta}{K}+(1-\eta)\mathbb{I}_{\{a=b\}}\right)~(P_C^a)^{d_a}(i_a'|i_a)~\mathbb{I}_{\{d_a'=1\text{ and }d'_{\tilde{a}}=d_{\tilde{a}}+1\text{ for all }\tilde{a}\neq a\}}~\mathbb{I}_{\{i_{\tilde{a}}'=i_{\tilde{a}}\text{ for all }\tilde{a}\neq a\}},
			\label{eq:MDP_TPM_2}
		\end{equation}
where $P_C^a$ in \eqref{eq:MDP_TPM_2} is as defined in \eqref{eq:P_C^a}. It follows that if $Q_C(\underline{d}', \underline{i}' \mid \underline{d}, \underline{i}, b)>0$, then it must be true that $d_a'=1$, $d'_{\tilde{a}}=d_{\tilde{a}}+1\text{ for all }\tilde{a}\neq a$, and $i_{\tilde{a}}'=i_{\tilde{a}}\text{ for all }\tilde{a}\neq a$ for some $a\in \mathcal{A}$. In this case,	
		\begin{equation}
			Q_{C}(\underline{d}', \underline{i}'\mid \underline{d}, \underline{i}, b)=\left(\frac{\eta}{K}+(1-\eta)\mathbb{I}_{\{a=b\}}\right)~(P_C^a)^{d_a}(i_a'|i_a).
			\label{eq:MDP_TPM_1}
		\end{equation}
		Because the term within brackets in \eqref{eq:MDP_TPM_1} is $\geq \frac{\eta}{K}>0$, it follows that any lower bound on $Q_{C}(\underline{d}', \underline{i}'\mid \underline{d}, \underline{i}, b)$ implies a lower bound on $(P_C^a)^{d_a}(i_a'|i_a)$ and vice-versa. This establishes the equivalence between Assumption \ref{assmptn:technical_assumption_on_TPMs} and \cite[Assumption I]{borkar1979adaptive}.

\subsection{Test Statistic}
We now introduce a test statistic and use it later to construct a policy based on certainty equivalence. The test statistic is based on a modification of the usual generalised likelihood ratio (GLR) test statistic in which the numerator of the usual GLR test statistic is replaced with an average likelihood computed with respect to an artificial prior. The details are as follows. Let $\mathcal{P}(\mathcal{S})$ denote the space of all probability distributions on the set $\mathcal{S}$, and let $\text{Dir}(\alpha_j:j\in \mathcal{S})$ denote the Dirichlet prior on $\mathcal{P}(\mathcal{S})$ with parameters $(\alpha_j:j\in \mathcal{S})$.  In particular, let $\text{Dir}(\mathbf{1})$ denote the Dirichlet distribution with $\alpha_j=1$ $\forall$ $j\in \mathcal{S}$. {\color{black} Notice that $\text{Dir}(\mathbf{1})$ is simply the uniform probability distribution on the probability simplex $\mathscr{P}(\mathcal{S})$.} Let $D$ denote the prior on $\mathscr{P}(\mathcal{S})$ induced by $\text{Dir}(\mathbf{1})$ when each row of $P\in \mathscr{P}(\mathcal{S})$ is sampled independently according to $\text{Dir}(\mathbf{1})$.

Given $C=(h, P, Q)$, let $f(B^n, A^n, \bar{X}^n|C)$ denote the likelihood of all the arm samples and observations up to time $n$ under the arms configuration $C$. Let $\bar{f}(B^n, A^n, \bar{X}^n|\mathcal{H}_h)$ denote the average likelihood of all the arm samples and observations up to time $n$ under the hypothesis $\mathcal{H}_h$, where the averaging is over $(P, Q) \stackrel{iid}{\sim} D$, i.e.,
\begin{equation}
	\bar{f}(B^n, A^n, \bar{X}^n|\mathcal{H}_h) = \int\limits_{\mathscr{P}(\mathcal{S})\times \mathscr{P}(\mathcal{S})}~f(B^n, A^n, \bar{X}^n|C=(h, P, Q))~D(P)~D(Q)~dP~dQ.
	\label{eq:average_likelihood}
\end{equation}
Further, let $\hat{f}(B^n, A^n, \bar{X}^n|\mathcal{H}_h)$ denote the maximum likelihood of all the arm samples and observations up to time $n$ under the hypothesis $\mathcal{H}_h$, i.e.,
\begin{equation}
	\hat{f}(B^n, A^n, \bar{X}^n|\mathcal{H}_h)=\sup\limits_{\substack{P, Q}}~f(B^n, A^n, \bar{X}^n|C=(h, P, Q)).
	\label{eq:maximum_likelihood}
\end{equation}
{\color{black} The supremum in \eqref{eq:maximum_likelihood} is over the compact set $\mathscr{P}(\mathcal{S})\times\mathscr{P}(\mathcal{S})$, and is attained because the likelihood is a continuous function of the TPMs.}
Let $(\hat{P}_{h, 1}(n),\hat{P}_{h, 2}(n))$ attain the supremum in \eqref{eq:maximum_likelihood}. These are the ML estimates of the TPMs under the hypothesis $\mathcal{H}_h$ (whose closed-form expressions are not available in the general case).
{\color{black} Notice that we do not impose the constraint $P\neq Q$ while computing the supremum in \eqref{eq:maximum_likelihood} for the following reason. We show later that under an arms configuration $C=(h, P_1, P_2)$, the convergences $\hat{P}_{h, 1}(n)\to P_1$, $\hat{P}_{h, 2}(n)\to P_2$ hold as $n\to \infty$ almost surely. Because $P_1\neq P_2$ by the definition of arms configuration, the constraint $\hat{P}_{h, 1}(n)\neq \hat{P}_{h, 2}(n)$ is almost surely automatically satisfied for all $n$ sufficiently large.}
%Due to the presence of arm delays in the likelihood function, closed form expressions for the ML estimates are not available. See \cite{zzz} for more details.

For $h, h'\in \mathcal{A}$ such that $h\neq h'$, our test statistic, which we denote by $M_{hh'}(n)$ at time $n$, is defined as
\begin{equation}
	M_{hh'}(n)\coloneqq \log\frac{\bar{f}(B^n, A^n, \bar{X}^n|\mathcal{H}_h)}{\hat{f}(B^n, A^n, \bar{X}^n|\mathcal{H}_{h'})}.
	\label{eq:modified_GLR_test_statistic}
\end{equation}
Before presenting the exact mathematical expression for \eqref{eq:modified_GLR_test_statistic}, we introduce a few notations. Given a policy $\pi$ and an arms configuration $C$, let $Z_C^\pi(n)$ denote the log-likelihood of all the arm samples and the observations up to time $n$ under the policy $\pi$ and the arms configuration $C$. That is,
\begin{align}
Z^\pi_C(n) &=\log f(B^n, A^n, \bar{X}^n|C)\nonumber\\
&= \sum\limits_{a=1}^{K} \log P^\pi(X_{a-1}^a|C) + \sum\limits_{t=K}^{n} \log P^\pi(B_t, A_t, \bar{X}_t\mid B^{t-1}, A^{t-1}, \bar{X}^{t-1}, C) \nonumber\\
&=  \sum\limits_{a=1}^{K} \log P^\pi(X_{a-1}^a|C) + \sum\limits_{t=K}^{n} \log P^\pi(B_t, A_t\mid B^{t-1}, A^{t-1}, \bar{X}^{t-1}, C) + \sum\limits_{t=K}^{n} \log P^\pi(\bar{X}_t\mid B^t, A^t, \bar{X}^{t-1}, C)\nonumber\\
&= \sum\limits_{a=1}^{K} \log P^\pi(X_{a-1}^a|C) + \sum\limits_{t=K}^{n} \log P^\pi(B_t, A_t\mid B^{t-1}, A^{t-1}, \bar{X}^{t-1}, C) + \sum\limits_{t=K}^{n} \log P^\pi(\bar{X}_t\mid A^t, \bar{X}^{t-1}, C).
\label{eq:Z^pi_C(n)_1}
\end{align}
In writing \eqref{eq:Z^pi_C(n)_1}, we use the fact that $\bar{X}_t$ is conditionally independent of $B^t$, conditioned on the actually sampled arm $A_t$, $A^{t-1}$ and $\bar{X}^{t-1}$. The last summation term in \eqref{eq:Z^pi_C(n)_1} may be written as
\begin{align}
\sum\limits_{t=K}^{n} \log P^\pi(\bar{X}_t\mid A^t, \bar{X}^{t-1}, C) &= \sum\limits_{(\underline{d}, \underline{i}) \in \mathbb{S}} ~\sum\limits_{a=1}^{K}~\sum\limits_{j\in \mathcal{S}}~\sum\limits_{t=K}^{n}  ~\mathbb{I}_{\{\underline{d}(t)=\underline{d}, \underline{i}(t)=\underline{i}, A_t=a, \bar{X}_t^a=j\}} ~\log P^\pi(\bar{X}_t\mid A^t, \bar{X}^{t-1}, C) \nonumber\\
&\stackrel{(a)}{=} \sum\limits_{(\underline{d}, \underline{i}) \in \mathbb{S}} ~\sum\limits_{a=1}^{K}~\sum\limits_{j\in \mathcal{S}}~\sum\limits_{t=K}^{n} ~ \mathbb{I}_{\{\underline{d}(t)=\underline{d}, \underline{i}(t)=\underline{i}, A_t=a, \bar{X}_t^a=j\}} ~\log \, (P_C^a)^{d_a}(j|i_a) \nonumber\\
&=  \sum\limits_{(\underline{d}, \underline{i}) \in \mathbb{S}} ~\sum\limits_{a=1}^{K}~\sum\limits_{j\in \mathcal{S}} N(n, \underline{d}, \underline{i}, a, j)~\log \, (P_C^a)^{d_a}(j|i_a),
\label{eq:Z^pi_C(n)_2}
\end{align}
where $(a)$ above follows by noting that when $\underline{d}(t)=\underline{d}$, $\underline{i}(t)=\underline{i}$, $A_t=a$ and $\bar{X}_t^a=j$,
\begin{align}
P^\pi(\bar{X}_t\mid A^t, \bar{X}^{t-1}, C) = P^\pi(X_t^a=j\mid X_{t-d_a}^a=i_a, C)=(P_C^a)^{d_a}(j|i_a).
\end{align}
Also, the term $N(n, \underline{d}, \underline{i}, a, j)$ in \eqref{eq:Z^pi_C(n)_2} is defined as
\begin{equation}
N(n, \underline{d}, \underline{i}, a, j) \coloneqq \sum\limits_{t=K}^{n} ~ \mathbb{I}_{\{\underline{d}(t)=\underline{d}, \underline{i}(t)=\underline{i}, A_t=a, \bar{X}_t^a=j\}} ,
\label{eq:N(n, d, i, a, j)}
\end{equation}
and denotes the number of times up to time $n$ the controlled Markov process $\{(\underline{d}(t), \underline{i}(t)): t\geq K\}$ is observed to be in the state $(\underline{d}, \underline{i})$, arm $a$ is sampled subsequently, and the state $j\in \mathcal{S}$ is observed on arm $a$. Plugging \eqref{eq:Z^pi_C(n)_2} into \eqref{eq:Z^pi_C(n)_1}, we get
\begin{align}
	& Z^\pi_C(n)\nonumber\\
	& = \sum\limits_{a=1}^{K} \log P^\pi(X_{a-1}^a|C) + \sum\limits_{t=K}^{n} \log P^\pi(B_t, A_t\mid B^{t-1}, A^{t-1}, \bar{X}^{t-1}, C) + \sum\limits_{(\underline{d}, \underline{i}) \in \mathbb{S}} ~\sum\limits_{a=1}^{K}~\sum\limits_{j\in \mathcal{S}} N(n, \underline{d}, \underline{i}, a, j)~\log \, (P_C^a)^{d_a}(j|i_a).
	\label{eq:Z^pi_C(n)_3}
\end{align}
Under an arms configuration $C=(h, P, Q)$, \eqref{eq:Z^pi_C(n)_3} may be written more explicitly as
\begin{align}
	Z^\pi_C(n)
	& = \sum\limits_{t=K}^{n} \log P^\pi(B_t, A_t\mid B^{t-1}, A^{t-1}, \bar{X}^{t-1}, C)\label{eq:Z^pi_C(n)_4_1}\\
	& + \log P^\pi(X_{h-1}^h| C) + \sum\limits_{(\underline{d}, \underline{i}) \in \mathbb{S}}~\sum\limits_{j\in \mathcal{S}} N(n, \underline{d}, \underline{i}, h, j)~\log \, P^{d_h}(j|i_h)\label{eq:Z^pi_C(n)_4_2}\\
	& + \sum\limits_{a\neq h} \log P^\pi(X_{a-1}^a|C) + \sum\limits_{(\underline{d}, \underline{i}) \in \mathbb{S}} ~\sum\limits_{a\neq h}~\sum\limits_{j\in \mathcal{S}} N(n, \underline{d}, \underline{i}, a, j)~\log \, Q^{d_a}(j|i_a).\label{eq:Z^pi_C(n)_4_3}
\end{align}
{\color{black} Suppose that $X_0^a\sim \phi$ $\forall$ $a\in \mathcal{A}$, where $\phi$ is a probability distribution $\phi$ on $\mathcal{S}$ that puts a strictly positive mass on each element of $\mathcal{S}$ (e.g., $\phi(i)=\frac{1}{|\mathcal{S}|}$ for all $i\in \mathcal{S}$) and is known to the decision entity beforehand}. Then, we have
\begin{align}
	Z^\pi_C(n)
	& = \sum\limits_{t=K}^{n} \log P^\pi(B_t, A_t\mid B^{t-1}, A^{t-1}, \bar{X}^{t-1}, C)\label{eq:Z^pi_C(n)_5_1}\\
	& + \log \left(\sum\limits_{i\in \mathcal{S}}~\phi(i)\, P^{h-1}(X_{h-1}^h|i)\right) + \sum\limits_{(\underline{d}, \underline{i}) \in \mathbb{S}}~\sum\limits_{j\in \mathcal{S}} N(n, \underline{d}, \underline{i}, h, j)~\log \, P^{d_h}(j|i_h)\label{eq:Z^pi_C(n)_5_2}\\
	& + \sum\limits_{a\neq h} \log \left(\sum\limits_{i\in \mathcal{S}}~\phi(i)\,Q^{a-1}(X_{a-1}^a|i)\right) + \sum\limits_{(\underline{d}, \underline{i}) \in \mathbb{S}} ~\sum\limits_{a\neq h}~\sum\limits_{j\in \mathcal{S}} N(n, \underline{d}, \underline{i}, a, j)~\log \, Q^{d_a}(j|i_a).\label{eq:Z^pi_C(n)_5_3}
\end{align}
With the above notations in place, \eqref{eq:modified_GLR_test_statistic} may be written as
\begin{equation}
	M_{hh'}(n)=T_1(n)+T_2(n)+T_3(n)+T_4(n),
	\label{eq:M_hh'(n)_as_sum_of_4_terms}
\end{equation}
where the terms $T_1(n)$, $T_2(n)$, $T_3(n)$, and $T_4(n)$ are as given below.
\begin{enumerate}
	\item The term $T_1(n)$ is given by
	\begin{equation}
 	T_1(n)=\log \mathbb{E}\bigg[\exp\bigg(\log \left(\sum\limits_{i\in \mathcal{S}}~\phi(i)\, P^{h-1}(X_{h-1}^h|i)\right) + \sum\limits_{(\underline{d}, \underline{i})\in \mathbb{S}}~\sum\limits_{j\in \mathcal{S}}N(n, \underline{d}, \underline{i}, h, j)~ \log P^{d_h}(j|i_h)\bigg)\bigg],
 	\label{eq:T_1(n)}
    \end{equation}
    where the expectation is with respect to $P\sim D$.
	
	\item The term $T_2(n)$ is given by
	\begin{equation}
		T_2(n) = \log \mathbb{E}\bigg[\exp\bigg(\sum\limits_{a\neq h} \log \left(\sum\limits_{i\in \mathcal{S}}~\phi(i)\,Q^{a-1}(X_{a-1}^a|i)\right) +\sum\limits_{a\neq h}~\sum\limits_{(\underline{d}, \underline{i})\in \mathbb{S}}~\sum\limits_{j\in \mathcal{S}}N(n, \underline{d}, \underline{i}, a, j)~ \log Q^{d_a}(j|i_a)\bigg)\bigg],
		\label{eq:T_2(n)}
	\end{equation}
	where the expectation is with respect to $Q\sim D$.

    \item The term $T_3(n)$ is given by
    \begin{equation}
    	T_3(n) = \log \frac{1}{\sum\limits_{i\in \mathcal{S}}~\phi(i)~\hat{P}_{h', 1}(n)(X_{h'-1}^{h'}|i)} + \sum\limits_{(\underline{d}, \underline{i})\in \mathbb{S}}~\sum\limits_{j\in \mathcal{S}}N(n, \underline{d}, \underline{i}, h', j)~ \log \frac{1}{(\hat{P}^n_{h',1})^{d_{h'}}(j|i_{h'})}.
    	\label{eq:T_3(n)}
    \end{equation}

    \item The term $T_4(n)$ is given by
    \begin{equation}
    	T_4(n) = \sum\limits_{a\neq h'}~\log \frac{1}{\sum\limits_{i\in \mathcal{S}}~\phi(i)~\hat{P}^n_{h',2}(X_{a-1}^{a}|i)} + \sum\limits_{a\neq h'}~\sum\limits_{(\underline{d}, \underline{i})\in \mathbb{S}}~\sum\limits_{j\in \mathcal{S}}N(n, \underline{d}, \underline{i}, a, j)~ \log \frac{1}{(\hat{P}^n_{h',2})^{d_a}(j|i_a)}.
    	\label{eq:T_4(n)}
    \end{equation}
\end{enumerate}
{\color{black} In writing the sum $T_1(n)+T_2(n)$ in \eqref{eq:M_hh'(n)_as_sum_of_4_terms}, we use the fact that $P$ and $Q$ are sampled independently according to the prior $D$}. Notice that the term \eqref{eq:Z^pi_C(n)_5_1} does not appear in the expression for $M_{hh'}(n)$. This is because $\pi$ is oblivious to the knowledge of the underlying arms configuration $C$, as a result of which at any given time $t$, the probability of selecting arm $B_t$ based on the history $(B^{t-1}, A^{t-1}, \bar{X}^{t-1})$ is independent of $C$. Therefore, \eqref{eq:Z^pi_C(n)_5_1}, which appears in both the numerator and the denominator of \eqref{eq:modified_GLR_test_statistic}, cancels out.

We shall refer to $M_{hh'}(n)$ as the {\em modified GLR test statistic} of hypothesis $\mathcal{H}_h$ with respect to $\mathcal{H}_{h'}$ at time $n$.
%The modification comes from replacing the maximum likelihood in numerator of the classical GLR test statistic with the average likelihood computed with respect to the artificial prior $D$.
Let $M_h(n)=\min\limits_{h'\neq h}~M_{hh'}(n)$ denote the modified GLR test statistic of hypothesis $\mathcal{H}_h$ with respect to its nearest alternative hypothesis.

\subsection{Policy Based on Certainty Equivalence}
%Given $\gamma>0$ and a probability distribution $\mu$ on $\mathcal{A}$, define the {\em $\gamma$-randomisation} of $\mu$, denoted $\mu^\gamma$, as
%	\begin{equation}
%		\mu^\gamma\coloneqq \texttt{Unif}(B(\mu, \gamma) \cap \mathcal{P}(\mathcal{A})).
%		\label{eq:gamma_randomization}
%	\end{equation}
%In \eqref{eq:gamma_randomization}, \texttt{Unif} denotes uniform distribution, $B(x, r)$ the Euclidean ball of radius $r$ centered at $x$, and $\mathcal	{P}(\mathcal{A})$ denotes the space of all probability distributions on the set $\mathcal{A}$.

Fix $L>1, \delta>0$. Our policy, which we denote by $\pi^\star(L, \delta)$, is as below with $L$ and $\delta$ as parameters.
\vspace{.1in}
\hrule

\vspace{.1in}

\noindent \textbf{\underline{\emph{Policy }$\pi^{\star}(L,\delta)$}}:
\vspace{.1in}\\
\noindent Without loss of generality, let $A_0=1$, $A_1=2$, and so on until $A_{K-1}=K$. Follow the below mentioned steps $\forall$ $n\geq K$.\\
\noindent (1) Compute $\theta(n)\in \arg\max\limits_{h\in \mathcal{A}}~ \min\limits_{h'\neq h}~M_{hh'}(n).$ Resolve ties, if any, uniformly at random.\\
\noindent (2) If $M_{\theta(n)}(n) \geq \log((K-1)L)$, stop further sampling and declare $\theta(n)$ as the odd arm.\\
\noindent (3) If $M_{\theta(n)}(n) < \log((K-1)L)$, sample arm $B_{n+1}$ according to $\lambda_{\theta(n), \hat{P}_{\theta(n), 1}(n), \hat{P}_{\theta(n), 2}(n), \delta}(\cdot\mid \underline{d}(n), \underline{i}(n))$.\\
\noindent (4) Update $n\leftarrow n+1$ and go back to (1).
\vspace{.1in}
\hrule
\vspace{0.1in}
In item (1) above, $\theta(n)$ denotes the guess of the odd arm at time $n$. In item (2), we check if $M_{\theta(n)}(n)$ has exceeded a certain fixed threshold $ (\geq\log((K-1)L))$. If this is the case, then the policy is confident that the true odd arm index is $\theta(n)$, in which case the policy terminates at time $n$ and outputs $\theta(n)$ as the odd arm index. If the condition in item (2) fails, the policy supposes that $C_n=(\theta(n), \hat{P}_{\theta(n), 1}(n), \hat{P}_{\theta(n), 2}(n))$ is the true arms configuration, and samples arm $B_{n+1}$ according to the $\delta$-optimal solution for $C_n$, thus following the principle of certainty equivalence. Observe that the policy does not rely on the knowledge of the constant $\bar{\varepsilon}^*$ from Assumption \ref{assmptn:technical_assumption_on_TPMs}.

{\color{black}\begin{remark}
\label{rem:incomputability_of_M_hh'(n)}
\begin{enumerate}
	\item The presence of arm delays in \eqref{eq:Z^pi_C(n)_5_1}-\eqref{eq:Z^pi_C(n)_5_3}, the expression for the log-likelihood function, makes obtaining the closed-form expressions for $(\hat{P}_{h, 1}(n), \hat{P}_{h, 2}(n))$ and the exact computation of $M_{hh'}(n)$ difficult. This difficulty can be circumvented by repeatedly sampling the Markov process of each arm and using the successive observations from each of the arms to estimate the TPMs. Because the Markov process of each arm is ergodic, these TPM estimates will converge to their true values asymptotically. However, this might be suboptimal in not achieving the lower bound \eqref{eq:lower_bound}. We shall see that $\pi^\star(L, \delta)$ samples the arms eventually according to $\lambda_{h, P_1, P_2, \delta}$, and therefore approaches the lower bound asymptotically as $L\to \infty$ and $\delta \rightarrow 0$.

%repeated sampling of the arms may not ensure that given $\delta>0$ and an arms configuration $C=(h, P_1, P_2)$, the arms are sampled eventually according to $\lambda_{h, P_1, P_2, \delta}$. This, as we shall see shortly, is crucial in order to achieve the lower bound \eqref{eq:lower_bound}. Whereas, in spite of the computational intractability of $M_{hh'}(n)$, we shall see that $\pi^\star(L, \delta)$ samples the arms eventually according to $\lambda_{h, P_1, P_2, \delta}$ and therefore meets the lower bound asymptotically as $L\to \infty$.
	
	\item Moreover, when sampling the arms repeatedly, it is not clear if the desired error probability criterion can be met. We shall see that the policy $\pi^\star(L, \delta)$ meets the desired error probability criterion for a suitable choice of $L$.

% [RS Comment: The following is too detailed at this stage.] A key step in the proof of this result is to upper bound the likelihood $f(B^n, A^n, \bar{X}^n|C)$ under an arms configuration $C=(h, P_1, P_2)$ by the maximum likelihood $\hat{f}(B^n, A^n, \bar{X}^n|\mathcal{H}_h)$. Clearly, this is not possible for a policy that does not use the maximum likelihood (for e.g., the policy that uses TPM estimates resulting from repeated sampling  of the arms).
\end{enumerate}
\end{remark}}

\subsection{Results on the Performance of the Policy}
We now present the results on the performance of the above described policy. The proofs are relegated to the Appendices.

Let Assumptions \ref{assmptn:continuous_selection} and \ref{assmptn:technical_assumption_on_TPMs} hold.
%The first result is on the convergence of the ML estimates of the TPMs converge to their true values.
\begin{prop}
	\label{prop:convergence_of_ML_estimates}
	Under the arms configuration $C=(h, P_1, P_2)$,
	\begin{equation}
		\hat{P}_{h, 1}(n)\longrightarrow P_1, \quad \hat{P}_{h, 2}(n)\longrightarrow P_2\quad \text{as }\quad n\to \infty \quad \text{almost surely.}
		\label{eq:convergence_of_ML_estimates_of_TPMs}
	\end{equation}
\end{prop}
\begin{IEEEproof}
See Appendix \ref{appndx:proof_of_prop_convergence_of_ML_estimates_of_TPMs}.	
\end{IEEEproof}
The proof of Proposition \ref{prop:convergence_of_ML_estimates} is based on verifying that the assumptions of \cite{borkar1982identification} hold in the context of this paper. The result then simply follows from \cite[Theorem 4.3]{borkar1982identification}. It is instructive to note here that the proof of \cite[Theorem 4.3]{borkar1982identification} is based on a notion of ``$\{\varepsilon_i\}$-randomisation'' which, for controlled Markov processes, ensures a strictly positive probability of choosing each  control at each time instant. The trembling hand model \eqref{eq:trembling_hand_relation} of this paper ensures that the probability of sampling an arm at any given time is $\geq \frac{\eta}{K}>0$, thus alleviating the need to consider $\{\varepsilon_i\}$-randomisations.

An immediate consequence of Proposition \ref{prop:convergence_of_ML_estimates} is the following: suppose $C=(h, P_1, P_2)$ is the underlying arms configuration. Then, for any $h'\in \mathcal{A}$ such that $h'\neq h$,
\begin{equation}
	\hat{P}_{h', 1}(n)\longrightarrow P_2, \quad \hat{P}_{h', 2}(n)\longrightarrow P\quad \text{as }\quad n\to \infty \quad \text{almost surely},
		\label{eq:convergence_of_ML_estimates_of_TPMs_under_h'}
\end{equation}
where $P$ in \eqref{eq:convergence_of_ML_estimates_of_TPMs_under_h'} is a transition probability matrix that is a function of $P_1$ and $P_2$, but whose closed-form expression is not available. It is worth noting here that when the arms are rested, the $(i, j)$th entry of $P$ is a convex combination of the $(i, j)$th entries of $P_1$ and $P_2$ whose closed-form expression given by \cite[Eq. (20)]{pnkarthik2019learning}.

The next result shows that the test statistic has a strictly positive drift under the non-stopping version of the policy (a policy, say $\pi_{\textsf{ns}}^\star(L, \delta)$, that never stops and picks an arm at time $n$ according to the rule in item (3) above).
\begin{prop}\label{prop:strict_positivity_of_drift_of_test_statistic}
	Fix $L > 1$ and $\delta>0$. Let $C=(h, P_1, P_2)$ be the underlying arms configuration. Under $\pi_{\textsf{ns}}^\star(L, \delta)$,  the non-stopping version of the policy $\pi^{\star}(L,\delta)$, $\forall$ $h'\neq h$, we have
	\begin{equation}
		\liminf\limits_{n \to \infty} \frac{M_{hh'}(n)}{n}>0 \quad \text{almost surely}.
		\label{eq:test_statistic_has_strictly_positive_drift}
	\end{equation}
\end{prop}
\begin{IEEEproof}
The proof uses the convergences in \eqref{eq:convergence_of_ML_estimates_of_TPMs}. See Appendix \ref{appndx:proof_of_prop_strict_positivity_of_drift_of_test_statistic} for the details.	
\end{IEEEproof}
An immediate consequence of Proposition \ref{prop:strict_positivity_of_drift_of_test_statistic} is that, almost surely, $\liminf\limits_{n\to\infty} M_h(n)>0$. For each $h'\in \mathcal{A}$, let $\pi^\star_{h'}(L,\delta)$ denote a version of the policy $\pi^\star(L, \delta)$ that waits until the event $M_{h'}(n)\geq \log((K-1)L)$ occurs, at which point it stops and always declares $h'$ as the index of the odd arm. Under the arms configuration $C=(h, P_1, P_2)$, it then follows that the stopping time of the policy $\pi^\star(L,\delta)$ may be upper bounded by that of $\pi^\star_h(L,\delta)$, i.e., $\tau(\pi^\star(L,\delta)) \leq \tau(\pi^\star_h(L,\delta))$ almost surely under $C=(h, P_1, P_2)$. As a consequence, under $C=(h, P_1, P_2)$, the following set of inequalities hold almost surely:
\begingroup\allowdisplaybreaks
\begin{align}
	\tau(\pi^\star(L,\delta)) &\leq \tau(\pi^\star_h(L,\delta))\nonumber\\
	&=\inf\{n\geq 1:M_h(n)\geq \log((K-1)L)\}\nonumber\\
	&\leq \inf\bigg\lbrace n\geq 1:M_{hh'}(n')\geq \log((K-1)L)\text{ for all }n'\geq n\text{ and for all }h'\neq h\bigg\rbrace\nonumber\\
	&<\infty,\label{eq:stopping_time_finite_almost_surely}
\end{align}\endgroup
where the last line above is due to Proposition \ref{prop:strict_positivity_of_drift_of_test_statistic}. This establishes that the policy $\pi^\star(L,\delta)$ stops in finite time almost surely.

For $h'\in \mathcal{A}$ such that $h'\neq h$, let
\begin{align}
	\textsf{GLR}_{hh'}(n)\coloneqq \log\frac{\hat{f}(B^n, A^n, \bar{X}^n|\mathcal{H}_h)}{\hat{f}(B^n, A^n, \bar{X}^n|\mathcal{H}_{h'})}
	\label{eq:GLR_test_statistic}
\end{align}
denote the GLR test statistic of hypothesis $\mathcal{H}_h$ with respect to the hypothesis $\mathcal{H}_{h'}$ at time $n$. Clearly, $\textsf{GLR}_{hh'}(n) \geq M_{hh'}(n)$ almost surely for all $n$. Also, $\textsf{GLR}_{h'h}(n)=-\textsf{GLR}_{hh'}(n)$. Then, almost surely,
\begingroup\allowdisplaybreaks\begin{align}
	\limsup\limits_{n\to\infty}M_{h'}(n)&=\limsup\limits_{n\to\infty}\min\limits_{a\neq h'}M_{h'a}(n)\nonumber\\
	&\leq \limsup\limits_{n\to\infty}M_{h'h}(n)\nonumber\\
	& \leq \limsup\limits_{n \to \infty} \textsf{GLR}_{h'h}(n)\nonumber\\
	&=\limsup\limits_{n\to\infty}-\textsf{GLR}_{hh'}(n)\nonumber\\
	&= -\liminf\limits_{n\to\infty}\textsf{GLR}_{hh'}(n)\nonumber\\
	& \leq -\liminf\limits_{n\to\infty}M_{hh'}(n)\nonumber\\
	&\leq -\liminf\limits_{n\to\infty}\min\limits_{a\neq h} M_{h,a}(n)\nonumber\\
	&= -\liminf\limits_{n\to\infty}M_{h}(n)\nonumber\\
	&<0,\label{eq:limsup_M_{h'}(n)_less_than_0}
\end{align}\endgroup
where \eqref{eq:limsup_M_{h'}(n)_less_than_0} follows from Proposition \ref{prop:strict_positivity_of_drift_of_test_statistic}. Under the arms configuration $C=(h, P_1, P_2)$ and under the policy $\pi_{\textsf{ns}}^\star(L, \delta)$, almost surely,
 \begin{equation}
 	\theta(n)=\arg\max\limits_{h\in\mathcal{A}}M_h(n)=h\quad \forall ~ n\text{ sufficiently large},\label{eq:h^*(n)_equal_to_h_almost_surely}
 \end{equation}
 which together with \eqref{eq:convergence_of_ML_estimates_of_TPMs} implies that
 \begin{equation}
 	\lambda_{\theta(n), \hat{P}_{\theta(n), 1}(n), \hat{P}_{\theta(n), 2}(n), \delta} \longrightarrow \lambda_{h, P_1, P_2, \delta} \quad \text{as }n\to \infty,
 	\label{eq:convergence_of_lambdas}
 \end{equation}
 where the convergence in \eqref{eq:convergence_of_lambdas} is with respect to the product topology on the space $\bigotimes\limits_{(\underline{d}, \underline{i})\in \mathbb{S}}\mathcal{P}(\mathcal{A})$, i.e.,
 \begin{equation}
 	\lambda_{\theta(n), \hat{P}_{\theta(n), 1}(n), \hat{P}_{\theta(n), 2}(n), \delta}(\cdot|\underline{d}, \underline{i})\to \lambda_{h, P_1, P_2, \delta}(\cdot|\underline{d}, \underline{i}) \quad \text{as }n\to \infty \quad \forall~(\underline{d}, \underline{i})\in \mathbb{S}.
 	\label{eq:convergence_of_lambdas_pointwise}
 \end{equation}
The property in \eqref{eq:convergence_of_lambdas} ensures that under the non-stopping policy $\pi_{\textsf{ns}}^\star(L, \delta)$, as more observations are collected from the arms and as the ML estimates of the TPMs approach their true values defined by the underlying arms configuration, the arms are eventually sampled according to the $\delta$-optimal selection for the underlying arms configuration. As we shall see shortly, it is this property of $\pi_{\textsf{ns}}^\star(L, \delta)$ that plays a key role in showing the asymptotic optimality of  the policy $\pi^\star(L, \delta)$.

Let us return to the policy $\pi^\star(L, \delta)$. The next result shows that any error probability $\epsilon>0$ can be met by setting the parameter $L=1/\epsilon$.
\begin{prop}
\label{prop:policy_satisfies_desried_error_probability}
	Fix an error probability $\epsilon>0$. If $L=1/\epsilon$, then $\pi^{\star}(L,\delta)\in\Pi(\epsilon)$ for all $\delta>0$.
\end{prop}
\begin{IEEEproof}
The proof uses the fact that the policy stops in finite time almost surely. The details are in Appendix \ref{appndx:proof_of_prop_policy_satisfies_desired_error_prob}.
\end{IEEEproof}

{\color{black} An important step in the proof of Proposition \ref{prop:policy_satisfies_desried_error_probability} is to upper bound the likelihood function under the arms configuration $C=(h, P_1, P_2)$ by the maximum likelihood under the hypothesis $\mathcal{H}_h$. Such an exercise of upper bounding the likelihood function is not possible if, for instance, the ML estimates in the expression for the maximum likelihood are replaced by the estimates of the TPMs computed by repeatedly sampling each of the arms.}

The next result improves upon Proposition \ref{prop:strict_positivity_of_drift_of_test_statistic} and shows that the modified GLR test statistic has the correct drift under the non-stopping version of the policy $\pi^\star(L, \delta)$.

\begin{prop}
	\label{prop:test_statistic_has_the_correct_drift}
	Suppose that $C=(h, P_1, P_2)$ is the underlying arms configuration. Fix $L> 1$ and $\delta>0$. For all $h'\in \mathcal{A}$ such that $h'\neq h$, under the non-stopping version of the policy $\pi^\star(L, \delta)$,
	\begin{equation}
		\liminf\limits_{n \to \infty} \frac{M_{hh'}(n)}{n}\geq \frac{R^*(h, P_1, P_2)}{(1+\delta)^2}\quad \text{almost surely}.
		\label{eq:test_statistic_has_the_right_drift}
	\end{equation}
\end{prop}
\begin{IEEEproof}
It is in proving this proposition that we make use of Assumption \ref{assmptn:continuous_selection}. The proof uses the fact that almost surely, $\theta(n)=h$ for all sufficiently large $n$ under the arms configuration $C=(h, P_1, P_2)$. As a result, $\hat{P}_{\theta(n),1}(n)=\hat{P}_{h,1}(n)$, $\hat{P}_{\theta(n),2}(n)=\hat{P}_{h,2}(n)$ for all sufficiently large $n$. The convergences in \eqref{eq:convergence_of_ML_estimates_of_TPMs} imply that the pair $(\hat{P}_{h,1}(n),\hat{P}_{h,2}(n))$ lies inside a neighbourhood (chosen based on the value of $\delta$) of $(P_1, P_2)$ for all $n$ sufficiently large. These together with Assumption \ref{assmptn:continuous_selection} and \eqref{eq:convergence_of_lambdas} imply that $\lambda_{\theta(n), \hat{P}_{\theta(n),1}(n), \hat{P}_{\theta(n),2}(n), \delta}$ is close to $\lambda_{h, P_1, P_2, \delta}$ for all $n$ sufficiently large. Using this, we arrive at \eqref{eq:test_statistic_has_the_right_drift}. For the details, see Appendix \ref{appndx:proof_of_prop_test_statistic_has_the_correct_drift}.
\end{IEEEproof}

Next, we show that the stopping time of the policy $\pi^\star(L, \delta)$ blows up as $L\to \infty$.
\begin{prop}
\label{prop:stopping_time_of_policy_blows_up_as_L_increases}
	For each $\delta>0$,
	\begin{equation}
		\liminf\limits_{L \to \infty} \tau(\pi^\star(L, \delta))=\infty \quad \text{almost surely}.
		\label{eq:stopping_time_of_policy_blows_up_as_L_to_infty}
	\end{equation}
\end{prop}
\begin{IEEEproof}
	See Appendix \ref{appndx:proof_of_prop_stopping_time_of_policy_blows_up_as_L_increases}.
\end{IEEEproof}
Combining the results of Proposition \ref{prop:test_statistic_has_the_correct_drift} and Proposition \ref{prop:stopping_time_of_policy_blows_up_as_L_increases}, it follows that for all $\delta > 0$,
\begin{equation}
	\liminf\limits_{L\to \infty}\frac{M_{hh'}(\tau(\pi^\star(L,\delta)))}{\tau(\pi^\star(L,\delta))}\geq\frac{R^*(h, P_1, P_2)}{(1+\delta)^2}\quad \text{almost surely}.
	\label{eq:combining_two_props}
\end{equation}

The following result shows that the stopping time of the policy $\pi^\star(L, \delta)$ satisfies an almost sure upper bound that is arbitrarily close to the lower bound in \eqref{eq:lower_bound}.
\begin{prop}
	\label{prop:almost_sure_upper_bound_for_policy}
	Suppose that $C=(h, P_1, P_2)$ is the underlying arms configuration. For all $\delta>0$, the stopping time of policy $\pi^\star(L,\delta)$ satisfies
	\begin{equation}
		\limsup\limits_{L\to\infty} \frac{\tau(\pi^\star(L,\delta))}{\log L} \leq \frac{(1+\delta)^2}{R^*(h, P_1, P_2)}\quad \text{almost surely.}
		\label{eq:almost_sure_upper_bound_for_policy}
	\end{equation}
\end{prop}
\begin{IEEEproof}[Proof of Proposition \ref{prop:almost_sure_upper_bound_for_policy}]
	By the definition of $\tau(\pi^\star(L,\delta))$, we know that under the arms configuration $C=(h, P_1, P_2)$, $$M_h(\tau(\pi^\star(L,\delta)))\geq \log ((K-1)L), \quad M_h(\tau(\pi^\star(L,\delta))-1) < \log ((K-1)L).$$ We then have, almost surely,
	\begin{align}
		1 &= \limsup\limits_{L \to \infty} \frac{\log ((K-1)L)}{\log L}\nonumber\\
		& \geq \limsup\limits_{L \to \infty} \frac{M_{h}(\tau(\pi^\star(L,\delta))-1)}{\log L}\nonumber\\
		& = \limsup\limits_{L \to \infty} \frac{M_{h}(\tau(\pi^\star(L,\delta))-1)}{\tau(\pi^\star(L,\delta))-1} \cdot \frac{\tau(\pi^\star(L,\delta))-1}{\log L} \nonumber\\
%		& \geq \limsup\limits_{L \to \infty} \frac{M_{h}(\tau(\pi^\star(L,\delta))-1)}{\tau(\pi^\star(L,\delta))-1} \cdot \limsup\limits_{L \to \infty}  \frac{\tau(\pi^\star(L,\delta))-1}{\log L}\nonumber\\
		& \geq \frac{R^*(h, P_1, P_2)}{(1+\delta)^2}\cdot \limsup\limits_{L \to \infty}  \frac{\tau(\pi^\star(L,\delta))}{\log L},
	\label{eq:proof_of_almost_sure_upper_bound_for_policy}
	\end{align}
where \eqref{eq:proof_of_almost_sure_upper_bound_for_policy} follows from \eqref{eq:combining_two_props} and Assumption \ref{assmptn:technical_assumption_on_TPMs}; here, Assumption \ref{assmptn:technical_assumption_on_TPMs} guarantees that the increment $M_{hh'}(n)-M_{hh'}(n-1)$ is bounded almost surely. The desired result follows by rearranging \eqref{eq:proof_of_almost_sure_upper_bound_for_policy}.
\end{IEEEproof}

The main result of this section on the expected stopping time of the policy is stated next.
\begin{prop}
	\label{prop:upper_bound}
	Fix $\delta>0$. Under the arms configuration $C=(h, P_1, P_2)$, the expected stopping time of the policy $\pi=\pi^{\star}(L,\delta)$ satisfies
	\begin{equation}
		\limsup\limits_{L \to \infty} \frac{E^\pi[\tau(\pi)|C]}{\log L} \leq \frac{(1+\delta)^2}{R^*(h, P_1, P_2)}.
	\label{eq:upper_bound_in_terms_of_delta}
	\end{equation}
\end{prop}
\begin{IEEEproof}
	In the proof, which we present in Appendix \ref{appndx:proof_of_prop_upper_bound}, we show that for each $\delta>0$, the family $\{\tau(\pi^\star(L,\delta))/\log L:L>1\}$ is uniformly integrable. Combining the almost sure upper bound in \eqref{eq:almost_sure_upper_bound_for_policy} with uniform integrability yields the desired upper bound in \eqref{eq:upper_bound_in_terms_of_delta} for the expected stopping time of the policy $\pi^\star(L, \delta)$.
\end{IEEEproof}

\section{Main Result}\label{sec:main_result}
With the above ingredients in place, the main result of this paper is as below.

\begin{thrm}\label{prop:main_result}
	Consider a multi-armed bandit with $K\geq 3$ arms in which each arm is a time homogeneous and ergodic Markov process on the finite state space $\mathcal{S}$. Fix an arms configuration $C=(h, P_1, P_2)$; here, $h$ is the odd arm, $P_1$ is the TPM of the odd arm $h$, and $P_2\neq P_1$ is the TPM of each non-odd arm. Fix $\eta\in (0,1]$, and suppose that a decision entity that wishes to identify the odd arm has a trembling hand with parameter $\eta$. When neither $P_1$ nor $P_2$ is known beforehand to the decision entity, the expected time required to identify the odd arm satisfies the asymptotic lower bound
	\begin{align}
		\liminf\limits_{\epsilon\downarrow 0}\inf\limits_{\pi\in\Pi(\epsilon)}\frac{E^\pi[\tau(\pi)|C]}{\log(1/\epsilon)}\geq \frac{1}{R^*(h, P_1, P_2)}.\label{eq:lower_bound_main_result}
	\end{align}
	Further, under Assumption \ref{assmptn:continuous_selection} and Assumption \ref{assmptn:technical_assumption_on_TPMs}, the policy $\pi^\star(L, \delta)$ satisfies the asymptotic upper bound
	\begin{align}		\limsup\limits_{\delta\downarrow 0} \limsup\limits_{L\to\infty}\frac{E^{\pi^\star(L, \delta)}[\tau(\pi^\star(L, \delta))|C]}{\log L}\leq \frac{1}{R^*(h, P_1, P_2)}.\label{eq:upper_bound_main_result}
	\end{align}
	Under these assumptions, we therefore have
		\begin{align}
		\lim\limits_{\epsilon\downarrow 0}\inf\limits_{\pi\in\Pi(\epsilon)}\frac{E^\pi[\tau(\pi)|C]}{\log(1/\epsilon)}&=\lim\limits_{\delta \downarrow 0}\lim\limits_{L\to\infty}\frac{E^{\pi^\star(L, \delta)}[\tau(\pi^\star(L,\delta))|C]}{\log L}=\frac{1}{R^*(h, P_1, P_2)}.\label{eq:main_result}
	\end{align}
\end{thrm}

\begin{IEEEproof}
The asymptotic lower bound in \eqref{eq:lower_bound_main_result} follows from Proposition \ref{prop:lower_bound}. Recall Assumption \ref{assmptn:continuous_selection} and Assumption \ref{assmptn:technical_assumption_on_TPMs}. Taking limits as $\delta \downarrow 0$ in \eqref{eq:upper_bound_in_terms_of_delta}, we arrive at \eqref{eq:upper_bound_main_result}. From Proposition \ref{prop:policy_satisfies_desried_error_probability}, we know that given any error tolerance parameter $\epsilon>0$, by setting $L=1/\epsilon$, we have $\pi^\star(L, \delta)\in \Pi(\epsilon)$ for all $\delta>0$. Therefore, it follows that for all $\epsilon, \delta>0$,
\begin{equation}
\inf\limits_{\pi \in \Pi(\epsilon)} \frac{E^\pi[\tau(\pi)|C]}{\log \left(1/\epsilon\right)} \leq \frac{E^{\pi^\star(L, \delta)}[\tau(\pi^\star(L, \delta))|C]}{\log L}.
\label{eq:proof_of_main_result_temp_1}
\end{equation}
Fixing $\delta>0$ and letting $\epsilon\downarrow 0$ (which is identical to letting $L\to \infty$) in \eqref{eq:proof_of_main_result_temp_1}, and using the upper bound in \eqref{eq:upper_bound_in_terms_of_delta}, we get
\begin{align}
\limsup\limits_{\epsilon\downarrow 0}\inf\limits_{\pi\in\Pi(\epsilon)}\frac{E^\pi[\tau(\pi)|C]}{\log(1/\epsilon)} \leq \limsup\limits_{L\to\infty}\frac{E^{\pi^\star(L, \delta)}[\tau(\pi^\star(L, \delta))|C]}{\log L}
\leq \frac{(1+\delta)^2}{R^*(h, P_1,  P_2)}.
\label{eq:proof_of_main_result_temp_2}
\end{align}
Letting $\delta\downarrow 0$ in \eqref{eq:proof_of_main_result_temp_2} and noting that the leftmost term in \eqref{eq:proof_of_main_result_temp_2} does not depend on $\delta$, we get
\begin{align}
\limsup\limits_{\epsilon\downarrow 0}\inf\limits_{\pi\in\Pi(\epsilon)}\frac{E^\pi[\tau(\pi)|C]}{\log(1/\epsilon)} &\leq \limsup\limits_{\delta \downarrow 0}\limsup\limits_{L\to\infty}\frac{E^{\pi^\star(L, \delta)}[\tau(\pi^\star(L, \delta))|C]}{\log L}\leq \frac{1}{R^*(h, P_1,P_2)}.
\label{eq:proof_of_main_result_temp_3}
\end{align}
Combining the result in \eqref{eq:proof_of_main_result_temp_3} with the lower bound in \eqref{eq:lower_bound}, we get
\begin{align}
\frac{1}{R^*(h,P_1,P_2)} &\leq \liminf\limits_{\epsilon\downarrow 0}\inf\limits_{\pi\in\Pi(\epsilon)}\frac{E^\pi[\tau(\pi)|C]}{\log(1/\epsilon)} \nonumber\\
&\leq \limsup\limits_{\epsilon\downarrow 0}\inf\limits_{\pi\in\Pi(\epsilon)}\frac{E^\pi[\tau(\pi)|C]}{\log(1/\epsilon)} \nonumber\\
&\leq \limsup\limits_{\delta \downarrow 0}\limsup\limits_{L\to\infty}\frac{E^{\pi^\star(L, \delta)}[\tau(\pi^\star(L, \delta))|C]}{\log L}\nonumber\\
&\leq \frac{1}{R^*(h, P_1,P_2)}.
\label{eq:proof_of_main_result_temp_4}
\end{align}
Thus, it follows that the limit infimum and the limit suprema in the chain of inequalities leading to \eqref{eq:proof_of_main_result_temp_4} are indeed limits, thereby yielding \eqref{eq:main_result}. This completes the proof of the theorem.
\end{IEEEproof}
Thus, under the arms configuration $C=(h, P_1, P_2)$, the asymptotic growth rate for the expected time required to identify the odd arm is $1/R^*(h, P_1, P_2)$. While the asymptotic lower bound holds for all arms configurations, the asymptotic upper bound is only established under a continuous selection assumption for those arms configurations whose TPMs satisfy a regularity condition. The continuous selection assumption and the regularity condition are used to prove  identifiability. The trembling hand model \eqref{eq:trembling_hand_relation} ensures that at any given time, each arm is selected with a strictly positive probability under every policy, and plays a key role in deriving the lower and the upper bounds.

\section{Concluding Remarks}
\label{sec:conclusions}
\begin{enumerate}
	\item The main results of this paper are the following. (a) We gave an asymptotic lower bound on the expected time required to find the odd arm, subject to an upper bound on the error probability (PAC setting). The asymptotics is as the error probability vanishes. The growth rate of the expected time to find the odd arm, see \eqref{eq:lower_bound}, is $1/R^*(h,P_1, P_2)$, where $R^*(h, P_1, P_2)$ is defined in \eqref{eq:R_delta^*(h,P_1,P_2)}. (b) We gave a policy (called $\pi^*(L,\delta)$) based on the principle of certainty equivalence that uses maximum likelihood (ML) estimates and achieves the lower bound asymptotically. 
	\item The achievability analysis relied crucially on showing that (a) the ML estimates of the TPMs converge to their true values asymptotically, and (b) given $\delta>0$ and an arms configuration $C$, the policy $\pi^\star(L, \delta)$ eventually samples the arms according to the $\delta$-optimal solution for $C$ as given by the lower bound. These conditions were established under Assumption \ref{assmptn:continuous_selection}, which is that there exists a continuous selection of $\delta$-optimal solutions, and Assumption \ref{assmptn:technical_assumption_on_TPMs}, which is that there exists $\bar{\varepsilon}^*\in (0, 1)$ such that for every arms configuration $C=(h, P, Q)$, the TPMs $P, Q\in \mathscr{P}(\bar{\varepsilon}^*)$. Both the assumptions were crucial in establishing that the ML estimates of the TPMs converge to their true values. In \cite{vaidhiyan2017learning, prabhu2017optimal, pnkarthik2019learning}, an analogue of the continuous selection property was established for the maximisers instead of $\delta$-optimal solutions. However, for more general settings such as the setting of this paper or the setting considered in \cite{prabhu2020sequential}, a continuous selection assumption seems inevitable and difficult to do away with.
	\item The expression for $R^*(h, P_1, P_2)$ contains relative entropies of the corresponding rows of the TPMs $P_1$ and $P_2$. The closer the rows of $P_1$ and $P_2$ are to each other, the smaller the value of $R^*(h,P_1,P_2)$ and therefore the larger the growth rate. The computability of $R^*(h, P_1, P_2)$ is an issue because it contains an outer supremum over all SRS policies which is difficult to evaluate. An algorithm such as $Q$-learning may need to be employed to compute $R^*(h, P_1, P_2)$.
	\item The policy $\pi^\star(L, \delta)$ is based a modified GLR test statistic that consists of an average average likelihood evaluated with respect to an artificial prior (the uniform distribution on the probability simplex $\mathscr{P}(\mathcal{S})$) in the numerator, and the maximum likelihood in the denominator. The computability of the maximum likelihood is an issue because closed-form expressions for the ML estimates of the TPMs are not available. At best, the ML estimates can be evaluated numerically.
	\item Repeatedly sampling each arm and using the consecutive observations from the arms to estimate the TPMs may lead to simple closed-form expressions for the TPM estimates. Because the Markov process of each arm is ergodic, by virtue of the ergodic theorem, these estimates converge to their true values. However, in this case, it is not clear if the arms will be sampled eventually according to the $\delta$-optimal solution for the underlying arms configuration (which, to recall, is crucial for achieving the asymptotic lower bound in \eqref{eq:lower_bound}). Also, it is not clear if the desired error probability can be met. Policies that sample the arms repeatedly are known to perform well for the problem of minimising regret \cite{liu2012learning}. However, it is not clear if such policies perform well for optimal stopping problems such as that studied in this paper.
	\item In writing \eqref{eq:M_hh'(n)_as_sum_of_4_terms}, the expression for the modified GLR test statistic $M_{hh'}(n)$, we assume that for each $a\in \mathcal{A}$, the initial state $X_0^a$ of arm $a$ is sampled according to the distribution $\phi$ that puts a strictly positive mass on each element of $\mathcal{S}$. However, this may not actually be the case. For instance, even before the decision entity begins the sampling of the arms at time $t=0$, if the Markov process of each arm has evolved for a sufficiently long duration of time and reached stationarity, then under the configuration $C=(h, P_1, P_2)$, we have $X_0^a\sim \mu_C^a$ where $\mu_C^a$ is the stationary distribution of arm $a$ under the configuration $C$. Because the TPMs $P_1$ and $P_2$ are unknown, the distributions $\mu_C^a$ too are unknown. In this case, there may be a mismatch between the average and the maximum likelihoods computed using $\phi$ (as in our policy $\pi^\star(L, \delta)$) and the actual values of these likelihoods (based on $\mu_C^a$). Let $\bar{M}_{hh'}(n)$ denote the denote the value of \eqref{eq:M_hh'(n)_as_sum_of_4_terms} with $\phi$ replaced by $\mu_C^a$ for each arm $a$. Because the Markov process of each arm is irreducible and positive recurrent, we have $\mu_{\textsf{min}}(C)\coloneqq\min\{\mu_C^a(i):i\in \mathcal{S}, ~a\in \mathcal{A}\}>0$. Letting $\phi_{\textsf{min}}=\min\{\phi(i):i\in \mathcal{S}\}$, we see that $$ \phi_{\textsf{min}}\leq \frac{\phi(i)}{\mu_h^a(i)}\leq \frac{1}{\mu_{\textsf{min}}(C)}\quad \text{for all }i\in \mathcal{S}.  $$ In this case, it can be shown that $\phi_{\textsf{min}}\leq |M_{hh'}(n)-\bar{M}_{hh'}(n)|\leq 1/\mu_{\textsf{min}}(C)$, as a result of which we have $$\lim\limits_{n\to\infty} \bigg|\frac{M_{hh'}(n)}{n} - \frac{\bar{M}_{hh'}(n)}{n}\bigg|=0.$$ That is, the asymptotic drift of $M_{hh'}(n)$ is identical to that of $\bar{M}_{hh'}(n)$. Therefore, our assumption about $X_0^a\sim \phi$ does not affect the asymptotic analysis in any way.
	\item It would be interesting to study the case when the trembling hand parameter $\eta=0$. Preliminary analysis in \cite[Section VII]{karthik2021detecting} for the case when the TPMs $P_1$ and $P_2$ are known beforehand reveals that for each $\eta>0$, the upper and the lower bounds match (as in this paper). However, the limiting values of these bounds as $\eta\downarrow 0$ may exhibit a gap. We anticipate that a similar behaviour holds when neither $P_1$ nor $P_2$ is known beforehand.
\end{enumerate}

%\section*{Acknowledgments}
%This work was supported in part by the Science and Engineering Research Board, Department of Science and Technology (grant no. EMR/2016/002503), by the Robert Bosch Centre for Cyber Physical Systems and the Centre for Networked Intelligence at the Indian Institute of Science.

\appendices

\section{Proof of Proposition \ref{prop:lower_bound}}\label{appndx:proof_of_prop_lower_bound}
Given a policy $\pi$ and two arm configurations $C=(h, P_1, P_2)$ and $C'=(h', P_1', P_2')$ such that $h\neq h'$, let $Z^\pi_{CC'}(n)\coloneqq Z^\pi_C(n)-Z^\pi_{C'}(n)$ denote the log-likelihood ratio, under the policy $\pi$, of all the arm samples and observations up to time $n$ under the arms configuration $C$ with respect to that under $C'$. Note that \eqref{eq:Z^pi_C(n)_5_1}, being independent of the arms configuration and common to the expressions for $Z^\pi_C(n)$ and $Z^\pi_{C'}(n)$, cancels out when writing the expression for $Z^\pi_C(n)-Z^\pi_{C'}(n)$. It then follows from \eqref{eq:Z^pi_C(n)_3} that
\begin{align}
	Z^\pi_{CC'}(n) = \sum\limits_{a=1}^{K} \log \frac{P^\pi_C(X_{a-1}^a)}{P^\pi_{C'}(X_{a-1}^a)} + \sum\limits_{(\underline{d}, \underline{i}) \in \mathbb{S}} ~\sum\limits_{a=1}^{K}~\sum\limits_{j\in \mathcal{S}}~N(n, \underline{d}, \underline{i}, a, j)~\log \, \frac{(P_C^a)^{d_a}(j|i_a)}{(P_{C'}^a)^{d_a}(j|i_a)}.
\label{eq:Z_pi_CC'(n)_2}
\end{align}

We organise the proof of Proposition \ref{prop:lower_bound} as follows. Given an error probability $\epsilon>0$, we first obtain a lower bound for $E^\pi[Z^\pi_{CC'}(\tau(\pi))|C]$ for all $\pi\in\Pi(\epsilon)$ using a change of measure argument of Kaufmann et al. \cite{Kaufmann2016}. Following this, we obtain an upper bound for $E^\pi[Z^\pi_{CC'}(\tau(\pi))|C]$ in terms of $E^\pi[\tau(\pi)|C]$ using a Wald-type lemma for the setting of restless Markov arms. Combining the upper and the lower bounds, and letting $\epsilon\downarrow 0$, we arrive at the desired result. The ergodicity property established in \cite[Lemma 1]{karthik2021detecting} for SRS policies plays a crucial role in deriving the final lower bound \eqref{eq:lower_bound}.

\subsection{A Lower Bound on $E^\pi[Z^\pi_{CC'}(\tau(\pi))|C]$ for $\pi\in\Pi(\epsilon)$}
As a first step towards deriving the lower bound, we use a result of Kaufmann et al. \cite{Kaufmann2016} to obtain a lower bound for $E^\pi[Z^\pi_{CC'}(\tau(\pi))|C]$ in terms of the error probability parameter $\epsilon$. This is based on a generalisation of \cite[Lemma 18]{Kaufmann2016}, a change of measure argument for iid observations from the arms, to the setting of restless arms with Markov observations. We present this generalisation in the following lemma.

\begin{lemma}\label{lem:change_of_measure}
	Let $(\Omega, \mathcal{F})$ be a measurable space. Let the filtration $\{\mathcal{F}_t:t\geq 0\}$ be defined as follows: $\mathcal{F}_0=\sigma(\Omega, \emptyset)$ and $\mathcal{F}_t=\sigma(B^t,A^t,\bar{X}^t)$ for all $t\geq 1$. Fix $\pi\in\Pi(\epsilon)$, and let $\tau(\pi)$ be the stopping time of policy $\pi$. Let $\mathcal{F}_{\tau(\pi)}$ be the $\sigma$-algebra
	\begin{equation}
		\mathcal{F}_{\tau(\pi)}=\{E\in\mathcal{F}:E\cap \{\tau(\pi)= t\}\in\mathcal{F}_t\text{ for all }t\geq 0\}.\label{eq:F_tau}
	\end{equation}
Then, for all $C=(h, P_1, P_2)$ and $C'=(h', P_1', P_2')$ such that $h' \neq h$,
	\begin{equation}
		P^\pi(E|C')=E^\pi[1_{E}\,\exp(-Z^\pi_{CC'}(\tau(\pi)))|C]\label{eq:change_of_measure}
	\end{equation}
	for all $E\in\mathcal{F}_{\tau(\pi)}$.
\end{lemma}

\begin{IEEEproof}[Proof of Lemma \ref{lem:change_of_measure}]
	We prove the lemma by demonstrating, through mathematical induction, that the relation
	\begin{equation}
		E^\pi[g(B^t,A^t,\bar{X}^t)|C']=E^\pi[g(B^t,A^t,\bar{X}^t)\,\exp(-Z^\pi_{CC'}(t))|C]\label{eq:change_of_measure_equiv}
	\end{equation}
	holds for all $t\geq 0$ and for all measurable functions $g:\mathcal{A}^{t+1}\times\mathcal{A}^{t+1}\times\mathcal{S}^{t+1}\to\mathbb{R}$. The proof for the case $t=0$ may be obtained as follows. For any measurable $g:\mathcal{A}\times\mathcal{A}\times\mathcal{S}\to\mathbb{R}$, we have
	\begingroup \allowdisplaybreaks\begin{align}
		E^\pi[g(B_0,A_0,\bar{X}_0)|C']&=\sum\limits_{b=1}^K\sum\limits_{a=1}^{K}\sum\limits_{i\in\mathcal{S}}~g(b,a,i)~P^\pi(B_0=b, A_0=a,\bar{X}_0=i|C')\nonumber\\
		&=\sum\limits_{b=1}^K\sum\limits_{a=1}^{K}\sum\limits_{i\in\mathcal{S}}~g(b,a,i)~P^\pi(B_0=b|C')~P^\pi(A_0=a|B_0=b, C')~P^\pi(\bar{X}_0=i|B_0=b,A_0=a,C')\nonumber\\
		&\stackrel{(a)}{=}\sum\limits_{b=1}^K\sum\limits_{a=1}^{K}\sum\limits_{i\in\mathcal{S}}~g(b,a,i)~P^\pi(B_0=b|C)~P^\pi(A_0=a|B_0=b,C)~P^\pi(\bar{X}_0=i|A_0=a,C')\nonumber\\
		&=\sum\limits_{b=1}^K\sum\limits_{a=1}^{K}\sum\limits_{i\in\mathcal{S}}~g(b,a,i)~P^\pi(B_0=b|C)~P^\pi(A_0=a|B_0=b,C)~P^\pi(X_0^a=i|C'),\label{eq:change_of_measure_1}
	\end{align}\endgroup
	where $(a)$ follows by noting that
	\begin{itemize}
		\item $P^\pi(B_0=b|C')=P^\pi(B_0=b|C)$ because $\pi$ is oblivious to the underlying arms configuration, and
		\item $P^\pi(A_0=a|B_0=b, C')=P^\pi(A_0=a|B_0=b,C)=\frac{\eta}{K}+(1-\eta)\,\mathbb{I}_{\{a=b\}}$.	
	\end{itemize}
Assuming that $X_0^a\sim \phi$ under $\pi$, where $\phi$ is a probability distribution on $\mathcal{S}$ that is independent of the underlying arms configuration  (which is not known to $\pi$), we have
	\begingroup \allowdisplaybreaks\begin{align}
		E^\pi[g(B_0,A_0,\bar{X}_0)|C']&=\sum\limits_{b=1}^{K}\sum\limits_{a=1}^{K}\sum\limits_{i\in\mathcal{S}}~g(b,a,i)~P^\pi(B_0=b|C)~P^\pi(A_0=a|B_0=b,C)~\phi(i)\\
		&=\sum\limits_{b=1}^{K}\sum\limits_{a=1}^{K}\sum\limits_{i\in\mathcal{S}}~g(b,a,i)~P^\pi(B_0=b|C)~P^\pi(A_0=a|B_0=b,C)~P^\pi(X_0^a=i|A_0=a,C)\\
		&=\sum\limits_{b=1}^{K}\sum\limits_{a=1}^{K}\sum\limits_{i\in\mathcal{S}}~g(b,a,i)~P^\pi(B_0=b|C)~P^\pi(A_0=a|B_0=b,C)~P^\pi(X_0^a=i|A_0=a, B_0=b,C).\label{eq:change_of_measure_2}
	\end{align}\endgroup
	Noting that
	\begingroup \allowdisplaybreaks\begin{align}
		Z^\pi_{CC'}(0)&=\log \frac{P^\pi(B_0,A_0,\bar{X}_0|C)}{P^\pi(B_0,A_0,\bar{X}_0|C')}=0,\label{eq:change_of_measure_3}
	\end{align}\endgroup
	and combining \eqref{eq:change_of_measure_2} with \eqref{eq:change_of_measure_3}, we get $$
	E^\pi[g(B_0,A_0,\bar{X}_0)|C']=E^\pi[g(B_0,A_0,\bar{X}_0|C)\exp (-Z^\pi_{CC'}(0))].
	$$ This proves \eqref{eq:change_of_measure_equiv} for $t=0$.
	
	Assume now that \eqref{eq:change_of_measure_equiv} is true for some $t>0$. We shall demonstrate that \eqref{eq:change_of_measure_equiv} holds for $t+1$. By the law of iterated expectations,
	\begingroup \allowdisplaybreaks\begin{align}
		E^\pi[g(B^{t+1},A^{t+1},\bar{X}^{t+1})|C']=E^\pi[E^\pi[g(B^{t+1},A^{t+1},\bar{X}^{t+1})|\mathcal{F}_t,C']|C'].\label{eq:change_of_measure_4}
	\end{align}\endgroup
	Because $E^\pi[g(B^{t+1},A^{t+1},\bar{X}^{t+1})|\mathcal{F}_t,C']$ is a measurable function of $(B^t,A^t,\bar{X}^t)$, by the induction hypothesis, we have
	\begingroup \allowdisplaybreaks\begin{align}
		E^\pi[g(B^{t+1},A^{t+1},\bar{X}^{t+1})|\mathcal{F}_t,C']&=E^\pi[E^\pi[g(B^{t+1},A^{t+1},\bar{X}^{t+1})|\mathcal{F}_t,C']\,\exp(-Z^\pi_{CC'}(t))|C].\label{eq:change_of_measure_5}
	\end{align}\endgroup
	We now note that
	\begingroup \allowdisplaybreaks\begin{align}
		& E^\pi[g(B^{t+1},A^{t+1},\bar{X}^{t+1})|\mathcal{F}_t,C']\,\exp(-Z^\pi_{CC'}(t))\nonumber\\
		&\stackrel{(a)}{=}E^\pi[g(B^{t+1},A^{t+1},\bar{X}^{t+1})\,\exp(-Z^\pi_{CC'}(t))|\mathcal{F}_t,C']\,\nonumber\\
		&=\sum\limits_{b=1}^{K}\sum\limits_{a=1}^{K}\sum\limits_{i\in\mathcal{S}}\bigg[g(B^t,A^t,\bar{X}^t,b, a, i)\cdot P^\pi(B_{t+1}=b|B^t,A^t,\bar{X}^t,C')\cdot P^\pi(A_{t+1}=a|B^{t+1}=b,B^t,A^t,\bar{X}^t,C')\nonumber\\
		&\hspace{7cm} \cdot P^\pi(\bar{X}_{t+1}=i|B^{t+1}=b,A_{t+1}=a,B^t,A^t,\bar{X}^t,C')\cdot \exp(-Z^\pi_{CC'}(t))\bigg]\nonumber\\
		&\stackrel{(b)}{=}\sum\limits_{b=1}^{K}\sum\limits_{a=1}^{K}\sum\limits_{i\in\mathcal{S}}\bigg[g(B^t,A^t,\bar{X}^t,b, a, i)\cdot P^\pi(B_{t+1}=b|B^t,A^t,\bar{X}^t,C)\cdot P^\pi(A_{t+1}=a|B^{t+1}=b,B^t,A^t,\bar{X}^t,C)\nonumber\\
		&\hspace{7cm} \cdot P^\pi(\bar{X}_{t+1}=i|A_{t+1}=a,A^t,\bar{X}^t,C')\cdot \exp(-Z^\pi_{CC'}(t))\bigg],\label{eq:change_of_measure_6}
	\end{align}\endgroup
	where $(a)$ follows by observing that $Z^\pi_{CC'}(t)$ is a measurable function of $(B^t, A^t,\bar{X}^t)$, and in writing $(b)$, we use the following facts: for any $t$,
	\begin{itemize}
		\item $P^\pi(B_{t+1}=b|B^t,A^t,\bar{X}^t,C')=P^\pi(B_{t+1}=b|B^t,A^t,\bar{X}^t,C)$ because $\pi$ is oblivious to the underlying arms configuration,
		\item $P^\pi(A_{t+1}=a|B_{t+1}=b,B^t,A^t,\bar{X}^t,C')=P^\pi(A_{t+1}=a|B_{t+1}=b,B^t,A^t,\bar{X}^t,C)=\frac{\eta}{K}+(1-\eta)\,\mathbb{I}_{\{a=b\}}$, and
		\item $P^\pi(\bar{X}_{t+1}=i|B_{t+1}=b,A_{t+1}=a,B^t,A^t,\bar{X}^t,C')=P^\pi(\bar{X}_{t+1}=i|A_{t+1}=a,A^t,\bar{X}^t,C')$ because the observation obtained from arm $a$ at time $t$ is a function only of the delay and the last observed state of arm $a$ as measured at time $t$, both of which may be deduced from $(A^t, \bar{X}^t)$.
	\end{itemize}
	Also, we have
	\begingroup \allowdisplaybreaks\begin{align}
		& \sum\limits_{i\in\mathcal{S}}P^\pi(\bar{X}_{t+1}=i|A_{t+1}=a,A^t,\bar{X}^t,C')\,\exp(-Z^\pi_{CC'}(t))\nonumber\\
		&=\sum\limits_{i\in\mathcal{S}}\frac{P^\pi(\bar{X}_{t+1}=i|A_{t+1}=a,A^t,\bar{X}^t, C')}{P^\pi(\bar{X}_{t+1}=i|A_{t+1}=a,A^t,\bar{X}^t, C)}\,\exp(-Z^\pi_{CC'}(t))\,P^\pi(\bar{X}_{t+1}=i|A_{t+1}=a,A^t,\bar{X}^t,C)\nonumber\\
		&=\sum\limits_{i\in\mathcal{S}}\exp(-Z^\pi_{CC'}(t+1,a,i))\,P^\pi(\bar{X}_{t+1}=i|A_{t+1}=a,A^t,\bar{X}^t,C),\label{eq:change_of_measure_7}
	\end{align}\endgroup
	since $$Z^\pi_{CC'}(t+1,a,i)= Z^\pi_{CC'}(t)+\log\frac{P_h(\bar{X}_{t+1}=i|A_{t+1}=a,A^t,\bar{X}^t)}{P_{h'}(\bar{X}_{t+1}=i|A_{t+1}=a,A^t,\bar{X}^t)}.$$
	Substituting \eqref{eq:change_of_measure_7} in \eqref{eq:change_of_measure_6} and simplifying, we get
	\begingroup \allowdisplaybreaks\begin{align}
		& E^\pi[g(B^{t+1},A^{t+1},\bar{X}^{t+1})|\mathcal{F}_t,C']\,\exp(-Z^\pi_{CC'}(t))\nonumber\\
		&=\sum\limits_{b=1}^{K}\sum\limits_{a=1}^{K}\sum\limits_{i\in\mathcal{S}} \bigg[g(B^t,A^t,\bar{X}^t, b, a, i)\cdot P^\pi(B_{t+1}=b|B^t,A^t,\bar{X}^t,C)\cdot P^\pi(A_{t+1}=a|B_{t+1}=b,B^t,A^t,\bar{X}^t,C)\nonumber\\
		&\hspace{1cm}\cdot P^\pi(\bar{X}_{t+1}=i|B_{t+1}=b,A_{t+1}=a,B^t,A^t,\bar{X}^t,C)\cdot \exp(-Z^\pi_{CC'}(t+1,a,i))\bigg]\\
		&=E^\pi[g(B^{t+1},A^{t+1},\bar{X}^{t+1})~\exp(-Z^\pi_{CC'}(t+1))|\mathcal{F}_t, C].\label{eq:change_of_measure_8}
	\end{align}\endgroup
	Thus, we have
	\begin{align}
		E^\pi[g(B^{t+1},A^{t+1},\bar{X}^{t+1})|\mathcal{F}_t,C']\,\exp(-Z^\pi_{CC'}(t))=E^\pi[g(B^{t+1},A^{t+1},\bar{X}^{t+1})\,\exp(-Z^\pi_{CC'}(t+1))|\mathcal{F}_t, C].
		\label{eq:change_of_measure_8_1}
	\end{align}
	Applying $E^\pi[\cdot|C]$ to both sides of \eqref{eq:change_of_measure_8_1}, plugging it into the right hand side of \eqref{eq:change_of_measure_5}, and using the law of iterated expectations, we arrive at the desired relation for $t+1$. This proves \eqref{eq:change_of_measure_equiv} for all $t\geq 0$.
	
	Finally, for any $E\in\mathcal{F}_{\tau(\pi)}$, we have
	\begingroup \allowdisplaybreaks\begin{align}
		P^\pi(E|C')&=E^\pi[1_{E}|C']\nonumber\\
		&=E^\pi\left[\sum\limits_{t\geq 0} 1_{E\cap \{\tau(\pi)= t\}}\bigg|C'\right]\nonumber\\
		&\stackrel{(a)}{=}\sum\limits_{t\geq 0}E^\pi\left[1_{E\cap \{\tau(\pi)= t\}}\right|C']\nonumber\\
		&\stackrel{(b)}{=}\sum\limits_{t\geq 0}E^\pi\left[1_{E\cap \{\tau(\pi)= t\}}\,\exp(-Z^\pi_{CC'}(t))\right|C]\nonumber\\
		&=\sum\limits_{t\geq 0}E^\pi\left[1_{E\cap \{\tau(\pi)= t\}}\,\exp(-Z^\pi_{CC'}(\tau(\pi)))\right|C]\nonumber\\
		&=E^\pi\left[1_{E}\,\exp(-Z^\pi_{CC'}(\tau(\pi)))\right|C],\label{eq:change_of_measure_9}
	\end{align}\endgroup
	where $(a)$ is due to monotone convergence theorem, and $(b)$ above follows from \eqref{eq:change_of_measure_equiv} and the fact that $E\cap \{\tau(\pi)=t\}\in\mathcal{F}_t$ for all $t\geq 0$ since $E\in\mathcal{F}_{\tau(\pi)}$. This completes the proof of the lemma.
\end{IEEEproof}

Lemma \ref{lem:change_of_measure} in conjunction with \cite[Lemma 19]{Kaufmann2016} implies that when $C=(h, P_1, P_2)$ is the underlying arms configuration, the following inequality holds for all $\pi\in \Pi(\epsilon)$ and $C'=(h', P_1', P_2')$ such that $h'\neq h$:
\begin{equation}
	E^\pi[Z^\pi_{CC'}(\tau(\pi))|C]\geq \sup\limits_{E\in\mathcal{F}_{\tau(\pi)}} d(P^\pi(E|C),P^\pi(E|C')),\label{eq:Kaufmann_DPI_bound}
\end{equation}
where for any $x,y\in [0,1]$, $$d(x,y)\coloneqq x\log (x/y)+(1-x)\log ((1-x)/(1-y))$$ is the binary relative entropy function. We now note that for any $\pi\in\Pi(\epsilon)$, when $C$ is the actual (underlying) arms configuration, $$P^\pi(\theta(\tau(\pi))=h|C)\geq 1-\epsilon,\quad P^\pi(\theta(\tau(\pi))=h|C')\leq \epsilon.$$  As noted in \cite{Kaufmann2016}, the mapping $x\mapsto d(x,y)$ is monotone increasing for $x<y$, and the mapping $y\mapsto d(x,y)$ is monotone decreasing for any fixed $x$. Using these facts in \eqref{eq:Kaufmann_DPI_bound}, we get
\begin{align}
	E^\pi[Z^\pi_{CC'}(\tau(\pi))|C] & \geq d(P^\pi(\theta(\tau(\pi))=h|C),P^\pi(\theta(\tau(\pi))=h|C'))\nonumber\\
	&\geq d(\epsilon, 1-\epsilon)
\end{align}
for all $\pi\in\Pi(\epsilon)$ and for all $C'=(h', P_1', P_2')$ such that $h' \neq h$, from which it follows that
\begin{equation}
	\inf\limits_{\substack{C'=(h', P_1', P_2'):\\h'\neq h,~P_1'\neq P_2'}}E^\pi[Z^\pi_{CC'}(\tau(\pi))|C] \geq d(\epsilon, 1-\epsilon).
	\label{eq:Kaufmann_DPI_bound_epsilon}
\end{equation}
\subsection{An Upper Bound for  $E^\pi[Z^\pi_{CC'}(\tau(\pi))|C]$ in Terms of $ E^\pi[\tau(\pi)|C]$}

We now obtain an upper bound for the left-hand side of \eqref{eq:Kaufmann_DPI_bound_epsilon}. From \eqref{eq:Z_pi_CC'(n)_2}, we have
\begingroup \allowdisplaybreaks\begin{align}
	E^\pi[Z^\pi_{CC'}(\tau(\pi))|C]&=E^\pi\bigg[\sum\limits_{a=1}^{K}\log\frac{P^\pi_C(X_{a-1}^a)}{P^\pi_{C'}(X_{a-1}^a)}\bigg\vert C\bigg]+E^\pi\bigg[\sum\limits_{(\underline{d},\underline{i})\in\mathbb{S}}~\sum\limits_{a=1}^{K}\sum\limits_{j\in\mathcal{S}}N(\tau(\pi),\underline{d},\underline{i},a,j)\log\frac{(P_C^{a})^{d_a}(j|i_a)}{(P_{C'}^{a})^{d_a}(j|i_a)}\bigg\vert C \bigg].\label{eq:lower_bound_1}
\end{align}\endgroup
To simplify the second expectation term on the right-hand side of \eqref{eq:lower_bound_1}, we use the following result.
\begin{lemma}\label{lem:relation_between_N(tau,d,i,j,a)_and_N(tau,d,i,a)}
	{\color{black} For every $(\underline{d},\underline{i})\in \mathbb{S}$, $a\in\mathcal{A}$ and $j\in \mathcal{S}$,}
	\begin{equation}
		E^\pi[E^\pi[N(\tau(\pi),\underline{d},\underline{i},a,j)|X_{a-1}^a,C]|\tau(\pi),C]=E^\pi[E^\pi[N(\tau(\pi),\underline{d},\underline{i},a)|X_{a-1}^a,C]|\tau(\pi),C]\,(P_C^a)^{d_a}(j|i_a).\label{eq:relation_between_N(tau,d,i,j,a)_and_N(tau,d,i,a)}
	\end{equation}
\end{lemma}

\begin{IEEEproof}[Proof of Lemma \ref{lem:relation_between_N(tau,d,i,j,a)_and_N(tau,d,i,a)}]
	Substituting $n=\tau(\pi)$ in \eqref{eq:N(n, d, i, a, j)}, we have
	\begingroup \allowdisplaybreaks\begin{align}
		E^\pi[E^\pi[N(\tau(\pi),\underline{d},\underline{i},a,j)|X_{a-1}^a,C]|\tau(\pi),C]&=E^\pi\bigg[E^\pi\bigg[\sum\limits_{t=K}^{\tau(\pi)} 1_{\{\underline{d}(t)=\underline{d},\underline{i}(t)=\underline{i},A_{t}=a,X_{t}^a=j\}}\bigg|X_{a-1}^a,C\bigg]\bigg|\tau(\pi),C\bigg]\nonumber\\
		&=E^\pi\bigg[\sum\limits_{t=K}^{\tau(\pi)}P^\pi(\underline{d}(t)=\underline{d},\underline{i}(t)=\underline{i},A_{t}=a,X_{t}^a=j|X_{a-1}^a,C)\,\bigg|\tau(\pi),C\bigg].\label{eq:relation_btw_two_qty_1}
	\end{align}\endgroup
	For each $t$ in the range of the summation in \eqref{eq:relation_btw_two_qty_1}, the conditional probability term for $t$ may be expressed as
	\begingroup \allowdisplaybreaks\begin{align}
		& P^\pi(\underline{d}(t)=\underline{d},\underline{i}(t)=\underline{i},A_{t}=a,X_{t}^a=j|X_{a-1}^a,C)\nonumber\\
		&=P^\pi(\underline{d}(t)=\underline{d},\underline{i}(t)=\underline{i},A_{t}=a|X_{a-1}^a,C)\cdot P^\pi(X_{t}^a=j|A_{t}=a,\underline{d}(t)=\underline{d},\underline{i}(t)=\underline{i},X_{a-1}^a,C)\nonumber\\
		&=P^\pi(\underline{d}(t)=\underline{d},\underline{i}(t)=\underline{i},A_{t}=a|X_{a-1}^a,C)\cdot (P_C^a)^{d_a}(j|i_a).\label{eq:relation_btw_two_qty_2}
	\end{align}\endgroup
	Plugging \eqref{eq:relation_btw_two_qty_2} back in \eqref{eq:relation_btw_two_qty_1} and simplifying, we arrive at the desired relation in \eqref{eq:relation_between_N(tau,d,i,j,a)_and_N(tau,d,i,a)}.
\end{IEEEproof}

Using Lemma \ref{lem:relation_between_N(tau,d,i,j,a)_and_N(tau,d,i,a)}, the second expectation term on the right-hand side of \eqref{eq:lower_bound_1} can be simplified as follows.
\begingroup \allowdisplaybreaks\begin{align}
	& E^\pi\bigg[\sum\limits_{(\underline{d},\underline{i})\in\mathbb{S}}~\sum\limits_{a=1}^{K}~\sum\limits_{j\in\mathcal{S}}~N(\tau(\pi),\underline{d},\underline{i},a,j)\log\frac{(P_C^{a})^{d_a}(j|i_a)}{(P_{C'}^{a})^{d_a}(j|i_a)}\bigg\vert C\bigg]\nonumber\\
	&=E^\pi\left[\sum\limits_{(\underline{d},\underline{i})\in\mathbb{S}}~\sum\limits_{a=1}^{K}~\sum\limits_{j\in\mathcal{S}}~E^\pi[E^\pi[N(\tau(\pi),\underline{d},\underline{i},a,j)|X_{a-1}^a,C]|\tau(\pi),C]\log\frac{(P_C^{a})^{d_a}(j|i_a)}{(P_{C'}^{a})^{d_a}(j|i_a)}\bigg|C\right]\nonumber\\
	&\stackrel{(a)}{=} E^\pi\left[\sum\limits_{(\underline{d},\underline{i})\in\mathbb{S}}~\sum\limits_{a=1}^{K}~\sum\limits_{j\in\mathcal{S}}~E^\pi[E^\pi[N(\tau(\pi),\underline{d},\underline{i},a)|X_{a-1}^a,C]|\tau(\pi),C]\cdot (P_C^a)^{d_a}(j|i)\cdot\log\frac{(P_C^{a})^{d_a}(j|i_a)}{(P_{C'}^{a})^{d_a}(j|i_a)}\right]\nonumber\\
	&= E^\pi\left[\sum\limits_{(\underline{d},\underline{i})\in\mathbb{S}}~\sum\limits_{a=1}^{K}~E^\pi[E^\pi[N(\tau(\pi),\underline{d},\underline{i},a)|X_{a-1}^a,C]|\tau(\pi),C]\cdot D((P_C^a)^{d_a}(\cdot|i_a)\|(P_{C'}^a)^{d_a}(\cdot|i_a))\right]\nonumber\\
	&=\sum\limits_{(\underline{d},\underline{i})\in\mathbb{S}}~\sum\limits_{a=1}^{K}~E^\pi[N(\tau(\pi),\underline{d},\underline{i},a)|C]\cdot D((P_C^a)^{d_a}(\cdot|i_a)\|(P_{C'}^a)^{d_a}(\cdot|i_a)),\label{eq:lower_bound_2}
\end{align}\endgroup
where in the above set of equations, $(a)$ follows from Lemma \ref{lem:relation_between_N(tau,d,i,j,a)_and_N(tau,d,i,a)}, $$ N(n, \underline{d}, \underline{i}, a)\coloneqq \sum\limits_{j\in \mathcal{S}} N(n, \underline{d}, \underline{i}, a, j) $$ for all $n\geq K$, $(\underline{d}, \underline{i})\in \mathbb{S}$ and $a\in \mathcal{A}$,
and \eqref{eq:lower_bound_2} is due to monotone convergence theorem and the fact that $$E^\pi[E^\pi[E^\pi[N(\tau(\pi),\underline{d},\underline{i},a)|X_{a-1}^a,C]|\tau(\pi),C]|C]=E^\pi[N(\tau(\pi),\underline{d},\underline{i},a)|C].$$

Plugging \eqref{eq:lower_bound_2} back in \eqref{eq:lower_bound_1},
%and noting that the first term on the right-hand side of \eqref{eq:lower_bound_1} is zero,
we get
\begingroup \allowdisplaybreaks\begin{align}
	& E^\pi[Z^\pi_{CC'}(\tau(\pi))|C]\nonumber\\
	&=E^\pi\left[\sum\limits_{a=1}^{K} \log\frac{P^\pi_C(X_{a-1}^a)}{P^\pi_{C'}(X_{a-1}^a)}\right]
	+\sum\limits_{(\underline{d},\underline{i})\in\mathbb{S}}~\sum\limits_{a=1}^{K}~E^\pi[N(\tau(\pi),\underline{d},\underline{i},a)|C]\cdot D((P_C^a)^{d_a}(\cdot|i_a)\|(P_{C'}^a)^{d_a}(\cdot|i_a)).\label{eq:lower_bound_3}
\end{align}\endgroup
Noting that
\begingroup \allowdisplaybreaks\begin{align}
	\sum\limits_{(\underline{d},\underline{i})\in\mathbb{S}}~\sum\limits_{a=1}^{K}~E^\pi[N(\tau(\pi),\underline{d},\underline{i},a)|C]&\stackrel{(a)}{=}E^\pi\bigg[\sum\limits_{(\underline{d},\underline{i})\in\mathbb{S}}~\sum\limits_{a=1}^{K}~N(\tau(\pi),\underline{d},\underline{i},a)\bigg|C\bigg]\nonumber\\
	&=E^\pi\bigg[\sum\limits_{(\underline{d},\underline{i})\in\mathbb{S}}~\sum\limits_{a=1}^{K}\sum\limits_{t=K}^{\tau(\pi)}1_{\{\underline{d}(t)=\underline{d},\underline{i}(t)=\underline{i},A_{t}=a\}}\bigg|C\bigg]\nonumber\\
	&=E^\pi\bigg[\sum\limits_{t=K}^{\tau(\pi)}1\bigg|C\bigg]\\
	&=E^\pi[\tau(\pi)-K+1|C],
\end{align}\endgroup
where $(a)$ above is due to monotone convergence theorem,
we write \eqref{eq:lower_bound_3} as
\begingroup \allowdisplaybreaks\begin{align}
	& E^\pi[Z^\pi_{CC'}(\tau(\pi))|C]\nonumber\\
	&=E^\pi\left[\sum\limits_{a=1}^{K} \log\frac{P^\pi_C(X_{a-1}^a)}{P^\pi_{C'}(X_{a-1}^a)}\right]\nonumber\\
	&\hspace{2cm}+\bigg(E^\pi[\tau(\pi)-K+1|C]\bigg) \sum\limits_{(\underline{d},\underline{i})\in\mathbb{S}}~\sum\limits_{a=1}^{K}~\frac{E^\pi[N(\tau(\pi),(\underline{d},\underline{i}),a)|C]}{E^\pi[\tau(\pi)-K+1|C]}\cdot D((P_C^a)^{d_a}(\cdot|i_a)\|(P_{C'}^a)^{d_a}(\cdot|i_a)).\label{eq:lower_bound_4}
\end{align}\endgroup
Combining \eqref{eq:Kaufmann_DPI_bound_epsilon} and \eqref{eq:lower_bound_4}, and noting that \eqref{eq:lower_bound_4} holds for all $C'=(h', P_1', P_2')$ such that $h' \neq h$, we get
\begingroup \allowdisplaybreaks\begin{align}
	d(\epsilon,1-\epsilon)&\leq \inf\limits_{\substack{C'=(h', P_1', P_2'):\\h'\neq h, ~P_1'\neq P_2'}}\bigg\lbrace E^\pi\left[\sum\limits_{a=1}^{K} \log\frac{P^\pi_C(X_{a-1}^a)}{P^\pi_{C'}(X_{a-1}^a)}\right]\nonumber\\
	&\hspace{2cm}+ \bigg(E^\pi[\tau(\pi)-K+1|C]\bigg)\cdot\,\sum\limits_{(\underline{d},\underline{i})\in\mathbb{S}}~\sum\limits_{a=1}^{K}~\frac{E^\pi[N(\tau(\pi),\underline{d},\underline{i},a)|C]}{E^\pi[\tau(\pi)-K+1|C]}\cdot D((P_C^a)^{d_a}(\cdot|i_a)\|(P_{C'}^a)^{d_a}(\cdot|i_a))\bigg\rbrace\nonumber\\
	&\leq \sup\limits_{\nu}\,\inf\limits_{\substack{C'=(h', P_1', P_2'):\\h'\neq h, ~P_1'\neq P_2'}}\bigg\lbrace E^\pi\left[\sum\limits_{a=1}^{K} \log\frac{P^\pi_C(X_{a-1}^a)}{P^\pi_{C'}(X_{a-1}^a)}\right]+\bigg(E^\pi[\tau(\pi)-K+1|C]\bigg)\cdot\,\sum\limits_{(\underline{d},\underline{i})\in\mathbb{S}}~\sum\limits_{a=1}^{K}~\nu(\underline{d},\underline{i},a)\, k_{CC'}(\underline{d}, \underline{i}, a)\bigg\rbrace,\label{eq:lower_bound_5}
\end{align}\endgroup
where the supremum in \eqref{eq:lower_bound_5} is over all state-action occupancy measures satisfying
\begin{align}
	\sum\limits_{a=1}^{K}\nu(\underline{d}',\underline{i}',a)&=\sum\limits_{(\underline{d},\underline{i})\in\mathbb{S}}~\sum\limits_{a=1}^{K}\,\nu(\underline{d},\underline{i},a)\,Q(\underline{d}',\underline{i}'|\underline{d},\underline{i},a)\quad \text{for all }(\underline{d}',\underline{i}')\in\mathbb{S},\label{eq:lower_bound_6_1}\\
	&\sum\limits_{(\underline{d},\underline{i})\in\mathbb{S}}~\sum\limits_{a=1}^{K}\,\nu(\underline{d},\underline{i},a)=1,\label{eq:lower_bound_6_2}\\
	&\nu(\underline{d},\underline{i},a)\geq 0\quad \text{for all }(\underline{d},\underline{i},a)\in\mathbb{S}\times\mathcal{A}.\label{eq:lower_bound_6_3}
\end{align}
Recall that $Q$ in \eqref{eq:lower_bound_6_1} denotes the transition probability matrix given by	 \eqref{eq:MDP_transition_probabilities}. The left-hand side of \eqref{eq:lower_bound_6_1} represents the long-term probability of leaving the state $(\underline{d},\underline{i})$, while the right-hand side of \eqref{eq:lower_bound_6_2} represents the long-term probability of entering into the state $(\underline{d},\underline{i})$. Thus, \eqref{eq:lower_bound_6_1} is the \emph{global balance equation} for the controlled Markov process $\{(\underline{d}(t),\underline{i}(t)):t\geq K\}$. Equations \eqref{eq:lower_bound_6_2} and \eqref{eq:lower_bound_6_3} together imply that $\nu$ is a probability measure on $\mathbb{S}\times\mathcal{A}$.

As outlined in Section \ref{sec:notations}, the controlled Markov process $\{(\underline{d}(t),\underline{i}(t)):t\geq K\}$, together with the sequence $\{B_t:t\geq 0\}$ of intended arm selections (or equivalently the sequence $\{A_t:t\geq 0\}$ of actual arm selections), defines a Markov decision problem (MDP) with state space $\mathbb{S}$ and action space $\mathcal{A}$. From \cite[Lemma 1]{karthik2021detecting}, we know that $\{(\underline{d}(t), \underline{i}(t)): t\geq K\}$ is an ergodic Markov process under every SRS policy. This suffices to apply \cite[Theorem 2]{karthik2021detecting} to deduce a one-one correspondence between feasible solutions to \eqref{eq:lower_bound_6_1}-\eqref{eq:lower_bound_6_3} and policies in $\Pi_{\textsf{SRS}}$. In other words, \cite[Theorem 2]{karthik2021detecting} implies that for any given $\nu$ satisfying \eqref{eq:lower_bound_6_1}-\eqref{eq:lower_bound_6_3}, we can find an SRS policy $\pi^\lambda\in\Pi_{\textsf{SRS}}$ such that $\nu^\lambda(\underline{d},\underline{i},a)=\nu(\underline{d},\underline{i},a)$ for all $(\underline{d},\underline{i},a)\in\mathbb{S}\times\mathcal{A}$. (Recall that under the SRS policy $\pi^\lambda$, the stationary distribution of the Markov process $\{(\underline{d}(t),\underline{i}(t)):t\geq K\}$ is $\mu^\lambda$. The associated ergodic state occupancy measure, $\nu^\lambda$, is then defined according to \eqref{eq:ergodic_state_action_occupancy_measure}.)

Using \cite[Theorem 2]{karthik2021detecting} , we may replace the supremum in \eqref{eq:lower_bound_5} by a supremum over all SRS policies. Doing so leads us to the relation
\begin{align}
	& d(\epsilon,1-\epsilon)\nonumber\\
	&\leq \sup\limits_{\pi^\lambda\in \Pi_{\textsf{SRS}}}\,\inf\limits_{\substack{C'=(h', P_1', P_2'):\\h'\neq h, ~P_1'\neq P_2'}} \bigg\lbrace E^\pi\left[\sum\limits_{a=1}^{K} \log\frac{P^\pi_C(X_{a-1}^a)}{P^\pi_{C'}(X_{a-1}^a)}\right]+\bigg(E^\pi[\tau(\pi)-K+1|C]\bigg)\cdot\,\sum\limits_{(\underline{d},\underline{i})\in\mathbb{S}}~\sum\limits_{a=1}^{K}\nu^\lambda(\underline{d},\underline{i},a)\, k_{CC'}(\underline{d}, \underline{i}, a)\bigg\rbrace.\label{eq:lower_bound_7}
\end{align}
for all $\pi\in\Pi(\epsilon)$. Observe that the constant term multiplying $E^\pi[\tau(\pi)-K+1|C]$ in \eqref{eq:lower_bound_7} is finite; further, it is not a function of either $\epsilon$ or of $\pi\in\Pi(\epsilon)$. The finiteness of this constant follows from the following observation: denote by $\mu_C^a$ the stationary distribution of the transition probability matrix $P_C^a$ (i.e., $\mu_C^a=\mu_1$ for $a=h$ and $\mu_C^a=\mu_2$ for all $a\neq h$). An application of the ergodic theorem to the Markov process of arm $a$ yields
\begin{equation}
	D((P_C^a)^{d_a}(\cdot|i_a)\|(P_{C'}^a)^{d_a}(\cdot|i_a))\longrightarrow D(\mu_C^a\|\mu_{C'}^a)<\infty \quad \text{as }d_a\to\infty.
\end{equation}
Since every convergent sequence is bounded, we may write $D((P_C^a)^{d_a}(\cdot|i_a)\|(P_{C'}^a)^{d_a}(\cdot|i_a))\leq M$ for all $(\underline{d},\underline{i},a)\in\mathbb{S}\times\mathcal{A}$, where $0<M<\infty$. Using \eqref{eq:lower_bound_6_2}, it follows that the constant term multiplying $E^\pi[\tau(\pi)-K+1|C]$ in \eqref{eq:lower_bound_7} is bounded above by $M$. We also note that the first term inside the braces in \eqref{eq:lower_bound_7} does not depend on $\epsilon$. Since $d(\epsilon,1-\epsilon)\to d(0,1)=+\infty$ as $\epsilon\downarrow 0$, the boundedness of the constant multiplying $E^\pi[\tau(\pi)-K+1|C]$ implies that $\epsilon\downarrow 0$ is equivalent to $E^\pi[\tau(\pi)|C]\to \infty$ for all $\pi\in\Pi(\epsilon)$.
%Since \eqref{eq:lower_bound_7} holds for all $\pi\in\Pi(\epsilon)$, we get
%\begin{align}
%	\frac{1}{R^*(P_1,P_2)}\leq \inf\limits_{\pi\in\Pi(\epsilon)}\frac{E_h[\tau(\pi)-K+1]}{d(\epsilon,1-\epsilon)}.
%\end{align}
Letting $\epsilon\downarrow 0$, and using  $d(\epsilon,1-\epsilon)/\log(1/\epsilon)\longrightarrow 1$ as $\epsilon\downarrow 0$, we arrive at the lower bound in \eqref{eq:lower_bound}. This completes the proof of the proposition.

\section{Proof of Proposition \ref{prop:convergence_of_ML_estimates}}
\label{appndx:proof_of_prop_convergence_of_ML_estimates_of_TPMs}
We first state the analogues of \cite[Assumptions A1-A5 and A6.1]{borkar1982identification} as applicable to the context of this paper and then verify each of the assumptions. Throughout this section, we let $C_0=(h, P_1, P_2)$ denote the underlying arms configuration.
\begin{itemize}
	\item The analogue of \cite[Assumption A1]{borkar1982identification} is that for all arms configuration $C$, $(\underline{d}, \underline{i}), (\underline{d}', \underline{i}')\in \mathbb{S}$, and $b\in \mathcal{A}$, the transition probabilities $Q_C(\underline{d}', \underline{i}'|\underline{d}, \underline{i},b)$ (defined in \eqref{eq:MDP_transition_probabilities}) are continuous in $b$ and $C$.
	\item The analogue of \cite[Assumption A2]{borkar1982identification} simply states that the actual transition probabilities are those corresponding to the underlying arms configuration $C_0$.
	\item The analogue of \cite[Assumption A3]{borkar1982identification} is that if $Q_{C_0}(\underline{d}', \underline{i}'|\underline{d}, \underline{i},b)=0$ for some $(\underline{d}, \underline{i}), (\underline{d}', \underline{i}')\in \mathbb{S}$ and $b\in \mathcal{A}$, then $Q_{C}(\underline{d}', \underline{i}'|\underline{d}, \underline{i},b')=0$ for all $C$ and $b'\in \mathcal{A}$. Further, if $Q_{C_0}(\underline{d}', \underline{i}'|\underline{d}, \underline{i},b)>0$ for some $(\underline{d}, \underline{i}), (\underline{d}', \underline{i}')\in \mathbb{S}$ and $b\in \mathcal{A}$, then there exists $\bar{\epsilon}>0$ independent of $(\underline{d}, \underline{i}), (\underline{d}', \underline{i}')\in \mathbb{S}$ and $b\in \mathcal{A}$ such that for all $C$, the relation $$ \bar{\epsilon}~\leq~ \frac{Q_{C}(\underline{d}', \underline{i}' \mid \underline{d}, \underline{i}, b)}{Q_{C_0}(\underline{d}', \underline{i}' \mid \underline{d}, \underline{i}, b)} ~\leq~ \frac{1}{\bar{\epsilon}} $$ holds.
	\item The analogue of \cite[Assumption A4]{borkar1982identification} is that for all  $b\in \mathcal{A}$ and arms configuration $C$, the controlled Markov process $\{(\underline{d}(t), \underline{i}(t)): t\geq K\}$ with the control process $\{B_t:t\geq 0\}$ given by $B_t=b$ for all $t$, and transition probabilities given by $Q_C(\underline{d}', \underline{i}' \mid \underline{d}, \underline{i}, b)$, is a positive recurrent Markov process whose state space $\mathbb{S}$ is a single communicating class.
	\item Instead of stating the analogue of \cite[Assumption A5]{borkar1982identification}, we state the analogue of its equivalent form \cite[Condition C2]{federgruen1978note}; for a proof of this equivalence, we refer the reader to \cite{federgruen1978note} and the references therein. The analogue of \cite[Condition C2]{federgruen1978note} as applicable to the context of this paper is that there is a finite set $K$, an integer $d\geq 1$, and a number $\rho>0$ such that under every SRS policy $\pi^\lambda$, the probability of the process $\{(\underline{d}(t), \underline{i}(t)): t\geq K\}$ starting from any state $(\underline{d}, \underline{i})\in \mathbb{S}$ and hitting the set $K$ after $d$ time steps is lower bounded by $\rho$.
	\item The analogue of \cite[Assumption A6.1]{borkar1982identification} is that for any two arm configurations $C\neq C'$, there exists $(\underline{d}, \underline{i})\in \mathbb{S}$ (possibly depending on $C$ and $C'$) such that for every $b\in \mathcal{A}$, $$ Q_C(\cdot|\underline{d}, \underline{i}, b) \neq Q_{C'}(\cdot|\underline{d}, \underline{i}, b). $$
\end{itemize}

 We now proceed to verify each of the above stated assumptions. From \eqref{eq:MDP_transition_probabilities}, it is clear that the transition probabilities $Q_C(\underline{d}', \underline{i}'|\underline{d}, \underline{i}, b)$ are continuous in $C$ and $b$. This verifies the analogue of \cite[Assumption A1]{borkar1982identification}.

 Next, we note that if $Q_{C_0}(\underline{d}', \underline{i}' \mid \underline{d}, \underline{i}, b)=0$ for some $(\underline{d}', \underline{i}'), (\underline{d}, \underline{i}) \in \mathbb{S}$ and $b\in \mathcal{A}$, then it follows from \eqref{eq:MDP_TPM_2} that one of the following conditions must hold:
\begin{itemize}
	\item For all $a\in \mathcal{A}$, $$\mathbb{I}_{\{d_a'=1\text{ and }d'_{\tilde{a}}=d_{\tilde{a}}+1\text{ for all }\tilde{a}\neq a\}} \cdot \mathbb{I}_{\{i_{\tilde{a}}'=i_{\tilde{a}}\text{ for all }\tilde{a}\neq a\}}=0.$$ That is, $(\underline{d}', \underline{i}')$ is not a valid state that can be reached in one step from the state $(\underline{d}, \underline{i})$. Clearly, then, we have $Q_{C}(\underline{d}', \underline{i}' \mid \underline{d}, \underline{i}, b)=0$ for all $C$ and $b\in \mathcal{A}$.
	
	\item $\mathbb{I}_{\{d_a'=1\text{ and }d'_{\tilde{a}}=d_{\tilde{a}}+1\text{ for all }\tilde{a}\neq a\}} \cdot \mathbb{I}_{\{i_{\tilde{a}}'=i_{\tilde{a}}\text{ for all }\tilde{a}\neq a\}}=1$ for some $a\in \mathcal{A}$. In this case, $(\underline{d}', \underline{i}')$ is a valid state that can be reached in one step from the state $(\underline{d}, \underline{i})$. From \eqref{eq:MDP_transition_probabilities}, it follows that $$ Q_{C_0}(\underline{d}', \underline{i}' \mid \underline{d}, \underline{i}, b)=0 \Longleftrightarrow (P_{C_0}^a)^{d_a}(i_a'|i_a)=0. $$ Using Assumption \ref{assmptn:technical_assumption_on_TPMs}, we have $(P_{C_0}^a)^{d_a}(i_a'|i_a)=0 \Longleftrightarrow (P_{C}^a)^{d_a}(i_a'|i_a)=0$ for all $C$. Therefore, it follows that if $Q_{C}(\underline{d}', \underline{i}' \mid \underline{d}, \underline{i}, b)=0$, then $Q_{C}(\underline{d}', \underline{i}' \mid \underline{d}, \underline{i}, b')=0$ for all $C$ and $b'\in \mathcal{A}$.
\end{itemize}
Now, suppose that $Q_{C_0}(\underline{d}', \underline{i}' \mid \underline{d}, \underline{i}, b)>0$ for some $(\underline{d}', \underline{i}'), (\underline{d}, \underline{i})\in \mathbb{S}$ and $b\in \mathcal{A}$. From \eqref{eq:MDP_TPM_1}, this implies that there exists $a\in \mathcal{A}$ such that $P_{C_0}^{d_a}(i_a'|i_a)>0$. From Assumption \ref{assmptn:technical_assumption_on_TPMs}, we then have $P_{C_0}^{d_a}(i_a'|i_a)>\bar{\varepsilon}^*$, which together with \eqref{eq:MDP_TPM_1} gives us that $Q_{C_0}(\underline{d}', \underline{i}' \mid \underline{d}, \underline{i}, b)\geq \frac{\eta}{K}\cdot \bar{\varepsilon}^*$. Setting $\bar{\epsilon}=\frac{\eta}{K}\cdot \bar{\varepsilon}^*$, it follows that $$ \bar{\epsilon}~\leq~ \frac{Q_C(\underline{d}', \underline{i}' \mid \underline{d}, \underline{i}, b)}{Q_{C_0}(\underline{d}', \underline{i}' \mid \underline{d}, \underline{i}, b)} ~\leq~ \frac{1}{\bar{\epsilon}}.$$ This verifies the analogue of \cite[Assumption A3]{borkar1982identification}.

Next, from \cite[Lemma 1]{karthik2021detecting}, we know that under the trembling hand model and under every SRS policy $\pi^\lambda$, the process $\{(\underline{d}(t), \underline{i}(t)): t\geq K\}$ is ergodic, i.e., irreducible, aperiodic and positive recurrent. This verifies the analogue of \cite[Assumption A4]{borkar1982identification}.

To verify the analogue of \cite[Condition C2]{federgruen1978note}, let $\underline{d}'=(K, \ldots, 1)$ and $\underline{i}'=(i, \ldots, i)$, where $i\in \mathcal{S}$ is a fixed state. We now show that given any SRS policy $\pi^\lambda$, the probability of reaching the state $(\underline{d}', \underline{i}')$ starting from any state $(\underline{d}, \underline{i})$ is lower bounded by a strictly positive constant, say $\rho>0$, that is independent of $(\underline{d}, \underline{i})$ and the underlying arms configuration $C_0$. Note that the Markov process of each arm evolves on the common, finite state space $\mathcal{S}$. Therefore, from \cite[Proposition 1.7]{milgrom2002envelope}, we have that there exists $M$ sufficiently large such that $$ P_1^M(j|i)>0, \quad P_2^M(j|i)>0\quad \text{for all }i,j\in \mathcal{S}. $$ Fix an arbitrary $(\underline{d}, \underline{i})$, and suppose that $\underline{d}(t)=\underline{d}$ and $\underline{i}(t)=\underline{i}$ at some time $t=T_0$. Let the following sequence of arm selections and observations be obtained under $\pi^\lambda$: arm $1$ is pulled at time $t=T_0+	1$, arm $2$ is pulled at time $t=T_0+2$ and so on until arm $K$ is pulled at time $t=T_0+K$. Thereafter, arm $1$ is pulled at time $t=T_0+M+1$ and the state $i$ is observed on arm $1$, arm $2$ is pulled at time $t=T_0+M+2$ and the state $i$ is observed on arm $2$, and so on until arm $K$ is pulled at time $t=T_0+M+K$ and the state $i$ is observed on arm $K$.

Clearly, then, we have $\underline{d}(T_0+M+K+1)=\underline{d}'$, $\underline{i}(T_0+M+K+1)=\underline{i}'$. Suppose that the observations obtained from the arms $1, \ldots, K$ at times $T_0+1, \ldots, T_0+K$ are $s_1, \ldots, s_K \in \mathcal{S}$ respectively. Let $\underline{s}=(s_1, \ldots, s_K)$. Then,
\begin{align}
	& P^{\pi^\lambda}(\underline{d}(T_0+M+K+1)=\underline{d}', \underline{i}(T_0+M+K+1)=\underline{i}'\mid \underline{d}(T_0)=\underline{d}, \underline{i}(T_0)=\underline{i}, C_0) \nonumber\\
	& \geq \prod\limits_{t=T_0+1}^{T_0+K} P^{\pi^\lambda}(A_t|\underline{d}(t), \underline{i}(t), C_0)\cdot  \prod\limits_{t=T_0+M+1}^{T_0+M+K+1} P^{\pi^\lambda}(A_t|\underline{d}(t), \underline{i}(t), C_0) \cdot \prod\limits_{a=1}^{K} (P_{C_0}^a)^M(i|s_a)\nonumber\\
	& \stackrel{(a)}{\geq}  \prod\limits_{t=T_0+1}^{T_0+K} \frac{\eta}{K} \cdot  \prod\limits_{t=T_0+M+1}^{T_0+M+K+1} \frac{\eta}{K} \cdot \prod\limits_{a=1}^{K} (P_{C_0}^a)^M(i|s_a)\nonumber\\
	& = \left(\frac{\eta}{K}\right)^{2K} \prod\limits_{a=1}^{K} (P_{C_0}^a)^M(i|s_a),
	\label{eq:verification_of_assumption_a5_1}
\end{align}
where $(a)$ above follows by observing that for any $t$,
\begin{align*}
	P^{\pi^\lambda}(A_t|\underline{d}(t), \underline{i}(t), C_0) &= \frac{\eta}{K} + (1-\eta)\,\lambda(A_t|\underline{d}(t), \underline{i}(t)) \\
	& \geq \frac{\eta}{K}.
\end{align*}

From Assumption \ref{assmptn:technical_assumption_on_TPMs} we know that $P_1, P_2\in \mathscr{P}(\bar{\varepsilon}^*)$ for some $\bar{\varepsilon}^*>0$. This implies that $(P_{C_0}^a)^M(i|s_a)>\bar{\varepsilon}^*$ for all $a\in \mathcal{A}$. Using this in \eqref{eq:verification_of_assumption_a5_1}, and setting $\rho = \left(\frac{\eta}{K}\right)^{2K} (\bar{\varepsilon}^*)^K$, we have
\begin{equation}
	P^{\pi^\lambda}(\underline{d}(T_0+M+K+1)=\underline{d}', \underline{i}(T_0+M+K+1)=\underline{i}'\mid \underline{d}(T_0)=\underline{d}, \underline{i}(T_0)=\underline{i}, C_0) \geq \rho\quad \forall~(\underline{d}, \underline{i})\in \mathbb{S}.
	\label{eq:verification_of_assumption_a5_2}
\end{equation}
This verifies the analogue of \cite[Condition C2]{federgruen1978note}.

Lastly, in order to verify the analogue of \cite[Assumption A6.1]{borkar1982identification}, we show that a condition stronger than \cite[Assumption A6.1]{borkar1982identification}, one that is the equivalent of Mandl's identifiability condition for countable-state controlled Markov processes, holds in the context of this paper. This is demonstrated in the following lemma.
\begin{lemma}
\label{lemma:mandl_identifiability_condition}
	For all arm configurations $C=(h, P_1, P_2)$ and $C'=(h', P_1', P_2')$ such that $C\neq C'$ (i.e., $C$ and $C'$ differ in at least one component), there exists $(\underline{d}, \underline{i})\in \mathbb{S}$, possibly depending on $C$ and $C'$, such that
\begin{equation}
	Q_C(\cdot \mid \underline{d}, \underline{i}, b) \neq Q_{C'}(\cdot \mid \underline{d}, \underline{i}, b)\quad \text{for all }b\in \mathcal{A}.
	\label{eq:Mandl_identifiability_condition}
\end{equation}
\end{lemma}
\begin{IEEEproof}[Proof of Lemma \ref{lemma:mandl_identifiability_condition}]
We present the proof of the lemma under various cases.
\subsection{Case 1: $C=(h, P_1, P_2)$, $C'=(h', P_1, P_2)$, where $h'\neq h$}
We first consider the case when $C$ and $C'$ differ only in their first components. Recall that the transition probability matrices $P_1$ and $P_2$ satisfy the condition $P_2\neq P_1$. This means that there exist $i^*, j^*\in \mathcal{S}$ such that $P_1(j^*|i^*) \neq P_2(j^*|i^*)$. Fix $a^*=h$. Let $(\underline{d}, \underline{i})\in \mathbb{S}$ be such that $d_{a^*}=1$ and $i_{a^*}=i^*$. Also, let $(\underline{d}', \underline{i}')$ be such that $d_{a^*}'=1$, $d_a'=d_a+1$ for all $a\neq a^*$, $i_{a^*}'=j^*$ and $i_a'=i_a$ for all $a\neq a^*$. Then, for the above choices of $(\underline{d}, \underline{i})$ and $(\underline{d}', \underline{i}')$, it follows from \eqref{eq:MDP_transition_probabilities} that for all $b\in \mathcal{A}$,
\begin{align}
	Q_C(\underline{d}', \underline{i}'\mid \underline{d}, \underline{i}, b)=\left(\frac{\eta}{K}+(1-\eta)\,\mathbb{I}_{\{b=a^*\}}\right)~P_1(j^*|i^*),\nonumber\\
	Q_{C'}(\underline{d}', \underline{i}'\mid \underline{d}, \underline{i}, b)=\left(\frac{\eta}{K}+(1-\eta)\,\mathbb{I}_{\{b=a^*\}}\right)~P_2(j^*|i^*).
	\label{eq:proof_of_mandl_identifiability_1}
\end{align}
Because $P_1(j^*|i^*) \neq P_2(j^*|i^*)$, it follows that $Q_C(\underline{d}', \underline{i}'\mid \underline{d}, \underline{i}, b) \neq Q_{C'}(\underline{d}', \underline{i}'\mid \underline{d}, \underline{i}, b)$ for all $b\in \mathcal{A}$. This establishes \eqref{eq:Mandl_identifiability_condition}.

\subsection{Case 2: $C=(h, P_1, P_2)$, $C'=(h, P_1', P_2)$, where $P_1\neq P_1'$}
The proof for this case follows along the lines of that for Case $1$, with $a^*=h$ and the following modifications:
\begin{align}
	Q_C(\underline{d}', \underline{i}'\mid \underline{d}, \underline{i}, b)=\left(\frac{\eta}{K}+(1-\eta)\,\mathbb{I}_{\{b=a^*\}}\right)~P_1(j^*|i^*),\nonumber\\
	Q_{C'}(\underline{d}', \underline{i}'\mid \underline{d}, \underline{i}, b)=\left(\frac{\eta}{K}+(1-\eta)\,\mathbb{I}_{\{b=a^*\}}\right)~P_1'(j^*|i^*)
	\label{eq:proof_of_mandl_identifiability_2}
\end{align}
for all $b\in \mathcal{A}$, where $i^*, j^*\in \mathcal{S}$ are chosen such that $P_1(j^*|i^*) \neq P_1'(j^*|i^*)$.

\subsection{Case 3: $C=(h, P_1, P_2)$, $C'=(h, P_1, P_2')$, where $P_2\neq P_2'$}
The proof for this case follows along the lines of that for Case 1, with $a^*=h$ and the following modifications:
\begin{align}
	Q_C(\underline{d}', \underline{i}'\mid \underline{d}, \underline{i}, b)=\left(\frac{\eta}{K}+(1-\eta)\,\mathbb{I}_{\{b=a^*\}}\right)~P_2(j^*|i^*),\nonumber\\
	Q_{C'}(\underline{d}', \underline{i}'\mid \underline{d}, \underline{i}, b)=\left(\frac{\eta}{K}+(1-\eta)\,\mathbb{I}_{\{b=a^*\}}\right)~P_2'(j^*|i^*)
	\label{eq:proof_of_mandl_identifiability_3}
\end{align}
for all $b\in \mathcal{A}$, where $i^*, j^*\in \mathcal{S}$ are chosen such that $P_2(j^*|i^*) \neq P_2'(j^*|i^*)$.

\subsection{Case 4: $C=(h, P_1, P_2)$, $C'=(h', P_1', P_2)$, where $h'\neq h$, $P_1\neq P_1'$}
In this case, because $C'$ is a valid arms configuration, it follows that $P_1'\neq P_2$. The proof for this case then follows along the lines of that for Case $1$, with $a^*=h'$ and the following modifications:
\begin{align}
	Q_C(\underline{d}', \underline{i}'\mid \underline{d}, \underline{i}, b)=\left(\frac{\eta}{K}+(1-\eta)\,\mathbb{I}_{\{b=a^*\}}\right)~P_2(j^*|i^*),\nonumber\\
	Q_{C'}(\underline{d}', \underline{i}'\mid \underline{d}, \underline{i}, b)=\left(\frac{\eta}{K}+(1-\eta)\,\mathbb{I}_{\{b=a^*\}}\right)~P_1'(j^*|i^*)
	\label{eq:proof_of_mandl_identifiability_4}
\end{align}
for all $b\in \mathcal{A}$, where $i^*, j^*\in \mathcal{S}$ are chosen such that $P_1'(j^*|i^*) \neq P_2(j^*|i^*)$.

\subsection{Case 5: $C=(h, P_1, P_2)$, $C'=(h', P_1, P_2')$, where $h'\neq h$, $P_2\neq P_2'$}
The proof for this case follows along the lines of that for Case $1$, with $a^*\neq h,h'$ and the following modifications:
\begin{align}
	Q_C(\underline{d}', \underline{i}'\mid \underline{d}, \underline{i}, b)=\left(\frac{\eta}{K}+(1-\eta)\,\mathbb{I}_{\{b=a^*\}}\right)~P_2(j^*|i^*),\nonumber\\
	Q_{C'}(\underline{d}', \underline{i}'\mid \underline{d}, \underline{i}, b)=\left(\frac{\eta}{K}+(1-\eta)\,\mathbb{I}_{\{b=a^*\}}\right)~P_2'(j^*|i^*)
	\label{eq:proof_of_mandl_identifiability_5}
\end{align}
for all $b\in \mathcal{A}$, where $i^*, j^*\in \mathcal{S}$ are chosen such that $P_2(j^*|i^*) \neq P_2'(j^*|i^*)$.

\subsection{Case 6: $C=(h, P_1, P_2)$, $C'=(h, P_1', P_2')$, where $P_1\neq P_1'$, $P_2\neq P_2'$}
The proof for this case follows along the lines of that for Case $1$, with $a^*=h$ and the following modifications:
\begin{align}
	Q_C(\underline{d}', \underline{i}'\mid \underline{d}, \underline{i}, b)=\left(\frac{\eta}{K}+(1-\eta)\,\mathbb{I}_{\{b=a^*\}}\right)~P_1(j^*|i^*),\nonumber\\
	Q_{C'}(\underline{d}', \underline{i}'\mid \underline{d}, \underline{i}, b)=\left(\frac{\eta}{K}+(1-\eta)\,\mathbb{I}_{\{b=a^*\}}\right)~P_1'(j^*|i^*)
	\label{eq:proof_of_mandl_identifiability_6}
\end{align}
for all $b\in \mathcal{A}$, where $i^*, j^*\in \mathcal{S}$ are chosen such that $P_1(j^*|i^*) \neq P_1'(j^*|i^*)$.

\subsection{Case 7: $C=(h, P_1, P_2)$, $C'=(h', P_1', P_2')$, where $h'\neq h$, $P_1\neq P_1'$, $P_2\neq P_2'$}
The proof for this case follows along the lines of that for Case $1$, with $a^*=h$ and the following modifications:
\begin{align}
	Q_C(\underline{d}', \underline{i}'\mid \underline{d}, \underline{i}, b)=\left(\frac{\eta}{K}+(1-\eta)\,\mathbb{I}_{\{b=a^*\}}\right)~P_1(j^*|i^*),\nonumber\\
	Q_{C'}(\underline{d}', \underline{i}'\mid \underline{d}, \underline{i}, b)=\left(\frac{\eta}{K}+(1-\eta)\,\mathbb{I}_{\{b=a^*\}}\right)~P_1'(j^*|i^*)
	\label{eq:proof_of_mandl_identifiability_7}
\end{align}
for all $b\in \mathcal{A}$, where $i^*, j^*\in \mathcal{S}$ are chosen such that $P_1(j^*|i^*) \neq P_1'(j^*|i^*)$.
This completes the proof of the lemma and also the verification of the analogue of \cite[Assumption A6.1]{borkar1982identification}.
\end{IEEEproof}

Finally, with the analogues of \cite[Assumptions A1-A5 and A6.1]{borkar1982identification} being verified, we apply \cite[Theorem 4.3]{borkar1982identification} (which simply states that under \cite[Assumptions A1-A5 and A6.1]{borkar1982identification}, the ML estimates converge to their true values almost surely) to deduce that under the arms configuration $C_0$, $$\hat{P}_{h, 1}(n)\longrightarrow P_1,\quad \hat{P}_{h, 2}(n)\longrightarrow P_2 \quad\text{as }n\to\infty ,\quad\text{almost surely},$$ thus proving  Proposition \ref{prop:convergence_of_ML_estimates}. We note here that in addition to \cite[Assumptions A1-A5 and A6.1]{borkar1982identification}, the proof of \cite[Theorem 4.3]{borkar1982identification} uses the notion of ``$\{\varepsilon_i\}$-randomisation" which, for controlled Markov processes, ensures that the probability of selecting any control at any given time is strictly positive. The trembling hand model \eqref{eq:trembling_hand_relation} of our paper guarantees that the probability of sampling any arm at any given time is $\geq \frac{\eta}{K}>0$, thus alleviating the need to consider $\{\varepsilon_i\}$ randomisation in our work.

\section{Proof of Proposition \ref{prop:strict_positivity_of_drift_of_test_statistic}}
\label{appndx:proof_of_prop_strict_positivity_of_drift_of_test_statistic}
We first note the following important points.
\begin{itemize}
	\item From the exposition in \cite[Appendix C]{karthik2021detecting}, we note that
	\begin{equation}
		\liminf\limits_{n \to \infty} \frac{N(n, \underline{d}, \underline{i}, a, j)}{n} > 0, \quad \liminf\limits_{n \to \infty} \frac{N(n, \underline{d}, \underline{i}, a)}{n} > 0 \quad \text{almost surely}
		\label{eq:liminf_of_two_qts_strictly_positive}
	\end{equation}
	for all $(\underline{d}, \underline{i})\in \mathbb{S}$, $a\in \mathcal{A}$ and $j\in \mathcal{S}$.
	 Therefore, by the ergodic theorem, it follows that for all $(\underline{d}, \underline{i})\in \mathbb{S}$, $a\in \mathcal{A}$ and $j\in \mathcal{S}$ and arms configuration $C$, the following pointwise convergence holds almost surely:
	\begin{equation}
		\frac{N(n, \underline{d}, \underline{i}, a, j)}{N(n, \underline{d}, \underline{i}, a)} \longrightarrow (P_C^a)^{d_a}(j\mid i_a)\quad \text{as }n\to \infty.
		\label{eq:convergence_for_each_d_i_ergodic_theorem}
	\end{equation}

    \item For each $(\underline{d}, \underline{i})\in \mathbb{S}$ and $a\in \mathcal{A}$, the sequence
    \begin{equation}
    	\bigg\lbrace \frac{N(n, \underline{d}, \underline{i}, a)}{n} \bigg \rbrace_{n\geq K}
    	\label{eq:sequence_M(n,d,i,a)/n}
    \end{equation}
    is bounded almost surely. This implies that there exists a null set $B$  such that for every $\omega \notin B$ and $(\underline{d}, \underline{i})\in \mathbb{S}$, there exists a subsequence, say $\{n_k(\omega, \underline{d}, \underline{i}): k\geq 1 	\}$, along which the above sequence converges. Because each term of the sequence in \eqref{eq:sequence_M(n,d,i,a)/n} lies in the compact interval $[0, 1]$, an application Tychonoff's theorem \cite[Theorem 37.3]{munkres2000topology} gives us that $[0, 1]^{|\mathbb{S}|}$ is compact; here, $[0, 1]^{|\mathbb{S}|}$ denotes the countable product of $[0, 1]$ with itself. Using an Arzela-Ascoli type diagonalization argument, there exists a non-empty subsequence,  call it $\{n_k(\omega):k\geq 1\}$, along which the sequence in \eqref{eq:sequence_M(n,d,i,a)/n} converges for all $(\underline{d}, \underline{i})\in \mathbb{S}$. Fixing attention to the subsequence $\{n_k(\omega): k\geq 	1\}$, $\omega \notin B$, let
    $$\frac{N(n_k(\omega), \underline{d}, \underline{i}, a)(\omega)}{n_k(\omega)} \longrightarrow \alpha(\underline{d}, \underline{i}, a, \omega) \quad \text{as}\quad k\to \infty,$$ where the limit $\alpha(\underline{d}, \underline{i}, a, \omega)>0$ for all $(\underline{d}, \underline{i})\in \mathbb{S}$, $ a\in \mathcal{A}$, and  $\omega\notin B$, thanks to \eqref{eq:liminf_of_two_qts_strictly_positive}.
\end{itemize}

Combining the above mentioned points, it follows that for all $\omega\notin B$, when $C$ is the underlying arms configuration,
    \begin{equation}
    	\frac{N(n_k(\omega), \underline{d}, \underline{i}, a, j)(\omega)}{n_k(\omega)} \longrightarrow \alpha(\underline{d}, \underline{i}, a, \omega)\cdot  (P_C^a)^{d_a}(j\mid i_a)\quad \text{as}\quad k\to \infty.
    	\label{eq:convergence_outside_of_a_null_set}
    \end{equation}
The following lemma shows that the convergence in  \eqref{eq:convergence_outside_of_a_null_set} is, in fact, uniform in $(\underline{d}, \underline{i})$. We omit the proof of the lemma, noting that it is a simple exercise in real analysis.
\begin{lemma}
\label{lemma:uniform_convergence}
	Let $\mathscr{S}$ be a countable set, and let $f_k, f:\mathscr{S}\to [0,1]$, $k\geq 1$, be such that $$\sum\limits_{s\in \mathscr{S}} f_k(s)<\infty\quad \forall ~k\geq 1,\quad \sum\limits_{s\in \mathscr{S}} f(s)<\infty.$$
	If $f_k\to f$ pointwise as $k\to \infty$, then $f_k\to f$ uniformly as $k\to \infty$.
\end{lemma}

We now begin the proof of Proposition \ref{prop:strict_positivity_of_drift_of_test_statistic}.
%Let $C=(h, P_1, P_2)$, and suppose that $C$ is the underlying arms configuration. Obviously, the policy $\pi^\star(L,\delta)$ is oblivious to the knowledge of $C$.
Fix an arbitrary null set $B$. Then, for every $\omega \notin B$, there exists a subsequence $\{n_k(\omega):k \geq 1\}$ such that
$$\liminf\limits_{n \to \infty} \frac{M_{hh'}(n)( \omega)}{n}=\lim\limits_{k \to \infty} \frac{M_{hh'}(n_k(\omega))( \omega)}{n_k(\omega)}.$$
Let us restrict our attention on a further subsequence (obtained as described earlier using Tychonoff's  theorem), and without loss of generality, let $\{n_k(\omega):k \geq 1\}$ be this subsequence. We then have
\begin{equation}
	\lim\limits_{k \to \infty} \frac{M_{hh'}(n_k(\omega))( \omega)}{n_k(\omega)} =\lim\limits_{k \to \infty}  \frac{T_1(n_k(\omega))+T_2(n_k(\omega))+T_3(n_k(\omega))+T_4(n_k(\omega))}{n_k(\omega)}.
	\label{eq:proof_of_strict_positivity_1}
\end{equation}
We shall demonstrate that the right hand side of \eqref{eq:proof_of_strict_positivity_1} is strictly positive for all $\omega \notin B$. Fix an arbitrary $\varepsilon\in (0,1)$.
\begin{itemize}
	\item {\color{black} From \eqref{eq:T_1(n)}, we have
	\begin{align}
		&\frac{T_1(n_k(\omega))}{n_k(\omega)}\nonumber\\
		&=\frac{1}{n_k(\omega)}~\log \mathbb{E}\bigg[\exp\bigg\lbrace\log\bigg(\sum\limits_{i\in \mathcal{S}}~\phi(i)\,P^{h-1}(X_{h-1}^h|i)\bigg)+\sum\limits_{(\underline{d}, \underline{i})\in \mathbb{S}}~\sum\limits_{j\in \mathcal{S}}\frac{N(n_k(\omega), \underline{d}, \underline{i}, h, j, \omega)}{n_k(\omega)}~ \log P^{d_h}(j|i_h)\bigg\rbrace\bigg]\nonumber\\
		&=\frac{1}{n_k(\omega)}~\log \mathbb{E}\bigg[\exp\bigg\lbrace\log\bigg(\sum\limits_{i\in \mathcal{S}}~\phi(i)\,P^{h-1}(X_{h-1}^h|i)\bigg)+\sum\limits_{(\underline{d}, \underline{i})\in \mathbb{S}}~\sum\limits_{j\in \mathcal{S}}\frac{N(n_k(\omega), \underline{d}, \underline{i}, h, j, \omega)}{n_k(\omega)}~ \log \frac{P^{d_h}(j|i_h)}{P_1^{d_h}(j|i_h)}\bigg\rbrace\bigg]\nonumber\\
		&\hspace{3cm}+\sum\limits_{(\underline{d}, \underline{i})\in \mathbb{S}}~\sum\limits_{j\in \mathcal{S}}\frac{N(n_k(\omega), \underline{d}, \underline{i}, h, j, \omega)}{n_k(\omega)}~ \log P_1^{d_h}(j|i_h).
		\label{eq:proof_of_strict_positivity_2}
\end{align}
Noting that the expectation in the first term of \eqref{eq:proof_of_strict_positivity_2} is of a non-negative (exponential) function, we may lower bound this expectation term by
\begin{align}
	&\mathbb{E}\bigg[\exp\bigg\lbrace\log \left(\sum\limits_{i\in \mathcal{S}}~\phi(i)\, P^{h-1}(X_{h-1}^{h}|i)\right) + \sum\limits_{(\underline{d}, \underline{i})\in \mathbb{S}}~\sum\limits_{j\in \mathcal{S}}\frac{N(n_k(\omega), \underline{d}, \underline{i}, h, j, \omega)}{n_k(\omega)}~ \log \frac{P^{d_{h }}(j|i_{h})}{P_1^{d_{h }}(j|i_{h})}\bigg\rbrace\cdot \mathbb{I}\bigg(P\in \mathscr{P}(\bar{\varepsilon}^*)\bigg)\bigg]\nonumber\\
	&\geq \exp\bigg\lbrace\log (\phi_{\textsf{min}}\cdot \bar{\varepsilon}^*)+(\log \bar{\varepsilon}^*)\cdot \sum\limits_{(\underline{d}, \underline{i})\in \mathbb{S}}~\sum\limits_{j\in \mathcal{S}}\frac{N(n_k(\omega), \underline{d}, \underline{i}, h, j, \omega)}{n_k(\omega)} \bigg\rbrace\cdot \mathbb{P}(P\in \mathscr{P}(\bar{\varepsilon}^*))\nonumber\\
	&\geq \exp\bigg\lbrace\log \phi_{\textsf{min}}+\log \bar{\varepsilon}^*+\log \bar{\varepsilon}^*\bigg\rbrace\cdot \mathbb{P}(P\in \mathscr{P}(\bar{\varepsilon}^*)),
	\label{eq:proof_of_strict_positivity_2_new_1}
\end{align} 
where the first inequality follows by noting that on the set $\{P\in \mathscr{P}(\bar{\varepsilon}^*)\}$, we have\footnote{Note that when $P\sim D$, where $D$ is the prior on $\mathscr{P}(\mathcal{S})$, we have $D(P^d(j|i)>0$ for all $d\geq 1$, $i, j\in \mathcal{S})=1$.} 
\begin{align*}
	P^d(j|i)>0,~ P_1^d(j|i)>0 \implies \frac{P^d(j|i)}{P_1^d(j|i)}\geq \bar{\varepsilon}^* \quad \text{for all }d\geq 1, ~i, j\in \mathcal{S}. 
\end{align*}
 Also, in the first inequality above, $\phi_{\textsf{min}}=\min\limits_{i\in\mathcal{S}}\phi(i)>0$. The second inequality above follows by using $$ \sum\limits_{(\underline{d}, \underline{i})\in \mathbb{S}}~\sum\limits_{j\in \mathcal{S}}\frac{N(n_k(\omega), \underline{d}, \underline{i}, h, j, \omega)}{n_k(\omega)} \leq 1. $$
 Applying logarithm to both sides of \eqref{eq:proof_of_strict_positivity_2_new_1}, we get
 \begin{align}
 	\frac{T_1(n_k(\omega))}{n_k(\omega)}\geq \frac{\log \phi_{\textsf{min}}+2\log \bar{\varepsilon}^*+\log \mathbb{P}(P\in \mathscr{P}(\bar{\varepsilon}^*))}{n_k(\omega)}+\sum\limits_{(\underline{d}, \underline{i})\in \mathbb{S}}~\sum\limits_{j\in \mathcal{S}}\frac{N(n_k(\omega), \underline{d}, \underline{i}, h, j, \omega)}{n_k(\omega)}~ \log P_1^{d_h}(j|i_h).
 	\label{eq:proof_of_strict_positivity_2_new_2}
 \end{align}
 We now claim that $\mathbb{P}(P\in \mathscr{P}(\bar{\varepsilon}^*))>0$, and therefore the limiting value of the first term in \eqref{eq:proof_of_strict_positivity_2_new_2} is equal to zero. To verify the claim, note that\footnote{In order for the set $\mathscr{B}(\bar{\varepsilon}^*)$ to be non-empty, it is important that $\bar{\varepsilon}^* < \frac{1}{|\mathcal{S}|}$. If this is not true, $\bar{\varepsilon}^*$ may be replaced by $\frac{\bar{\varepsilon}^*}{|\mathcal{S}|}$ without altering the proof.} $$ \mathscr{B}(\bar{\varepsilon}^*)\coloneqq \bigg\lbrace P\in \mathscr{P}(\mathcal{S}):P(j|i)> \bar{\varepsilon}^*\text{ for all }i, j\in \mathcal{S}\bigg\rbrace \subset \mathscr{P}(\bar{\varepsilon}^*)$$ is an open set (with respect to the relative topology on $\mathscr{P}(\mathcal{S})$ induced by the topology arising from the Euclidean metric on $\mathbb{R}^{|\mathcal{S}|(|\mathcal{S}|-1)}$). Also, $D(\mathscr{B}(\bar{\varepsilon}^*))>0$\footnote{Recall that the prior $D$ is induced by sampling each row of $P\in \mathscr{P}(\mathcal{S})$ independently according to $\text{Dir}(\mathbf{1})$. Because $\text{Dir}(\mathbf{1})$ is simply the uniform probability distribution on the probability simplex $\mathscr{P}(\mathcal{S})$, every open set has a strictly positive probability under $D$.}. Therefore, we have
 \begin{align}
 	\mathbb{P}(P\in \mathscr{P}(\bar{\varepsilon}^*)) &\geq \mathbb{P}(\mathscr{B}(\bar{\varepsilon}^*))
 	\nonumber\\
 	&=D(\mathscr{B}(\bar{\varepsilon}^*))\nonumber\\
 	&>0.
 \end{align}
 Thus, there exists $K_{11}=K_{11}(\varepsilon, \omega)$ such that for all $k\geq K_{11}$, we have
 \begin{align}
 	\frac{\log \phi_{\textsf{min}}+2\log \bar{\varepsilon}^*+\log \mathbb{P}(P\in \mathscr{P}(\bar{\varepsilon}^*))}{n_k(\omega)}\geq -\varepsilon.
 	\label{eq:proof_of_strict_positivity_2_new_3}
 \end{align}
 
 We now turn our attention to the second term in \eqref{eq:proof_of_strict_positivity_2_new_2}. From the convergence in \eqref{eq:convergence_outside_of_a_null_set}, we know that given any $\varepsilon'>0$, there exists $K_{12}=K_{12}(\varepsilon', \omega)$ (the dependence of $K_{12}$ on $(\underline{d}, \underline{i})\in \mathbb{S}$ is removed thanks to Lemma \ref{lemma:uniform_convergence}) such that under the arms configuration $C=(h, P_1, P_2)$,
 \begin{align}
 	\frac{N(n_k(\omega), \underline{d}, \underline{i}, h, j, \omega)}{n_k(\omega)}\leq \alpha(\underline{d}, \underline{i}, h, \omega)\cdot P_1^{d_h}(j|i_h)\cdot (1+\varepsilon')
 	\label{eq:proof_of_strict_positivity_2_new_4}
 \end{align}
 for all $k\geq K_{12}$. Combining \eqref{eq:proof_of_strict_positivity_2_new_3} and \eqref{eq:proof_of_strict_positivity_2_new_4}, we get that for all $k\geq K_1=\max\{K_{11}, K_{12}\}$, 
\begin{align}
	\frac{T_1(n_k(\omega))}{n_k(\omega)} \geq (1+\varepsilon') \sum\limits_{(\underline{d}, \underline{i})\in \mathbb{S}}~\sum\limits_{j\in\mathcal{S}}~ \alpha(\underline{d}, \underline{i}, h, \omega)\,\, P_1^{d_h}(j|i_h) ~\log P_1^{d_h}(j|i_h) - \varepsilon.
	\label{eq:proof_of_strict_positivity_2_new_5}
\end{align}
Choose $\varepsilon'$ so that
\begin{align}
	\frac{T_1(n_k(\omega))}{n_k(\omega)} \geq  \sum\limits_{(\underline{d}, \underline{i})\in \mathbb{S}}~\sum\limits_{j\in\mathcal{S}}~ \alpha(\underline{d}, \underline{i}, h, \omega)\,\, P_1^{d_h}(j|i_h) ~\log P_1^{d_h}(j|i_h) - 2\,\varepsilon.
	\label{eq:liminf_T_1(n)/n_final}
\end{align}

}
	\item Similar arguments as above can be used to show that there exists $K_2=K_2(\varepsilon, \omega)$ such that for all $k\geq K_2$, under the arms configuration $C=(h, P_1, P_2)$,
		\begin{equation}
			\frac{T_2(n_{k}(\omega))}{n_k(\omega)} \geq  \sum\limits_{a\neq h}~\sum\limits_{(\underline{d}, \underline{i})\in \mathbb{S}}~\sum\limits_{j\in\mathcal{S}}~ \alpha(\underline{d}, \underline{i}, a, \omega)\,\, P_2^{d_a}(j|i_a) ~\log P_2^{d_a}(j|i_a) -2\, \varepsilon.
			\label{eq:liminf_T_2(n)/n_final}
		\end{equation}
	
	\item We now handle the term $\frac{T_3(n_k(\omega))}{n_k(\omega)}$. Note that under the arms configuration $C=(h, P_1, P_2)$, it follows from \eqref{eq:convergence_of_ML_estimates_of_TPMs_under_h'} that there exists $K_3=K_3(\varepsilon, \omega)$ such that
		\begin{equation}
			\frac{T_3(n_k(\omega))}{n_k(\omega)} \geq \sum\limits_{(\underline{d}, \underline{i})\in \mathbb{S}}~\sum\limits_{j\in \mathcal{S}}\alpha(\underline{d}, \underline{i}, h', j, \omega)~ P_2^{d_{h'}}(j|i_{h'})~ \log \frac{1}{P_2^{d_{h'}}(j|i_{h'})} - \varepsilon
			\label{eq:liminf_T_3(n)/n_final}
		\end{equation}
		for all $k\geq K_3$ (in arriving at \eqref{eq:liminf_T_3(n)/n_final}, the first term in \eqref{eq:T_3(n)}, being non-negative, is lower bounded by $0$). Also, there exists $K_4=K_4(\varepsilon, \omega)$ such that
		{\color{black} \begin{equation}
			\frac{T_4(n_k(\omega))}{n_k(\omega)} \geq \sum\limits_{(\underline{d}, \underline{i})\in \mathbb{S}}~\sum\limits_{j\in \mathcal{S}}\alpha(\underline{d}, \underline{i}, h, \omega)~ P_1^{d_{h}}(j|i_{h})~ \log \frac{1}{P^{d_{h}}(j|i_{h})}+\sum\limits_{a\neq h, h'}\alpha(\underline{d}, \underline{i}, a, \omega)~ P_2^{d_{a}}(j|i_{a})~ \log \frac{1}{P^{d_{a}}(j|i_{a})} - \varepsilon.
			\label{eq:liminf_T_4(n)/n_final}
		\end{equation}}
		for all $k\geq K_4$.
\end{itemize}
Combining the inequalities in \eqref{eq:liminf_T_1(n)/n_final}-\eqref{eq:liminf_T_4(n)/n_final}, we see that for all $\varepsilon>0$ and for all $\omega\notin B$,
\begin{align}
	\frac{M_{hh'}(n_k(\omega))}{n_k(\omega)}\geq
\sum\limits_{(\underline{d}, \underline{i})\in \mathbb{S}}\alpha(\underline{d}, \underline{i}, h, \omega)\cdot  D(P_1^{d_h}(\cdot|i_h) \| P^{d_h}(\cdot|i_h)) + \sum\limits_{a\neq h, h'} \alpha(\underline{d}, \underline{i}, a, \omega) \cdot D(P_2^{d_a}(\cdot|i_a) \| P^{d_a}(\cdot|i_a))-6\varepsilon
\end{align}
for all $k\geq \max\{K_1, \ldots, K_4\}$, from which it follows that
\begin{align}
	\liminf\limits_{n\to\infty}\frac{M_{hh'}(n)( \omega)}{n}
	&=\lim\limits_{k\to \infty}\frac{M_{hh'}(n_k(\omega))}{n_k(\omega)}\nonumber\\
	&\geq
\sum\limits_{(\underline{d}, \underline{i})\in \mathbb{S}}\alpha(\underline{d}, \underline{i}, h, \omega)\cdot  D(P_1^{d_h}(\cdot|i_h) \| P^{d_h}(\cdot|i_h)) + \sum\limits_{a\neq h, h'} \alpha(\underline{d}, \underline{i}, a, \omega) \cdot D(P_2^{d_a}(\cdot|i_a) \| P^{d_a}(\cdot|i_a))-6\varepsilon.
\label{eq:proof_of_strict_positivity_6}
\end{align}
Letting $\varepsilon\downarrow 0$ in \eqref{eq:proof_of_strict_positivity_6}, we get
\begin{align}
	\liminf\limits_{n\to\infty}\frac{M_{hh'}(n)( \omega)}{n}&\geq
\sum\limits_{(\underline{d}, \underline{i})\in \mathbb{S}}\alpha(\underline{d}, \underline{i}, h, \omega)\cdot  D(P_1^{d_h}(\cdot|i_h) \| P^{d_h}(\cdot|i_h)) + \sum\limits_{a\neq h, h'} \alpha(\underline{d}, \underline{i}, a, \omega) \cdot D(P_2^{d_a}(\cdot|i_a) \| P^{d_a}(\cdot|i_a))
\label{eq:proof_of_strict_positivity_7}
\end{align}
for all $\omega \notin B$.
%We now claim that the right hand side of \eqref{eq:proof_of_strict_positivity_7} is strictly positive. Indeed, from \cite[Eq. (103), pp. 30]{karthik2021detecting}, we know that $$\liminf\limits_{n\to \infty}\frac{N(n,\underline{d}, \underline{i}, a)}{n}>0\quad \text{almost surely.}$$
Because $\alpha(\underline{d}, \underline{i}, a, \omega)>0$ for all $(\underline{d}, \underline{i})\in \mathbb{S}$, $a\in \mathcal{A}$, and $\omega \notin B$, the right hand side of \eqref{eq:proof_of_strict_positivity_7} is strictly positive. This establishes \eqref{eq:test_statistic_has_strictly_positive_drift}.

\section{Proof of Proposition \ref{prop:policy_satisfies_desried_error_probability}}
\label{appndx:proof_of_prop_policy_satisfies_desired_error_prob}
  The policy $\pi^\star(L,\delta)$ commits error if one of the following events is true:
\begin{enumerate}
	\item The policy never stops in finite time.
	\item The policy stops in finite time and, when $C=(h, P_1, P_2)$ is the actual (underlying) arms configuration, declares $h'\neq h$ as the true index of the odd arm.
\end{enumerate}
The event in item $1$ above has zero probability thanks to Proposition \ref{prop:strict_positivity_of_drift_of_test_statistic}.
Thus, the probability of error of policy $\pi=\pi^{\star}(L,\delta,\gamma)$, which we denote by $P^\pi_e$, may be evaluated as follows: suppose $C=(h,P_1,P_2)$ is the underlying arms configuration. Then,
\begingroup\allowdisplaybreaks\begin{align}
	P^\pi_e =P^\pi(\theta(\tau(\pi))\neq h|C)
	=P^\pi\bigg(\exists~ n\text{ and }~h'\neq h\text{ such that }
	\theta(\tau(\pi))=h'\text{ and } \tau(\pi)=n\bigg\vert C\bigg).\label{eq:P_e_partial_1}
\end{align}\endgroup
Let $\mathcal{R}_{h'}(n)\coloneqq\{\omega:\tau(\pi)(\omega)=n,\,\theta(n,\omega)=h'\}\label{eq:R_{h'}(n)}$ denote the set of all sample paths for which the policy stops at time $n$ and declares $h'\neq h$ as the true index of the odd arm. Clearly, the collection $\{\mathcal{R}_{h'}(n):h'\neq h,\,n\geq 0\}$ is a collection of mutually disjoint sets. Therefore, we have
\begin{align}
	& P^\pi_e =P^\pi\left(\bigcup\limits_{h'\neq h}\,\bigcup\limits_{n=0}^{\infty}\mathcal{R}_{h'}(n)\bigg\vert C\right)\nonumber\\
	&= \sum\limits_{h'\neq h}\sum\limits_{n=0}^{\infty}P^\pi(\tau(\pi)=n,\theta(n)=h'|C)\nonumber\\
	&= \sum\limits_{h'\neq h}\sum\limits_{n=0}^{\infty}~\int\limits_{\mathcal{R}_{h'}(n)}\,dP^\pi(\omega|C)\nonumber\\
	&=\sum\limits_{h'\neq h}\sum\limits_{n=0}^{\infty}~\int\limits_{\mathcal{R}_{h'}(n)}f(B^n(\omega), A^n(\omega), \bar{X}^n(\omega)|C)~d(B^n(\omega), A^n(\omega),\bar{X}^n(\omega))\nonumber\\
	& \stackrel{(a)}{\leq} \sum\limits_{h'\neq h}\sum\limits_{n=0}^{\infty}~\int\limits_{\mathcal{R}_{h'}(n)}\hat{f}(B^n(\omega), A^n(\omega), \bar{X}^n(\omega)| H_h)~d(B^n(\omega), A^n(\omega),\bar{X}^n(\omega))\nonumber\\
	&= \sum\limits_{h'\neq h}\sum\limits_{n=0}^{\infty}~\int\limits_{\mathcal{R}_{h'}(n)}\frac{\hat{f}(B^n(\omega), A^n(\omega), \bar{X}^n(\omega)| H_h)}{\bar{f}(B^n(\omega), A^n(\omega), \bar{X}^n(\omega)| H_{h'})}\cdot \bar{f}(B^n(\omega), A^n(\omega), \bar{X}^n(\omega)| H_{h'})~d(B^n(\omega), A^n(\omega),\bar{X}^n(\omega))\nonumber\\
	&= \sum\limits_{h'\neq h}\sum\limits_{n=0}^{\infty}~\int\limits_{\mathcal{R}_{h'}(n)}e^{-M_{h'h}(n)(\omega)}~\bar{f}(B^n(\omega), A^n(\omega),\bar{X}^n(\omega)|H_{h'})~ d(B^n(\omega), A^n(\omega),\bar{X}^n(\omega))\nonumber\\
	&\stackrel{(b)}{\leq} \sum\limits_{h'\neq h}\sum\limits_{n=0}^{\infty}~\int\limits_{\mathcal{R}_{h'}(n)}\frac{1}{(K-1)L}~\bar{f}(B^n(\omega), A^n(\omega), \bar{X}^n(\omega)| \mathcal{H}_{h'})~d(B^n(\omega), A^n(\omega),\bar{X}^n(\omega))\nonumber\\
	&=\sum\limits_{h'\neq h}\frac{1}{(K-1)L}~\bar{P}^\pi\left(\bigcup\limits_{n=0}^{\infty}\mathcal{R}_{h'}(n)\bigg|\mathcal{H}_{h'}\right)\nonumber\\
	&\leq\frac{1}{L},
	\label{eq:proof_of_admissibility_of_policy}
\end{align}
where $(a)$ follows from the definition of the maximum likelihood $\hat{f}$, $(b)$ follows from noting that when $\pi^\star(L, \delta)$ stops at time $n$ and outputs $h'$ as the odd arm, we must have $M_{h'h}(n)\geq \log((K-1)L)$ almost surely, and $\bar{P}^\pi$ in \eqref{eq:proof_of_admissibility_of_policy} denotes the probability measure under the average likelihood $\bar{f}$. Setting $L=1/\epsilon$ yields the desired result.

\section{Proof of Proposition \ref{prop:test_statistic_has_the_correct_drift}}
\label{appndx:proof_of_prop_test_statistic_has_the_correct_drift}
The convergences in \eqref{eq:convergence_of_lambdas_pointwise} imply that for each $a\in \mathcal{A}$ and $(\underline{d}, \underline{i}) \in \mathbb{S}$, almost surely,
\begin{align}
	P(A_n=a \mid \underline{d}(n)=\underline{d},~ \underline{i}(n)=\underline{i},~ \mathcal{F}_{n-1}) &= \frac{\eta}{K} + (1-\eta)\, \lambda_{\theta(n), \hat{P}_{\theta(n), 1}(n), \hat{P}_{\theta(n), 2}(n), \delta}(a \mid \underline{d}, \underline{i})\nonumber\\
	&\longrightarrow \frac{\eta}{K} + (1-\eta)\, \lambda_{h, P_1, P_2, \delta}(a \mid \underline{d}, \underline{i})\quad \text{as }n \to \infty.
	\label{eq:convergence_of_arm_selection_probabilities}
\end{align}

Recall that for any conditional distribution $\lambda=\lambda(\cdot|\cdot)$, the unique stationary distribution of the ergodic Markov process $\{(\underline{d}(t), \underline{i}(t)): t\geq K\}$ under $\pi^\lambda \in \Pi_{\textsf{SRS}}$ is given by $\mu^\lambda=\{\mu^\lambda(\underline{d}', \underline{i}'): (\underline{d}', \underline{i}')\in \mathbb{S}\}$. We then have the following important result.
\begin{lemma}
	\label{lemma:continuity_of_mu_lambda_in_lambda}
	For each $(\underline{d}', \underline{i}')\in \mathbb{S}$, the mapping $\lambda \mapsto \mu^\lambda(\underline{d}', \underline{i}')$ is continuous.
\end{lemma}
\begin{IEEEproof}
Note that under $\pi^\lambda\in \Pi_{\textsf{SRS}}$, the process $\{(\underline{d}(t),\underline{i}(t)): t\geq K\}$ is an ergodic Markov process whose transition probabilities under the arms configuration $C=(h, P_1, P_2)$ are given by	
\begin{align}
	&P^{\pi^\lambda}(\underline{d}(t+1)=\underline{d}',~\underline{i}(t+1)=\underline{i}'\mid \underline{d}(t)=\underline{d},~\underline{i}(t)=\underline{i},~C)\nonumber\\
	&=\begin{cases}
		\left(\frac{\eta}{K}+(1-\eta)\,\lambda(a|\underline{d},\underline{i})\right)\,(P_C^a)^{d_a}(i_a'|i_a),&\text{if }d_a'=1\text{ and }d_{\tilde{a}}'=d_{\tilde{a}}+1\text{ for all }\tilde{a}\neq a,\\
		&i_{\tilde{a}}'=i_{\tilde{a}}\text{ for all }\tilde{a}\neq a,\\
		0,&\text{otherwise}.
	\end{cases}
	\label{eq:transition_probabilities_of_markov_process_under_pi^lambda}
\end{align}
It is clear that the one-step transition probabilities in \eqref{eq:transition_probabilities_of_markov_process_under_pi^lambda} are continuous in $\lambda$ for all $(\underline{d}, \underline{i}), (\underline{d}', \underline{i}')\in\mathbb{S}$. As a consequence, it follows that for each $t\geq 1$, the $t$-step transition probabilities derived from the one-step transition probabilities in \eqref{eq:transition_probabilities_of_markov_process_under_pi^lambda} are continuous in $\lambda$ for all $(\underline{d}, \underline{i}), (\underline{d}', \underline{i}')\in\mathbb{S}$. Denoting the $t$-step transition probability from the state $(\underline{d}, \underline{i})$ to the state $(\underline{d}', \underline{i}')$ under $\pi^\lambda$ by $P^t_{(\underline{d}, \underline{i}), ~ (\underline{d}', \underline{i}')}(\lambda)$, we invoke \cite[Assumption A5$^\prime$]{borkar1982identification}, an assumption that is equivalent to \cite[Assumption A5]{borkar1982identification}, to deduce that for each $(\underline{d}', \underline{i}')\in \mathbb{S}$,
\begin{equation}
	\lim\limits_{n\to \infty} \frac{1}{n}~\sum\limits_{t=1}^n P^t_{(\underline{d}, \underline{i}), ~ (\underline{d}', \underline{i}')}(\lambda) = \mu^\lambda(\underline{d}', \underline{i}')\quad \text{uniformly in }(\underline{d}, \underline{i})\in \mathbb{S}\text{ and }\lambda.
	\label{eq:uniform_convergence_of_n_step_transition_probabilities}
\end{equation}
Combining (a) the continuity of $P^t_{(\underline{d}, \underline{i}), ~ (\underline{d}', \underline{i}')}(\lambda)$ in $\lambda$ for all $t\geq 1$ and $(\underline{d}, \underline{i}), (\underline{d}', \underline{i}')\in\mathbb{S}$, and (b) the uniform convergence in \eqref{eq:uniform_convergence_of_n_step_transition_probabilities}, we get that $\mu^\lambda(\underline{d}', \underline{i}')$ is continuous in $\lambda$ for each $(\underline{d}', \underline{i}')\in \mathbb{S}$. This establishes the lemma.
\end{IEEEproof}
Lemma \ref{lemma:continuity_of_mu_lambda_in_lambda} implies that for every $(\underline{d}, \underline{i})\in \mathbb{S}$, there exists an open neighbourhood $O=O(\underline{d}, \underline{i})$
% (the dependence of $O$ on $(\underline{d}, \underline{i})$ is removed thanks to the uniform convergence)
 around $(P_1, P_2)$ such that for all $(P, Q)\in O$,
\begin{equation}
	\mu^{\lambda_{h, P, Q, \delta}}(\underline{d}, \underline{i}) \geq \frac{\mu^{\lambda_{h, P_1, P_2, \delta}}(\underline{d}, \underline{i})}{1+\delta}
%	 \quad \forall~(\underline{d}, \underline{i})\in \mathbb{S}
	 .
	\label{eq:stat_distributions_in_a_neighbourhood}
\end{equation}
Also, Lemma \ref{lemma:continuity_of_mu_lambda_in_lambda} together with Proposition \ref{prop:convergence_of_ML_estimates} implies that for each $(\underline{d}, \underline{i})\in \mathbb{S}$, under the arms configuration $C=(h, P_1, P_2)$,
\begin{equation}
	\mu^{\lambda_{\theta(n), \hat{P}_{\theta(n), 1}(n), \hat{P}_{\theta(n), 2}(n), \delta}}(\underline{d}, \underline{i}) \longrightarrow \mu^{\lambda_{h, P_1, P_2, \delta}}(\underline{d}, \underline{i})\quad \text{as}\quad n\to\infty\quad \text{almost surely}.
	\label{eq:stat_distributions_converge}
\end{equation}
%Using Lemma \ref{lemma:uniform_convergence}, we get that the convergence in \eqref{eq:stat_distributions_converge} is, in fact, uniform in $(\underline{d}, \underline{i})$.
Because $(\theta(n), \hat{P}_{\theta(n), 1}(n), \hat{P}_{\theta(n), 2}(n)) \longrightarrow (h, P_1, P_2)$ as $n\to \infty$ almost surely under the policy $\pi_{\textsf{ns}}^\star(L,\delta)$ and under the arms configuration $C=(h, P_1, P_2)$, it follows that for all $n$ sufficiently large, $(\hat{P}_{\theta(n), 1}(n), \hat{P}_{\theta(n), 2}(n))=(\hat{P}_{h, 1}(n), \hat{P}_{h, 2}(n))\in O$ almost surely.
%Let the parameter $\gamma$ be chosen in such a way that for all $t$ sufficiently large, the $\gamma$-randomization of $\lambda_{\theta(t), \hat{P}_{\theta(t), 1}(t), \hat{P}_{\theta(t), 2}(t), \delta}$ lies within $O$.
Invoking \cite[Lemma 2.10]{borkar1982identification} with the bounded cost function $c(x_m, x_{m+1}, z_m)$ therein set as $c(x_m, x_{m+1}, z_m)=1-\mathbb{I}_{\{x_m = (\underline{d}, \underline{i})\}}$, we get
\begin{align}
	\liminf\limits_{n\to \infty}\frac{N(n, \underline{d}, \underline{i})}{n} &\geq \inf\limits_{(P, Q)\in O} \mu^{\lambda_{h, P, Q, \delta}}(\underline{d}, \underline{i}) \nonumber\\
	 & \geq \frac{\mu^{\lambda_{h, P_1, P_2, \delta}}(\underline{d}, \underline{i})}{1+\delta}\quad \text{almost surely}.
	\label{eq:liminf_N(n,d,i)/n_has_the_right_drift}
\end{align}
Here, $N(n, \underline{d}, \underline{i})=\sum\limits_{a=1}^{K} N(n, \underline{d}, \underline{i}, a)$. Invoking the null set $B$ from the proof of Proposition \ref{prop:strict_positivity_of_drift_of_test_statistic} in Appendix \ref{appndx:proof_of_prop_strict_positivity_of_drift_of_test_statistic}, restricting the almost sure inequality in \eqref{eq:liminf_N(n,d,i)/n_has_the_right_drift} to outside the null set $B$, and using the convergence in \eqref{eq:convergence_of_arm_selection_probabilities}, we get that for all $\omega\notin B$, $(\underline{d}, \underline{i})\in \mathbb{S}$ and $a\in \mathcal{A}$,
\begin{align}
	\alpha(\underline{d}, \underline{i}, a, \omega)
	&\geq \left(\inf\limits_{(P, Q)\in O} \mu^{\lambda_{h, P, Q, \delta}}(\underline{d}, \underline{i}) \right)\left(\frac{\eta}{K} + (1-\eta)\, \lambda_{h, P_1, P_2, \delta}(a\mid \underline{d}, \underline{i})\right)\nonumber\\
	&\geq \frac{\mu^{\lambda_{h, P_1, P_2, \delta}}(\underline{d}, \underline{i})}{1+\delta}~\left(\frac{\eta}{K} + (1-\eta)\, \lambda_{h, P_1, P_2, \delta}(a\mid \underline{d}, \underline{i})\right)\nonumber\\
	&= \frac{\nu^{\lambda_{h, P_1, P_2, \delta}}(\underline{d}, \underline{i}, a)}{1+\delta}.
	\label{eq:alpha(d,i,a,omega)_has_the_right_lower_bound}
\end{align}
Plugging \eqref{eq:alpha(d,i,a,omega)_has_the_right_lower_bound} into \eqref{eq:proof_of_strict_positivity_7}, we see that under the arms configuration $C=(h, P_1, P_2)$, for all $\omega\notin B$,
\begin{align}
	&\liminf\limits_{n\to \infty}\frac{M_{hh'}(n)(\omega)}{n} \nonumber\\
	&\geq \frac{1}{1+\delta} \left[
\sum\limits_{(\underline{d}, \underline{i})\in \mathbb{S}}\nu^{\lambda_{h, P_1, P_2, \delta}}(\underline{d}, \underline{i}, h)\cdot  D(P_1^{d_h}(\cdot|i_h) \| P^{d_h}(\cdot|i_h)) + \sum\limits_{a\neq h, h'} \nu^{\lambda_{h, P_1, P_2, \delta}}(\underline{d}, \underline{i}, a) \cdot D(P_2^{d_a}(\cdot|i_a) \| P^{d_a}(\cdot|i_a))\right]\nonumber\\
&\geq \frac{1}{1+\delta} ~\inf\limits_{P_2'}\left[
\sum\limits_{(\underline{d}, \underline{i})\in \mathbb{S}}\nu^{\lambda_{h, P_1, P_2, \delta}}(\underline{d}, \underline{i}, h)\cdot  D(P_1^{d_h}(\cdot|i_h) \| (P_2')^{d_h}(\cdot|i_h)) + \sum\limits_{a\neq h, h'} \nu^{\lambda_{h, P_1, P_2, \delta}}(\underline{d}, \underline{i}, a) \cdot D(P_2^{d_a}(\cdot|i_a) \| (P_2')^{d_a}(\cdot|i_a))\right]\nonumber\\
&=\frac{1}{1+\delta} \inf\limits_{\substack{P_1', P_2':\\P_1'\neq P_2'}}\bigg[
\sum\limits_{(\underline{d}, \underline{i})\in \mathbb{S}}\nu^{\lambda_{h, P_1, P_2, \delta}}(\underline{d}, \underline{i}, h)\cdot  D(P_1^{d_h}(\cdot|i_h) \| (P_2')^{d_h}(\cdot|i_h)) + \nu^{\lambda_{h, P_1, P_2, \delta}}(\underline{d}, \underline{i}, h')\cdot  D(P_2^{d_{h'}}(\cdot|i_h) \| (P_1')^{d_{h'}}(\cdot|i_{h'}))\nonumber \\
&\hspace{6cm}+ \sum\limits_{a\neq h, h'} \nu^{\lambda_{h, P_1, P_2, \delta}}(\underline{d}, \underline{i}, a) \cdot D(P_2^{d_a}(\cdot|i_a) \| (P_2')^{d_a}(\cdot|i_a))\bigg]\nonumber\\
&\geq \frac{1}{1+\delta} \inf\limits_{\substack{C'=(h', P_1', P_2'):\\h'\neq h,\\P_1'\neq P_2'}}\bigg[
\sum\limits_{(\underline{d}, \underline{i})\in \mathbb{S}}\nu^{\lambda_{h, P_1, P_2, \delta}}(\underline{d}, \underline{i}, h)\cdot  D(P_1^{d_h}(\cdot|i_h) \| (P_2')^{d_h}(\cdot|i_h)) + \nu^{\lambda_{h, P_1, P_2, \delta}}(\underline{d}, \underline{i}, h')\cdot  D(P_2^{d_{h'}}(\cdot|i_h) \| (P_1')^{d_{h'}}(\cdot|i_{h'}))\nonumber \\
&\hspace{6cm}+ \sum\limits_{a\neq h, h'} \nu^{\lambda_{h, P_1, P_2, \delta}}(\underline{d}, \underline{i}, a) \cdot D(P_2^{d_a}(\cdot|i_a) \| (P_2')^{d_a}(\cdot|i_a))\bigg]\nonumber\\
	&\geq \frac{R^*(h, P_1, P_2)}{(1+\delta)^2},
\end{align}
%which implies that almost surely,
%\begin{align}
%	&\liminf\limits_{n\to \infty} \frac{M_{h}(n)}{n} \nonumber\\
%	&=\liminf\limits_{n\to\infty} \min\limits_{h'\neq h} \frac{M_{hh'}(n)}{n}\nonumber\\
%	&\geq \frac{1}{1+\delta} \inf\limits_{\substack{C'=(h', P_1', P_2'):\\h'\neq h,\\P_1'\neq P_2'}}\bigg[
%\sum\limits_{(\underline{d}, \underline{i})\in \mathbb{S}}\nu^{\lambda_{h, P_1, P_2, \delta}}(\underline{d}, \underline{i}, h)\cdot  D(P_1^{d_h}(\cdot|i_h) \| (P_2')^{d_h}(\cdot|i_h)) + \nu^{\lambda_{h, P_1, P_2, \delta}}(\underline{d}, \underline{i}, h')\cdot  D(P_2^{d_{h'}}(\cdot|i_h) \| (P_1')^{d_{h'}}(\cdot|i_{h'}))\nonumber \\
%&\hspace{6cm}+ \sum\limits_{a\neq h, h'} \nu^{\lambda_{h, P_1, P_2, \delta}}(\underline{d}, \underline{i}, a) \cdot D(P_2^{d_a}(\cdot|i_a) \| (P_2')^{d_a}(\cdot|i_a))\bigg]\nonumber\\
%	&\geq \frac{R^*(h, P_1, P_2)}{(1+\delta)^2},
%\end{align}
where the last line above follows from the definition of $\lambda_{h, P_1, P_2, \delta}$ in \eqref{eq:lambda_h_P1_P2_delta_defining_equation}.
The desired result is thus established.

\section{Proof of Proposition \ref{prop:stopping_time_of_policy_blows_up_as_L_increases}}
\label{appndx:proof_of_prop_stopping_time_of_policy_blows_up_as_L_increases}
It suffices to show that under the arms configuration $C=(h, P_1, P_2)$ and under the policy $\pi=\pi^\star(L, \delta)$,
\begin{equation}
	\limsup\limits_{L\to \infty} P^\pi(\tau(\pi)\leq m \mid C)=0 \quad \text{for all} \quad m\geq 1.
	\label{eq:proof_of_stopping_time_blowing_up_1}
\end{equation}
Because the policy $\pi^\star(L,\delta)$ selects arm $1$ at time $t=0$, arm $2$ at time $t=1$ and so on until arm $K$ at time $t=K$, it suffices to show that for each $m\geq K$,
\begin{equation}
	\limsup\limits_{L \to \infty} P^\pi(\tau(\pi) \leq m \mid C)=0.
	\label{eq:proof_of_stopping_time_blowing_up_2}
\end{equation}
{\color{black} Note that
\begin{align}
	P^\pi(\tau(\pi) \leq m \mid C)&=P^\pi(\exists~K\leq n\leq m\text{ and }\tilde{h}\text{ such that }M_{\tilde{h}}(n)\geq \log((K-1)L))\nonumber\\
	&\stackrel{(a)}{\leq} \sum\limits_{\tilde{h}\in \mathcal{A}}~\sum\limits_{n=K}^{m} P^\pi(M_{\tilde{h}}(n) \geq \log((K-1)L)\mid C)\nonumber\\
	 &=\sum\limits_{\tilde{h}\in \mathcal{A}}~\sum\limits_{n=K}^{m} E^\pi[\mathbb{I}_{\{M_{\tilde{h}}(n) \geq \log((K-1)L)\}}\mid C]\nonumber\\
	& \stackrel{(b)}{\leq} \sum\limits_{\tilde{h}\in \mathcal{A}}~\sum\limits_{n=K}^{m}E^\pi\bigg[\frac{M_{\tilde{h}}(n)}{\log((K-1)L)}\cdot \mathbb{I}_{\{M_{\tilde{h}}(n) \geq \log((K-1)L)\}}\bigg | C\bigg]\nonumber\\
	&=\frac{1}{\log ((K-1)L)}\sum\limits_{\tilde{h}\in \mathcal{A}}~\sum\limits_{n=K}^{m} \bigg(E^\pi[M_{\tilde{h}}(n)\mid C] - E^\pi[M_{\tilde{h}}(n)\cdot \mathbb{I}_{\{M_{\tilde{h}}(n) < \log((K-1)L)\}}\mid C]\bigg),
	\label{eq:proof_of_stopping_time_blowing_up_3}
\end{align}
where $(a)$ above is due to the union bound, $(b)$ follows from the fact that on the set $\{M_{\tilde{h}}(n)\geq \log ((K-1)L)\}$, we have $1\leq \frac{M_{\tilde{h}}(n)}{\log ((K-1)L)}$, and in writing \eqref{eq:proof_of_stopping_time_blowing_up_3}, we use the fact that $$ E^\pi[M_{\tilde{h}}(n)\mid C]= E^\pi[M_{\tilde{h}}(n)\cdot \mathbb{I}_{\{M_{\tilde{h}}(n) \geq \log((K-1)L)\}}\mid C] + E^\pi[M_{\tilde{h}}(n)\cdot \mathbb{I}_{\{M_{\tilde{h}}(n) < \log((K-1)L)\}}\mid C].$$ In order to handle the second term inside the brackets in \eqref{eq:proof_of_stopping_time_blowing_up_3}, we note that for any $\tilde{h}$,  
\begin{align}
	M_{\tilde{h}}(n)&=\min\limits_{h'\neq \tilde{h}} M_{\tilde{h}h'}(n).
	\label{eq:proof_of_stopping_time_blowing_up_3_new_1}
\end{align} 
Fixing an arbitrary $h'\neq \tilde{h}$ and using \eqref{eq:M_hh'(n)_as_sum_of_4_terms} noting that $T_3(n)$ and $T_4(n)$ are non-negative, we have 
\begin{align}
	M_{\tilde{h}h'}(n)&\geq \log \mathbb{E}\bigg[\exp\bigg(\log \left(\sum\limits_{i\in \mathcal{S}}~\phi(i)\, P^{\tilde{h}-1}(X_{\tilde{h}-1}^{\tilde{h}}|i)\right) + \sum\limits_{(\underline{d}, \underline{i})\in \mathbb{S}}~\sum\limits_{j\in \mathcal{S}}N(n, \underline{d}, \underline{i}, \tilde{h}, j)~ \log P^{d_{\tilde{h}}}(j|i_{\tilde{h}})\bigg)\bigg]
	\label{eq:proof_of_stopping_time_blowing_up_3_new_2}\\
	&+\log \mathbb{E}\bigg[\exp\bigg(\sum\limits_{a\neq \tilde{h}} \log \left(\sum\limits_{i\in \mathcal{S}}~\phi(i)\,Q^{a-1}(X_{a-1}^a|i)\right) +\sum\limits_{a\neq \tilde{h}}~\sum\limits_{(\underline{d}, \underline{i})\in \mathbb{S}}~\sum\limits_{j\in \mathcal{S}}N(n, \underline{d}, \underline{i}, a, j)~ \log Q^{d_a}(j|i_a)\bigg)\bigg].
	\label{eq:proof_of_stopping_time_blowing_up_3_new_3}
\end{align}
From an earlier exposition, we now note that the expectation in \eqref{eq:proof_of_stopping_time_blowing_up_3_new_2} can be lower bounded by 
\begin{align}
	&\mathbb{E}\bigg[\exp\bigg(\log \left(\sum\limits_{i\in \mathcal{S}}~\phi(i)\, P^{\tilde{h}-1}(X_{\tilde{h}-1}^{\tilde{h}}|i)\right) + \sum\limits_{(\underline{d}, \underline{i})\in \mathbb{S}}~\sum\limits_{j\in \mathcal{S}}N(n, \underline{d}, \underline{i}, h, j)~ \log P^{d_{\tilde{h}}}(j|i_{\tilde{h}})\bigg)\cdot \mathbb{I}\bigg(P\in \mathscr{P}(\bar{\varepsilon}^*)\bigg)\bigg]\nonumber\\
	&\geq \exp\bigg(\log (\phi_{\textsf{min}}\cdot \bar{\varepsilon}^*)+(\log \bar{\varepsilon}^*)\cdot \sum\limits_{(\underline{d}, \underline{i})\in \mathbb{S}}~\sum\limits_{j\in \mathcal{S}}N(n, \underline{d}, \underline{i}, h, j) \bigg)\cdot \mathbb{P}(P\in \mathscr{P}(\bar{\varepsilon}^*))\nonumber\\
	&\geq \exp\bigg(\log \phi_{\textsf{min}}+\log \bar{\varepsilon}^*+n\,\log \bar{\varepsilon}^*\bigg)\cdot \mathbb{P}(P\in \mathscr{P}(\bar{\varepsilon}^*)).
	\label{eq:proof_of_stopping_time_blowing_up_3_new_4}
\end{align} 
In writing the second inequality above, we use the fact that $ \sum\limits_{(\underline{d}, \underline{i})\in \mathbb{S}}~\sum\limits_{j\in \mathcal{S}}N(n, \underline{d}, \underline{i}, h, j) \leq n $. Taking logarithm on both sides of \eqref{eq:proof_of_stopping_time_blowing_up_3_new_4}, it follows that \eqref{eq:proof_of_stopping_time_blowing_up_3_new_2} may be lower bounded by 
\begin{align}
	 \log \phi_{\textsf{min}}+(n+1)\,\log\bar{\varepsilon}^*+  \log \mathbb{P}(P\in \mathscr{P}(\bar{\varepsilon}^*)).
	 \label{eq:proof_of_stopping_time_blowing_up_3_new_5}
\end{align} 
On similar lines, \eqref{eq:proof_of_stopping_time_blowing_up_3_new_3} may be lower bounded by
\begin{align}
	(K-1)\,\log \phi_{\textsf{min}}+(n+K-1)\log \bar{\varepsilon}^*+\log \mathbb{P}(Q\in \mathscr{P}(\bar{\varepsilon}^*)),
	\label{eq:proof_of_stopping_time_blowing_up_3_new_6}
\end{align}
Combining \eqref{eq:proof_of_stopping_time_blowing_up_3_new_5} and \eqref{eq:proof_of_stopping_time_blowing_up_3_new_6}, and noting that $\mathbb{P}(Q\in \mathscr{P}(\bar{\varepsilon}^*))=\mathbb{P}(P\in \mathscr{P}(\bar{\varepsilon}^*))$, we get
\begin{align}
	M_{\tilde{h}h'}(n)\geq K\,\log\phi_{\textsf{min}}+(n+K)\,\log \bar{\varepsilon}^* + 2\,\log \mathbb{P}(P\in \mathscr{P}(\bar{\varepsilon}^*)). 
	\label{eq:proof_of_stopping_time_blowing_up_3_new_7}
\end{align}
Noting that the lower bound in \eqref{eq:proof_of_stopping_time_blowing_up_3_new_7} holds for all $h'\neq \tilde{h}$ and is therefore a lower bound for $M_{\tilde{h}}(n)$,   \eqref{eq:proof_of_stopping_time_blowing_up_3} may be upper bounded by
\begin{align}
	&\frac{1}{\log ((K-1)L)}\sum\limits_{\tilde{h}\in \mathcal{A}}~\sum\limits_{n=K}^{m} E^\pi[M_{\tilde{h}}(n)\mid C] \nonumber\\
	&\hspace{3cm}+ \frac{1}{\log ((K-1)L)}\sum\limits_{\tilde{h}\in \mathcal{A}}~\sum\limits_{n=K}^{m} \bigg((n+K)\,\log\frac{1}{\bar{\varepsilon}^*}+2\,\log \frac{1}{\mathbb{P}(P\in\mathscr{P}(\bar{\varepsilon}^*))}+K\,\log\phi_{\textsf{min}}\bigg).
	\label{eq:proof_of_stopping_time_blowing_up_3_new_7_1}
\end{align}
 
Recall from an earlier exposition that $\mathbb{P}(P\in \mathscr{P}(\bar{\varepsilon}^*))>0$. Thus, it follows that
\begin{align}
	&\limsup\limits_{L\to\infty} P^\pi(\tau(\pi) \leq m \mid C)\nonumber\\
	&\leq \limsup\limits_{L\to\infty} \bigg\lbrace\frac{1}{\log ((K-1)L)}\sum\limits_{\tilde{h}\in \mathcal{A}}~\sum\limits_{n=K}^{m} E^\pi[M_{\tilde{h}}(n)\mid C] \nonumber\\
	&\hspace{3cm}+ \frac{1}{\log ((K-1)L)}\sum\limits_{\tilde{h}\in \mathcal{A}}~\sum\limits_{n=K}^{m} \bigg((n+K)\,\log\frac{1}{\bar{\varepsilon}^*}+2\,\log \frac{1}{\mathbb{P}(P\in\mathscr{P}(\bar{\varepsilon}^*))}+K\,\log \phi_{\textsf{min}}\bigg)\bigg\rbrace\nonumber\\
	&=\limsup\limits_{L\to\infty} \frac{1}{\log ((K-1)L)}\sum\limits_{\tilde{h}\in \mathcal{A}}~\sum\limits_{n=K}^{m} E^\pi[M_{\tilde{h}}(n)\mid C],
	\label{eq:proof_of_stopping_time_blowing_up_3_new_8}
\end{align}} 
where the last line above follows by noting that the limit supremum of the second term in \eqref{eq:proof_of_stopping_time_blowing_up_3_new_7_1} is equal to zero.

We now show that for each $n\in \{K, \ldots, m\}$, the expectation term $E^\pi[M_{\tilde{h}}(n)\mid C]$ is finite, from which the desired result follows.
We carry out the analysis by considering the cases $\tilde{h}=h$ and $\tilde{h}\neq h$ separately. Here, $h$ is the odd arm in the arms configuration $C=(h, P_1, P_2)$.

\subsection{Case $\tilde{h}=h$}
In this case, we have
\begin{align}
	&E^\pi[M_{\tilde{h}}(n)\mid C]=E^\pi[M_{h}(n)\mid C]\nonumber\\
	&=E^\pi\left[\min\limits_{h'\neq h}~\log \frac{\bar{f}(B^n, A^n, \bar{X}^n \mid \mathcal{H}_{h})}{\hat{f}(B^n, A^n, \bar{X}^n \mid \mathcal{H}_{h'})}\bigg \vert C \right]\nonumber\\
	&\leq E^\pi\left[\log \frac{\bar{f}(B^n, A^n, \bar{X}^n \mid \mathcal{H}_{h})}{\hat{f}(B^n, A^n, \bar{X}^n \mid \mathcal{H}_{h'})}\bigg \vert C\right]\quad \forall~h'\neq h\nonumber\\
	&\stackrel{\text{Jensen's}}{\leq} \log E^\pi\left[\frac{\bar{f}(B^n, A^n, \bar{X}^n \mid \mathcal{H}_{h})}{\hat{f}(B^n, A^n, \bar{X}^n \mid \mathcal{H}_{h'})}\bigg \vert C\right]\quad \forall ~ h'\neq h\nonumber\\
	&= \log  \int\limits_{\Omega} \frac{\bar{f}(B^n(\omega), A^n(\omega), \bar{X}^n(\omega) \mid \mathcal{H}_{h})}{\hat{f}(B^n(\omega), A^n(\omega), \bar{X}^n(\omega) \mid \mathcal{H}_{h'})} ~\cdot ~ f(B^n(\omega), A^n(\omega), \bar{X}^n(\omega) \mid C)~dP^\pi(\omega)\quad \forall~h'\neq h\nonumber\\
	& = \log \int\limits_{\Omega}\frac{\bar{f}(B^n(\omega), A^n(\omega), \bar{X}^n(\omega) \mid \mathcal{H}_{h})}{\hat{f}(B^n(\omega), A^n(\omega), \bar{X}^n(\omega) \mid \mathcal{H}_{h'})} ~\cdot~\exp(-Z_{hh'}(n) (\omega))~\cdot ~ f(B^n(\omega), A^n(\omega), \bar{X}^n(\omega) \mid C'=(h', P_1, P_2))~dP^\pi(\omega) \nonumber\\
	&\hspace{16.2cm} \forall~h'\neq h\nonumber\\
	&\leq \log \int\limits_{\Omega}\frac{\bar{f}(B^n(\omega), A^n(\omega), \bar{X}^n(\omega) \mid \mathcal{H}_{h})}{\hat{f}(B^n(\omega), A^n(\omega), \bar{X}^n(\omega) \mid \mathcal{H}_{h'})} ~\cdot~\exp(-Z_{hh'}(n)(\omega))~\cdot ~ \hat{f}(B^n(\omega), A^n(\omega), \bar{X}^n(\omega) \mid \mathcal{H}_{h'})~dP^\pi(\omega) \quad \forall~h'\neq h\nonumber\\
	&=\log \int\limits_{\Omega}\bar{f}(B^n(\omega), A^n(\omega), \bar{X}^n(\omega) \mid \mathcal{H}_{h}) ~\cdot~\exp(-Z_{hh'}(n)( \omega))~\cdot ~dP^\pi(\omega) \quad \forall~h'\neq h,
	\label{eq:proof_of_stopping_time_blowing_up_6}
\end{align}
where the last line follows from an application of the change of measure technique presented in \cite[Lemma 18]{Kaufmann2016}. Also,  $Z_{hh'}(n)$ in the above set of equations is given by
\begin{align}
	Z_{hh'}(n) &= \sum\limits_{(\underline{d}, \underline{i})\in \mathbb{S}}~\sum\limits_{j\in \mathcal{S}}~\sum\limits_{a=1}^{K} ~ N(n, \underline{d}, \underline{i}, a, j)~\log \frac{(P_h^a)^{d_a}(j|i_a)}{(P_{h'}^a)^{d_a}(j|i_a)}\nonumber\\
	&=\sum\limits_{(\underline{d}, \underline{i})\in \mathbb{S}}~\sum\limits_{j\in \mathcal{S}}~ N(n, \underline{d}, \underline{i}, h, j)~\log \frac{P_1^{d_h}(j|i_h)}{P_2^{d_h}(j|i_h)} + N(n, \underline{d}, \underline{i}, h', j)~\log \frac{P_2^{d_{h'}}(j|i_{h'})}{P_1^{d_{h'}}(j|i_{h'})}.
	\label{eq:Z_hh'(n)}
\end{align}

From Assumption \ref{assmptn:technical_assumption_on_TPMs}, we know that there exists $\bar{\varepsilon}^* \in (0,1)$ such that
\begin{equation}
	\log \frac{1}{\bar{\varepsilon}^*} \geq \log \frac{P_1^{d}(j|i)}{P_2^{d}(j|i)} \geq \log \bar{\varepsilon}^*\quad \forall~d\geq 1,~i,j\in \mathcal{S}.
	\label{eq:proof_of_stopping_time_blowing_up_7}
\end{equation}
Therefore, it follows that almost surely,
\begin{align}
	Z_{hh'}(n) &\geq (\log \bar{\varepsilon}^*) \cdot \left(\sum\limits_{(\underline{d}, \underline{i})\in \mathbb{S}}~\sum\limits_{j\in \mathcal{S}}~N(n, \underline{d}, \underline{i}, h, j)+N(n, \underline{d}, \underline{i}, h', j)\right)\nonumber\\
	&\geq (\log \bar{\varepsilon}^*)\cdot \left(\sum\limits_{(\underline{d}, \underline{i})\in \mathbb{S}}~\sum\limits_{j\in \mathcal{S}}~\sum\limits_{a=1}^{K}~N(n, \underline{d}, \underline{i}, a, j)\right)\\
	&\geq n\cdot (\log \bar{\varepsilon}^*).
	\label{eq:proof_of_stopping_time_blowing_up_8}
\end{align}
Using \eqref{eq:proof_of_stopping_time_blowing_up_8} in \eqref{eq:proof_of_stopping_time_blowing_up_6}, we get
\begin{align}
	E^\pi[M_{\tilde{h}}(n)\mid C]&\leq \log \int\limits_{\Omega}\bar{f}(B^n(\omega), A^n(\omega), \bar{X}^n(\omega) \mid \mathcal{H}_{h}) ~\cdot~\exp(-n\cdot (\log \bar{\varepsilon}^*))~\cdot ~dP^\pi(\omega)\nonumber\\
	 &= n~\log \left(\frac{1}{\bar{\varepsilon}^*}\right),
	 \label{eq:proof_of_stopping_time_blowing_up_case_1}
\end{align}
where the last line follows by noting that $\int\limits_{\Omega}\bar{f}(B^n(\omega), A^n(\omega), \bar{X}^n(\omega) \mid \mathcal{H}_{h})~dP^\pi(\omega)=1$ because $\bar{f}$ is an average likelihood.

\subsection{Case $\tilde{h} \neq h$}
In this case, we note that
\begin{align}
	E^\pi[M_{\tilde{h}}(n)\mid C]&=E^\pi\left[\min\limits_{h'\neq \tilde{h}}~\log \frac{\bar{f}(B^n, A^n, \bar{X}^n \mid \mathcal{H}_{\tilde{h}})}{\hat{f}(B^n, A^n, \bar{X}^n \mid \mathcal{H}_{h'})}\bigg \vert C \right]\nonumber\\
	&\leq E^\pi\left[\log \frac{\bar{f}(B^n, A^n, \bar{X}^n \mid \mathcal{H}_{\tilde{h}})}{\hat{f}(B^n, A^n, \bar{X}^n \mid \mathcal{H}_{h})}\bigg \vert C\right]\nonumber\\
	&\stackrel{\text{Jensen's}}{\leq} \log E^\pi\left[\frac{\bar{f}(B^n, A^n, \bar{X}^n \mid \mathcal{H}_{\tilde{h}})}{\hat{f}(B^n, A^n, \bar{X}^n \mid \mathcal{H}_{h})}\bigg \vert C\right]\nonumber\\
	&= \log  \int\limits_{\Omega} \frac{\bar{f}(B^n(\omega), A^n(\omega), \bar{X}^n(\omega) \mid \mathcal{H}_{\tilde{h}})}{\hat{f}(B^n(\omega), A^n(\omega), \bar{X}^n(\omega) \mid \mathcal{H}_{h})} ~\cdot ~ f(B^n(\omega), A^n(\omega), \bar{X}^n(\omega) \mid C)~dP^\pi(\omega)\nonumber\\
	& \leq \log \int\limits_{\Omega}\bar{f}(B^n(\omega), A^n(\omega), \bar{X}^n(\omega) \mid \mathcal{H}_{\tilde{h}})  ~dP^\pi(\omega)\nonumber\\
	& = 0,
	\label{eq:proof_of_stopping_time_blowing_up_case_2}
\end{align}
where the last line follows by noting, as before, that $\int\limits_{\Omega}\bar{f}(B^n(\omega), A^n(\omega), \bar{X}^n(\omega) \mid \mathcal{H}_{\tilde{h}})~dP^\pi(\omega)=1$ because $\bar{f}$ is an average likelihood.

Using \eqref{eq:proof_of_stopping_time_blowing_up_case_1} and \eqref{eq:proof_of_stopping_time_blowing_up_case_2} in \eqref{eq:proof_of_stopping_time_blowing_up_3_new_8}, it follows that for each $m\geq K$,
\begin{align}
	\limsup\limits_{L\to \infty}P(\tau(\pi^\star(L,\delta)) \leq m \mid C) &\leq \limsup\limits_{L \to \infty} \frac{1}{\log((K-1)L)}\sum\limits_{\tilde{h}\in \mathcal{A}}~\sum\limits_{n=K}^{m}\,\,n~\log \left(\frac{1}{\bar{\varepsilon}^*}\right)\nonumber\\
	&=0.
	\label{eq:proof_of_stopping_time_blowing_up_11}
\end{align}
This establishes the desired result.

\section{Proof of Proposition \ref{prop:upper_bound}}
\label{appndx:proof_of_prop_upper_bound}
Here, we demonstrate that for each $\delta>0$, the family $\{\tau(\pi^\star(L,\delta))/\log L:L>1\}$ is uniformly integrable. Clearly, it suffices to show that under the policy $\pi=\pi^\star(L,\delta)$ and under the arms configuration $C=(h, P_1, P_2)$,
\begin{equation}
	\limsup\limits_{L\to\infty}E^\pi\bigg[\bigg(\frac{\tau(\pi)}{\log L}\bigg)^2\bigg|C\bigg]<\infty.
	\label{eq:exp_moment_finite}
\end{equation}

\subsection{Showing that $P^\pi(M_h(n)<\log((K-1)L)|C)$ is $O(1/n^3)$}
Before showing that \eqref{eq:exp_moment_finite} holds, we record the following important result.
\begin{lemma}
	\label{lemma:exp_upper_bound_for_a_certain_probability_term}
	Fix $C=(h,P_1,P_2)$, $L> 1$ and $\delta> 0$. There exists $0<B<\infty$ independent of $L$ such that for all sufficiently large values of $n$, under the policy $\pi=\pi^\star(L,\delta)$ and under the arms configuration $C$,
	\begin{equation}
		P^\pi(M_h(n)<\log((K-1)L)|C)\leq \frac{B}{n^3}.\label{eq:1/n^4_bound}
	\end{equation}
\end{lemma}
\begin{IEEEproof}[Proof of Lemma \ref{lemma:exp_upper_bound_for_a_certain_probability_term}]
	Note that
	\begingroup\allowdisplaybreaks\begin{align}
	P^\pi(M_h(n)<\log((K-1)L)|C)
	&=P^\pi\left(\min\limits_{h'\neq h}M_{hh'}(n)<\log((K-1)L)~\bigg|~C\right)\nonumber\\
	&\stackrel{(a)}{\leq} \sum\limits_{h'\neq h}P^\pi\left(M_{hh'}(n)<\log((K-1)L)~\bigg|~C\right)\nonumber\\
	&=\sum\limits_{h'\neq h}P^\pi\left(\frac{M_{hh'}(n)}{n}<\frac{\log((K-1)L)}{n}~\bigg|~C\right)\nonumber\\
	&=\sum\limits_{h' \neq h}P^\pi\left(\frac{T_1(n)}{n}+\frac{T_2(n)}{n}+ \frac{T_3(n)}{n}+\frac{T_4(n)}{n}<\frac{\log((K-1)L}{n}~\bigg|~C\right),\label{eq:exp_bound_1}
\end{align}\endgroup
where $(a)$ above follows from the union bound, and the last line follows from \eqref{eq:M_hh'(n)_as_sum_of_4_terms}. In order to prove \eqref{eq:1/n^4_bound}, it suffices to prove that each term inside the summation in \eqref{eq:exp_bound_1} is $O(1/n^3)$.

By virtue of Assumption \ref{assmptn:technical_assumption_on_TPMs}, we know that $P_1, P_2\in \mathscr{P}(\bar{\varepsilon}^*)$ for some $\bar{\varepsilon}^*\in (0, 1)$. From \eqref{eq:T_1(n)}, it follows that
\begin{align}
	&\frac{T_1(n)}{n}-\sum\limits_{(\underline{d}, \underline{i})\in \mathbb{S}}~\sum\limits_{j\in \mathcal{S}}~\frac{N(n, \underline{d}, \underline{i}, h, j)}{n}~\log P_1^{d_h}(j|i_h)\nonumber\\
	&={\color{black}\frac{1}{n}\,\log\mathbb{E}\bigg[\exp\bigg\lbrace\log\bigg(\sum\limits_{i\in \mathcal{S}}~\phi(i)\,P^{h-1}(X_{h-1}^h|i)\bigg)+\sum\limits_{(\underline{d}, \underline{i})\in \mathbb{S}}~\sum\limits_{j\in \mathcal{S}}N(n, \underline{d}, \underline{i}, h, j)~ \log P^{d_h}(j|i_h)\bigg\rbrace\bigg]}\nonumber\\
	&\hspace{7cm}{\color{black} -\frac{1}{n}\,\log\exp\bigg\lbrace\sum\limits_{(\underline{d}, \underline{i})\in \mathbb{S}}~\sum\limits_{j\in \mathcal{S}}N(n, \underline{d}, \underline{i}, h, j)~ \log P_1^{d_h}(j|i_h)\bigg\rbrace}\nonumber\\
	&{\color{black} =\frac{1}{n}\,\log\mathbb{E}\bigg[\exp\bigg\lbrace\log\bigg(\sum\limits_{i\in \mathcal{S}}~\phi(i)\,P^{h-1}(X_{h-1}^h|i)\bigg) + \sum\limits_{(\underline{d}, \underline{i})\in \mathbb{S}}~\sum\limits_{j\in \mathcal{S}}N(n, \underline{d}, \underline{i}, h, j)~ \log \frac{P^{d_h}(j|i_h)}{P_1^{d_h}(j|i_h)}\bigg\rbrace\bigg]}\nonumber\\
	&{\color{black}\geq \frac{1}{n}\,\log\mathbb{E}\bigg[\exp\bigg\lbrace\log\bigg(\sum\limits_{i\in \mathcal{S}}~\phi(i)\,P^{h-1}(X_{h-1}^h|i)\bigg) + \sum\limits_{(\underline{d}, \underline{i})\in \mathbb{S}}~\sum\limits_{j\in \mathcal{S}}N(n, \underline{d}, \underline{i}, h, j)~ \log \frac{P^{d_h}(j|i_h)}{P_1^{d_h}(j|i_h)}\bigg\rbrace\cdot \mathbb{I}\bigg(P\in \mathscr{P}(\bar{\varepsilon}^*)\bigg)\bigg]}\nonumber\\
	&\stackrel{(a)}{\geq} \frac{1}{n}\log \bigg[\exp\bigg\lbrace\log (\phi_{\textsf{min}}\cdot \bar{\varepsilon}^*) + (\log \bar{\varepsilon}^*)\cdot \sum\limits_{(\underline{d}, \underline{i})\in \mathbb{S}}~\sum\limits_{j\in \mathcal{S}}N(n, \underline{d}, \underline{i}, h, j)\bigg\rbrace \cdot \mathbb{P}(P\in \mathscr{P}(\bar{\varepsilon}^*))\bigg]\nonumber\\
	&\stackrel{(b)}{=} \frac{1}{n}\,\log \phi_{\textsf{min}}+\frac{1}{n}\,\log \bar{\varepsilon}^* + \frac{N(n, h)}{n}(\log \bar{\varepsilon}^*) + \frac{1}{n}~\log \mathbb{P}(P\in \mathscr{P}(\bar{\varepsilon}^*)),
	\label{eq:exp_bound_2}
\end{align}
where $(a)$ above follows by noting that on the set $\{P\in \mathscr{P}(\bar{\varepsilon}^*)\}$, $$\forall~ d\geq 1 \text{ and }i, j\in \mathcal{S}, \quad P^d(j|i)\geq\bar{\varepsilon}^*  \text{ whenever } P^d(j|i)>0,$$ and in $(b)$ above, $$ N(n, h)\coloneqq \sum\limits_{(\underline{d}, \underline{i})\in \mathbb{S}}~\sum\limits_{j\in \mathcal{S}} ~N(n, \underline{d}, \underline{i}, h, j).$$ Let $N(n, a)$ be defined similarly for all $a\neq h$. It follows from \eqref{eq:exp_bound_2} that almost surely
\begin{equation}
	\frac{T_1(n)}{n} \geq \sum\limits_{(\underline{d}, \underline{i})\in \mathbb{S}}~\sum\limits_{j\in \mathcal{S}}~\frac{N(n, \underline{d}, \underline{i}, h, j)}{n}~\log P_1^{d_h}(j|i_h)+\frac{N(n, h)}{n}(\log \bar{\varepsilon}^*) + \frac{1}{n}\,\log\phi_{\textsf{min}} +  \frac{1}{n}~\log\bar{\varepsilon}^* + \frac{1}{n}~\log \mathbb{P}(P\in \mathscr{P}(\bar{\varepsilon}^*)).
	\label{eq:lower_bound_for_T_1(n)/n}
\end{equation}
Similarly, it can be shown that almost surely,
\begin{align}
	\frac{T_2(n)}{n} &\geq \sum\limits_{a\neq h}~\sum\limits_{(\underline{d}, \underline{i})\in \mathbb{S}}~\sum\limits_{j\in \mathcal{S}}~\frac{N(n, \underline{d}, \underline{i}, a, j)}{n}~\log P_2^{d_a}(j|i_a)+(\log \bar{\varepsilon}^*)\left(\sum\limits_{a\neq h}~\frac{N(n, a)}{n}\right) \nonumber\\
	&\hspace{4cm}+ \frac{K-1}{n}\,\log \phi_{\textsf{min}} + \frac{K-1}{n}~\log\bar{\varepsilon}^* +  \frac{1}{n}~\log \mathbb{P}(Q\in \mathscr{P}(\bar{\varepsilon}^*)).
	\label{eq:lower_bound_on_T_2(n)/n}
\end{align}
Next, we note from \eqref{eq:T_3(n)} that for each $h'\neq h$, almost surely,
\begin{align}
	\frac{T_3(n)}{n}&\geq \sum\limits_{(\underline{d}, \underline{i})\in \mathbb{S}}~\sum\limits_{j\in \mathcal{S}}~\frac{N(n, \underline{d}, \underline{i}, h', j)}{n}~\log\frac{1}{(\hat{P}_{h', 1}(n))^{d_{h'}}(j|i_{h'})}\nonumber\\
	&=\sum\limits_{(\underline{d}, \underline{i})\in \mathbb{S}}~\sum\limits_{j\in \mathcal{S}}~\frac{N(n, \underline{d}, \underline{i}, h', j)}{n}~\log\frac{1}{P_2^{d_{h'}}(j|i_{h'})}+\sum\limits_{(\underline{d}, \underline{i})\in \mathbb{S}}~\sum\limits_{j\in \mathcal{S}}~\frac{N(n, \underline{d}, \underline{i}, h', j)}{n}~\log\frac{P_2^{d_{h'}}(j|i_{h'})}{(\hat{P}_{h', 1}(n))^{d_{h'}}(j|i_{h'})}\nonumber\\
	&\geq \sum\limits_{(\underline{d}, \underline{i})\in \mathbb{S}}~\sum\limits_{j\in \mathcal{S}}~\frac{N(n, \underline{d}, \underline{i}, h', j)}{n}~\log\frac{1}{P_2^{d_{h'}}(j|i_{h'})} + (\log \bar{\varepsilon}^*)\sum\limits_{(\underline{d}, \underline{i})\in \mathbb{S}}~\sum\limits_{j\in \mathcal{S}}~\frac{N(n, \underline{d}, \underline{i}, h', j)}{n}\nonumber\\
	&=\sum\limits_{(\underline{d}, \underline{i})\in \mathbb{S}}~\sum\limits_{j\in \mathcal{S}}~\frac{N(n, \underline{d}, \underline{i}, h', j)}{n}~\log\frac{1}{P_2^{d_{h'}}(j|i_{h'})} + (\log \bar{\varepsilon}^*)\frac{N(n, h')}{n},
	\label{eq:lower_bound_for_T_3(n)/n}
\end{align}
where the inequality above follows by noting that because $P_2\in \mathscr{P}(\bar{\varepsilon}^*)$, we have $$ \forall~ d\geq 1 \text{ and }i, j\in \mathcal{S}, \quad P_2^d(j|i)\geq\bar{\varepsilon}^*  \text{ whenever } P_2^d(j|i)>0,$$ as a consequence of which we have $$ \frac{P_2^d(j|i)}{(\hat{P}_{h', 1}(n))^d(j|i)}\geq \bar{\varepsilon}^*\quad \text{whenever }(\hat{P}_{h', 1}(n))^d(j|i)>0.$$

Lastly, in order to lower bound the term $T_4(n)/n$, let us fix an arbitrary $P_*\in \mathscr{P}(\bar{\varepsilon}^*)$ such that $P_*\neq P_1,P_2$. Then, for each $h'\neq h$, almost surely,
\begin{align}
	\frac{T_4(n)}{n} &\geq\sum\limits_{a\neq h'}~\sum\limits_{(\underline{d}, \underline{i})\in \mathbb{S}}~\sum\limits_{j\in \mathcal{S}}~\frac{N(n, \underline{d}, \underline{i}, a, j)}{n}~\log\frac{1}{(\hat{P}_{h', 2}(n))^{d_{a}}(j|i_{a})}\nonumber\\
	&=\sum\limits_{a\neq h'}~\sum\limits_{(\underline{d}, \underline{i})\in \mathbb{S}}~\sum\limits_{j\in \mathcal{S}}~\frac{N(n, \underline{d}, \underline{i}, a, j)}{n}~\log\frac{1}{P_*^{d_{a}}(j|i_{a})}+\sum\limits_{a\neq h'}~\sum\limits_{(\underline{d}, \underline{i})\in \mathbb{S}}~\sum\limits_{j\in \mathcal{S}}~\frac{N(n, \underline{d}, \underline{i}, a, j)}{n}~\log\frac{P_*^{d_{a}}(j|i_{a})}{(\hat{P}_{h', 2}(n))^{d_{a}}(j|i_{a})}\nonumber\\
	&\geq \sum\limits_{a\neq h'}~\sum\limits_{(\underline{d}, \underline{i})\in \mathbb{S}}~\sum\limits_{j\in \mathcal{S}}~\frac{N(n, \underline{d}, \underline{i}, a, j)}{n}~\log\frac{1}{P_*^{d_{a}}(j|i_{a})} + (\log \bar{\varepsilon}^*)\sum\limits_{a\neq h'}~\sum\limits_{(\underline{d}, \underline{i})\in \mathbb{S}}~\sum\limits_{j\in \mathcal{S}}~\frac{N(n, \underline{d}, \underline{i}, a, j)}{n}\nonumber\\
	&=\sum\limits_{a\neq h'}~\sum\limits_{(\underline{d}, \underline{i})\in \mathbb{S}}~\sum\limits_{j\in \mathcal{S}}~\frac{N(n, \underline{d}, \underline{i}, a, j)}{n}~\log\frac{1}{P_*^{d_{a}}(j|i_{a})} + (\log \bar{\varepsilon}^*)\left(\sum\limits_{a\neq h'}~\frac{N(n, a)}{n}\right),
	\label{eq:lower_bound_for_T_4(n)/n}
\end{align}
Combining the lower bounds in \eqref{eq:lower_bound_for_T_1(n)/n}-\eqref{eq:lower_bound_for_T_4(n)/n}, it follows that for each $h'\neq h$, almost surely,
\begin{align}
	&\frac{M_{hh'}(n)}{n}\nonumber\\
	&\geq 2(\log \bar{\varepsilon}^*)\left(\sum\limits_{a=1}^{K}\frac{N(n, a)}{n}\right)+\frac{2}{n}~ \log \mathbb{P}(P\in \mathscr{P}(\bar{\varepsilon}^*))+\frac{K}{n}~ \log \bar{\varepsilon}^*\nonumber\\
	&+ \sum\limits_{(\underline{d}, \underline{i})\in \mathbb{S}}~\sum\limits_{j\in \mathcal{S}}~\frac{N(n, \underline{d}, \underline{i}, h, j)}{n}~\log \frac{P_1^{d_h}(j|i_h)}{P_*^{d_h}(j|i_h)} + \sum\limits_{a\neq h, h'}~\sum\limits_{(\underline{d}, \underline{i})\in \mathbb{S}}~\sum\limits_{j\in \mathcal{S}}~\frac{N(n, \underline{d}, \underline{i}, a, j)}{n}~\log \frac{P_2^{d_a}(j|i_a)}{P_*^{d_a}(j|i_a)}\nonumber\\
	&=~ 2(\log \bar{\varepsilon}^*)\left(\sum\limits_{a=1}^{K}\frac{N(n, a)}{n}\right)+\frac{2}{n}~ \log \mathbb{P}(P\in \mathscr{P}(\bar{\varepsilon}^*)) + \frac{K}{n}~ \log \bar{\varepsilon}^*\label{eq:first_term_to_be_handled}\\
	&+ \sum\limits_{(\underline{d}, \underline{i})\in \mathbb{S}}~\sum\limits_{j\in \mathcal{S}}~\left(\frac{N(n, \underline{d}, \underline{i}, h, j)}{n}-\frac{N(n, \underline{d}, \underline{i}, h)}{n}~P_1^{d_h}(j|i_h)\right)~\log \frac{P_1^{d_h}(j|i_h)}{P_*^{d_h}(j|i_h)}\label{eq:second_term_to_be_handled}\\
	&+ \sum\limits_{a\neq h, h'}~\sum\limits_{(\underline{d}, \underline{i})\in \mathbb{S}}~\sum\limits_{j\in \mathcal{S}}~\left(\frac{N(n, \underline{d}, \underline{i}, a, j)}{n}-\frac{N(n, \underline{d}, \underline{i}, a)}{n}~P_2^{d_a}(j|i_a)\right)~\log \frac{P_2^{d_a}(j|i_a)}{P_*^{d_a}(j|i_a)}\label{eq:third_term_to_be_handled}\\
	&+\sum\limits_{(\underline{d}, \underline{i})\in \mathbb{S}}~\sum\limits_{j\in \mathcal{S}}~\frac{N(n, \underline{d}, \underline{i}, h)}{n}~D(P_1^{d_h}(\cdot|i_h)\|P_*^{d_h}(\cdot|i_h))+\sum\limits_{a\neq h, h'}~\sum\limits_{(\underline{d}, \underline{i})\in \mathbb{S}}~\sum\limits_{j\in \mathcal{S}}~\frac{N(n, \underline{d}, \underline{i}, a)}{n}~D(P_2^{d_a}(\cdot|i_a)\|P_*^{d_a}(\cdot|i_a))\label{eq:fourth_term_to_be_handled}.
\end{align}

Using \eqref{eq:first_term_to_be_handled}-\eqref{eq:fourth_term_to_be_handled}, the probability term $P^\pi\left(\frac{M_{hh'}(n)}{n}<\frac{\log((K-1)L)}{n}~\bigg|~C\right)$ may be written as
\begin{equation}
	P^\pi\left(\frac{M_{hh'}(n)}{n}<\frac{\log((K-1)L)}{n}~\bigg|~C\right)\leq U_1(n)+U_2(n)+U_3(n)+U_4(n),
	\label{eq:exp_bound_3}
\end{equation}
where the terms $U_1(n)$-$U_4(n)$ are as follows:
\begin{itemize}
	\item The term $U_1(n)$ is given by
	\begin{equation}
		U_1(n)=P^\pi\left(\frac{K}{n}\,\log\phi_{\textsf{min}} + \frac{K}{n}~ \log\bar{\varepsilon}^* + \frac{2}{n}~ \log \mathbb{P}(P\in \mathscr{P}(\bar{\varepsilon}^*))< -\varepsilon ~\bigg|~C\right).
		\label{eq:U_1(n)}
	\end{equation}
	
	\item The term $U_2(n)$ is given by
	\begin{equation}
		U_2(n)=P^\pi\left(\sum\limits_{(\underline{d}, \underline{i})\in \mathbb{S}}~\sum\limits_{j\in \mathcal{S}}~\left(\frac{N(n, \underline{d}, \underline{i}, h, j)}{n}-\frac{N(n, \underline{d}, \underline{i}, h)}{n}~P_1^{d_h}(j|i_h)\right)~\log \frac{P_1^{d_h}(j|i_h)}{P_*^{d_h}(j|i_h)} < -\varepsilon ~\bigg|~C\right).
		\label{eq:U_2(n)}
	\end{equation}
	
	\item The term $U_3(n)$ is given by
	\begin{equation}
		U_3(n)=P^\pi\left(\sum\limits_{a\neq h, h'}~\sum\limits_{(\underline{d}, \underline{i})\in \mathbb{S}}~\sum\limits_{j\in \mathcal{S}}~\left(\frac{N(n, \underline{d}, \underline{i}, a, j)}{n}-\frac{N(n, \underline{d}, \underline{i}, a)}{n}~P_2^{d_a}(j|i_a)\right)~\log \frac{P_2^{d_a}(j|i_a)}{P_*^{d_a}(j|i_a)} < -\varepsilon  ~\bigg|~C\right).
		\label{eq:U_3(n)}
	\end{equation}
	
	\item The term $U_4(n)$ is given by
	\begin{align}
		U_4(n)&=P^\pi\bigg(\sum\limits_{(\underline{d}, \underline{i})\in \mathbb{S}}~\frac{N(n, \underline{d}, \underline{i}, h)}{n}~D(P_1^{d_h}(\cdot|i_h)\|P_*^{d_h}(\cdot|i_h))\nonumber\\
		&\hspace{3cm} +\sum\limits_{a\neq h, h'}~\sum\limits_{(\underline{d}, \underline{i})\in \mathbb{S}}~\frac{N(n, \underline{d}, \underline{i}, a)}{n}~D(P_2^{d_a}(\cdot|i_a)\|P_*^{d_a}(\cdot|i_a)) \nonumber\\
		&\hspace{6cm} + 2(\log \bar{\varepsilon}^*)\left(\sum\limits_{a=1}^{K}\frac{N(n, a)}{n}\right) + 3\varepsilon  < \frac{\log ((K-1)L)}{n}  ~\bigg|~C\bigg).
		\label{eq:U_4(n)}
	\end{align}
\end{itemize}
The relations in \eqref{eq:U_1(n)}-\eqref{eq:U_4(n)} hold for all $\varepsilon>0$. We shall shortly demonstrate how to choose $\varepsilon>0$.

We now show that each of the terms $U_1(n)$-$U_4(n)$ is either $0$ or $O(1/n^3)$. This will then establish \eqref{eq:1/n^4_bound}.

\subsubsection{Handling $U_1(n)$}
Because $\mathbb{P}(P\in \mathscr{P}(\bar{\varepsilon}^*))>0$, it follows that inside the probability term in \eqref{eq:U_1(n)},
% if $\mathbb{P}(P\in \mathscr{P}(\bar{\varepsilon}^*))>0$,
 the left hand side goes to zero as $n\to\infty$, whereas the right hand side is strictly negative. Thus, there exists $N_1=N_1(\varepsilon)$ such that $U_1(n)=0$ for all $n\geq N_1$.

\subsubsection{Handling $U_2(n)$}
Next, we note that $$\left\lbrace\sum\limits_{(\underline{d}, \underline{i})\in \mathbb{S}}~\sum\limits_{j\in \mathcal{S}}~\left(N(n, \underline{d}, \underline{i}, h, j)-N(n, \underline{d}, \underline{i}, h)~P_1^{d_h}(j|i_h)\right)~\log \frac{P_1^{d_h}(j|i_h)}{P_*^{d_h}(j|i_h)}\right\rbrace_{n\geq K}$$ is a martingale. Indeed, because $\log \bar{\varepsilon}^* \leq \log \frac{P_1^{d_h}(j|i_h)}{P_*^{d_h}(j|i_h)} \leq \log \frac{1}{\bar{\varepsilon}^*}$ (which follows as a consequence of Assumption \ref{assmptn:technical_assumption_on_TPMs}), using the dominated convergence theorem,
\begin{align}
	&E^\pi\left[\sum\limits_{(\underline{d}, \underline{i})\in \mathbb{S}}~\sum\limits_{j\in \mathcal{S}}~\left(N(n, \underline{d}, \underline{i}, h, j)-N(n, \underline{d}, \underline{i}, h)~P_1^{d_h}(j|i_h)\right)~\log \frac{P_1^{d_h}(j|i_h)}{P_*^{d_h}(j|i_h)}~\bigg|~B^{n-1}, A^{n-1}, \bar{X}^{n-1}, ~C\right]\nonumber\\
	&=E^\pi\left[\sum\limits_{t=K}^{n}~\sum\limits_{(\underline{d}, \underline{i})\in \mathbb{S}}~\sum\limits_{j\in \mathcal{S}}~\mathbb{I}_{\{\underline{d}(t)=\underline{d},~\underline{i}(t)=\underline{i},~A_t=h\}}~\left(\mathbb{I}_{\{\bar{X}_t=j\}}-P_1^{d_h}(j|i_h)\right)~\log \frac{P_1^{d_h}(j|i_h)}{P_*^{d_h}(j|i_h)}~\bigg|~B^{n-1}, A^{n-1}, \bar{X}^{n-1}, ~C\right]\nonumber\\
	&=\sum\limits_{(\underline{d}, \underline{i})\in \mathbb{S}}~\sum\limits_{j\in \mathcal{S}}~\left(N(n-1, \underline{d}, \underline{i}, h, j)-N(n-1, \underline{d}, \underline{i}, h)~P_1^{d_h}(j|i_h)\right)~\log \frac{P_1^{d_h}(j|i_h)}{P_*^{d_h}(j|i_h)}\nonumber\\
	&+\sum\limits_{(\underline{d}, \underline{i})\in \mathbb{S}}~\sum\limits_{j\in \mathcal{S}}~\mathbb{I}_{\{\underline{d}(n)=\underline{d},~\underline{i}(n)=\underline{i}\}}~\bigg(P^\pi(A_n=h,~\bar{X}_n=j|B^{n-1}, A^{n-1}, \bar{X}^{n-1}, ~C)\nonumber\\
	&\hspace{6cm}-P^\pi(A_n=h|B^{n-1}, A^{n-1}, \bar{X}^{n-1}, ~C)\cdot P_1^{d_h}(j|i_h)\bigg)~\log \frac{P_1^{d_h}(j|i_h)}{P_*^{d_h}(j|i_h)}\nonumber\\
		&=\sum\limits_{(\underline{d}, \underline{i})\in \mathbb{S}}~\sum\limits_{j\in \mathcal{S}}~\left(N(n-1, \underline{d}, \underline{i}, h, j)-N(n-1, \underline{d}, \underline{i}, h)~P_1^{d_h}(j|i_h)\right)~\log \frac{P_1^{d_h}(j|i_h)}{P_*^{d_h}(j|i_h)},
\end{align}
where the last line follows by noting that when $(\underline{d}(t), \underline{i}(t))=(\underline{d}, \underline{i})$, under the arms configuration $C=(h, P_1, P_2)$, $$  P^\pi(\bar{X}_t=j|B^{n-1}, A^{n-1}, \bar{X}^{n-1}, ~C)=P_1^{d_h}(j|i_h).$$ Further, above martingale is bounded, and its quadratic variation $\langle M_n \rangle$ satisfies
\begin{align}
	& \langle M_n\rangle \coloneqq \sum\limits_{t=K}^n ~E^\pi\left[\left(\sum\limits_{(\underline{d}, \underline{i})\in \mathbb{S}}~\sum\limits_{j\in \mathcal{S}}~\mathbb{I}_{\{\underline{d}(t)=\underline{d},~\underline{i}(t)=\underline{i},~A_t=h\}}~\left(\mathbb{I}_{\{\bar{X}_t=j\}}-P_1^{d_h}(j|i_h)\right)~\log \frac{P_1^{d_h}(j|i_h)}{P_*^{d_h}(j|i_h)}\right)^2~\bigg|~B^{t-1}, ~A^{t-1}, ~\bar{X}^{t-1},~C\right]\nonumber\\
	&\leq \sum\limits_{t=K}^n ~E^\pi\left[\sum\limits_{(\underline{d}, \underline{i})\in \mathbb{S}}~\sum\limits_{j\in \mathcal{S}}~\mathbb{I}_{\{\underline{d}(t)=\underline{d},~\underline{i}(t)=\underline{i},~A_t=h\}}~\left(\mathbb{I}_{\{\bar{X}_t=j\}}-P_1^{d_h}(j|i_h)\right)^2~\left(\log \frac{P_1^{d_h}(j|i_h)}{P_*^{d_h}(j|i_h)}\right)^2~\bigg|~B^{t-1}, ~A^{t-1}, ~\bar{X}^{t-1},~C\right]\nonumber\\
	&\stackrel{(a)}{\leq} 4~\left(\log \frac{1}{\bar{\varepsilon}^*}\right)^2~\sum\limits_{t=K}^n ~E^\pi\left[\sum\limits_{(\underline{d}, \underline{i})\in \mathbb{S}}~\sum\limits_{j\in \mathcal{S}}~\mathbb{I}_{\{\underline{d}(t)=\underline{d},~\underline{i}(t)=\underline{i},~A_t=h\}}~\bigg|~C\right]\nonumber\\
	&\leq n\left(4~\left(\log \frac{1}{\bar{\varepsilon}^*}\right)^2~|\mathcal{S}|\right)
	\label{eq:exp_bound_4}
\end{align}
almost surely.
In arriving at $(a)$ above, we use the fact that for all $d\geq 1$ and $i, j\in \mathcal{S}$, $$ \log \frac{P_1^{d}(j|i)}{P_*^{d}(j|i)}\leq \log \frac{1}{\bar{\varepsilon}^*}. $$
We then have
\begin{align}
	U_2(n)&=P^\pi\left(\sum\limits_{(\underline{d}, \underline{i})\in \mathbb{S}}~\sum\limits_{j\in \mathcal{S}}~\left(N(n, \underline{d}, \underline{i}, h, j)-N(n, \underline{d}, \underline{i}, h)~P_1^{d_h}(j|i_h)\right)~\log \frac{P_1^{d_h}(j|i_h)}{P_*^{d_h}(j|i_h)} < -n\varepsilon ~\bigg|~C\right)\nonumber\\
	&=P^\pi\left(\left\lvert\sum\limits_{(\underline{d}, \underline{i})\in \mathbb{S}}~\sum\limits_{j\in \mathcal{S}}~\left(N(n, \underline{d}, \underline{i}, h, j)-N(n, \underline{d}, \underline{i}, h)~P_1^{d_h}(j|i_h)\right)~\log \frac{P_1^{d_h}(j|i_h)}{P_*^{d_h}(j|i_h)}\right\rvert > n\varepsilon ~\bigg|~C\right)\nonumber\\
	&\leq P^\pi\left(\sup\limits_{K\leq t\leq n}~\left\lvert\sum\limits_{(\underline{d}, \underline{i})\in \mathbb{S}}~\sum\limits_{j\in \mathcal{S}}~\left(N(t, \underline{d}, \underline{i}, h, j)-N(t, \underline{d}, \underline{i}, h)~P_1^{d_h}(j|i_h)\right)~\log \frac{P_1^{d_h}(j|i_h)}{P_*^{d_h}(j|i_h)}\right\rvert > n\varepsilon ~\bigg|~C\right)\nonumber\\
	&\stackrel{(a)}{\leq} \frac{1}{n^6~\varepsilon^6}~E^\pi\left[\left(\sup\limits_{K\leq t\leq n}~\left\lvert\sum\limits_{(\underline{d}, \underline{i})\in \mathbb{S}}~\sum\limits_{j\in \mathcal{S}}~\left(N(t, \underline{d}, \underline{i}, h, j)-N(t, \underline{d}, \underline{i}, h)~P_1^{d_h}(j|i_h)\right)~\log \frac{P_1^{d_h}(j|i_h)}{P_*^{d_h}(j|i_h)}\right\rvert\right)^6 ~\bigg|~C\right]\nonumber\\
	&\stackrel{(b)}{\leq} \frac{A}{n^6~\varepsilon^6}~E^\pi[|\langle M_n \rangle|^{3}|C]\nonumber\\
	&\stackrel{(c)}{\leq }\frac{A}{n^6~\varepsilon^6}~n^{3}~\left(4~\left(\log \frac{1}{\bar{\varepsilon}^*}\right)^2~|\mathcal{S}|\right)^{3}\nonumber\\
	&= \frac{A'}{n^{3}},
	\label{eq:exp_bound_5}
\end{align}
where $(a)$ above is due to Markov's inequality, $(b)$ is due to Burkholder inequality \cite[p. 414]{chow2012probability}, and $(c)$ follows from \eqref{eq:exp_bound_4}. The constant $A$ in $(b)$ above comes from Burkholder inequality. We have thus shown that $U_2(n)$ is $O(1/n^3)$.

\subsubsection{Handling $U_3(n)$}
Following the arguments presented for handling $U_2(n)$, it can be shown that $U_3(n)=O(1/n^3)$. The details are omitted.

\subsubsection{Handling $U_4(n)$}
We first note that
\begin{align}
	U_4(n)&\leq P^\pi\bigg(\sum\limits_{(\underline{d},\underline{i})\in \mathbb{S}}\frac{N(n, \underline{d}, \underline{i}, h)}{n}~D(P_1^{d_h}(\cdot|i_h)\|P_*^{d_h}(\cdot|i_h))+\sum\limits_{(\underline{d},\underline{i})\in \mathbb{S}}~\sum\limits_{a\neq h, h'}~\frac{N(n, \underline{d}, \underline{i}, a)}{n}~D(P_2^{d_a}(\cdot|i_a)\|P_*^{d_a}(\cdot|i_a)) \nonumber\\
		&\hspace{6cm} + 2(\log \bar{\varepsilon}^*)\left(\sum\limits_{a=1}^{K}\frac{N(n, a)}{n}\right) + 3\varepsilon  < \frac{\log ((K-1)L)}{n}  ~\bigg|~C\bigg)\nonumber\\
		&\leq P^\pi\bigg(\frac{N(n, \underline{d}, \underline{i}, h)}{n}~D(P_1^{d_h}(\cdot|i_h)\|P_*^{d_h}(\cdot|i_h))+\sum\limits_{a\neq h, h'}~\frac{N(n, \underline{d}, \underline{i}, a)}{n}~D(P_2^{d_a}(\cdot|i_a)\|P_*^{d_a}(\cdot|i_a)) \nonumber\\
		&\hspace{6cm} + 2(\log \bar{\varepsilon}^*)+ 3\varepsilon  < \frac{\log ((K-1)L)}{n}  ~\bigg|~C\bigg)\quad \text{for all }(\underline{d}, \underline{i})\in \mathbb{S}.
		\label{eq:exp_bound_6}
\end{align}
In writing \eqref{eq:exp_bound_6}, we use the fact that $\sum\limits_{a=1}^{K}\frac{N(n, a)}{n}\leq 1$. Going further, let us fix $\underline{d}=(K, K-1,\ldots, 1)$ and $\underline{i}=(i^*, i^*, \ldots, i^*)$ for some $i^*\in \mathbb{S}$. Let $E_n$ be the event inside the probability term in \eqref{eq:exp_bound_6}. From \eqref{eq:alpha(d,i,a,omega)_has_the_right_lower_bound}, we know that almost surely,
\begin{align}
	\liminf\limits_{n\to \infty}~\frac{N(n, \underline{d}, \underline{i}, a)}{n}&\geq \frac{\nu^{\lambda_{h, P_1, P_2, \delta}}(\underline{d}, \underline{i}, a)}{1+\delta}.
	\label{eq:exp_bound_7}
\end{align}
Exploiting this, and denoting the right hand side of \eqref{eq:exp_bound_7} as $C(\underline{d}, \underline{i}, a)$, the probability term in \eqref{eq:exp_bound_6} may be upper bounded as
\begin{align}
	P^\pi(E_n|C)&\leq P^\pi\left(E_n~\cap~\left\lbrace\frac{N(n, \underline{d}, \underline{i}, a)}{n}\geq C(\underline{d}, \underline{i}, a)(1-\epsilon')~ \forall a\in \mathcal{A}\right\rbrace~\bigg|~C\right)\label{eq:P(E_n_and_good_event}\\
	&\hspace{4cm} + P^\pi\left(E_n~\cap~\left\lbrace\exists~a\in \mathcal{A}:\frac{~N(n, \underline{d}, \underline{i}, a)}{n}< C(\underline{d}, \underline{i}, a)(1-\epsilon')\right\rbrace~\bigg|~C\right)
	\label{eq:exp_bound_8}
\end{align}
for all $\epsilon'\in (0, 1)$. We shall soon specify how to choose $\epsilon'$.
The probability term in \eqref{eq:P(E_n_and_good_event} may then be upper bounded by
\begin{align}
	& P^\pi\bigg((1-\epsilon')\left(C(\underline{d}, \underline{i}, h)~D(P_1^{d_h}(\cdot|i_h)\|P_*^{d_h}(\cdot|i_h))+\sum\limits_{a\neq h, h'}~C(\underline{d}, \underline{i}, a)~D(P_2^{d_a}(\cdot|i_a)\|P_*^{d_a}(\cdot|i_a))\right)\nonumber\\
	&\hspace{10cm}+2(\log \bar{\varepsilon}^*)+3\varepsilon<\frac{\log((K-1)L)}{n}~\bigg|~C\bigg).
	\label{eq:exp_bound_9}
\end{align}
Recall that $P_*\in \mathscr{P}(\bar{\varepsilon}^*)$, $P_*\neq P_1, P_2$ as indicated prior to \eqref{eq:lower_bound_for_T_4(n)/n}. Let $\varepsilon>0$ be chosen such that
\begin{equation}
	3\varepsilon>2\,\log\frac{1}{\bar{\varepsilon}^*}-(1-\epsilon')\left(C(\underline{d}, \underline{i}, h)~D(P_1^{d_h}(\cdot|i_h)\|P_*^{d_h}(\cdot|i_h))+\sum\limits_{a\neq h, h'}~C(\underline{d}, \underline{i}, a)~D(P_2^{d_a}(\cdot|i_a)\|P_*^{d_a}(\cdot|i_a))\right).
	\label{eq:exp_bound_10}
\end{equation}
Such a choice of $\varepsilon$ is possible because the right hand side of \eqref{eq:exp_bound_10} is strictly positive by virtue of the fact that $$D(P_1^{d_h}(\cdot|i_h)\|P_*^{d_h}(\cdot|i_h))\leq \log \frac{1}{\bar{\varepsilon}^*}, \quad  D(P_2^{d_a}(\cdot|i_h)\|P_*^{d_a}(\cdot|i_a))\leq \log \frac{1}{\bar{\varepsilon}^*}$$ for all $a\neq h, h'$, and therefore
\begin{align}
	&(1-\epsilon')\left(C(\underline{d}, \underline{i}, h)~D(P_1^{d_h}(\cdot|i_h)\|P_*^{d_h}(\cdot|i_h))+\sum\limits_{a\neq h, h'}~C(\underline{d}, \underline{i}, a)~D(P_2^{d_a}(\cdot|i_a)\|P_*^{d_a}(\cdot|i_a))\right) \nonumber\\
	& \leq \left(\log\frac{1}{\bar{\varepsilon}^*}\right)~(1-\epsilon')~\sum\limits_{a\neq h'}C(\underline{d}, \underline{i}, a)\nonumber\\
	&\leq \log\frac{1}{\bar{\varepsilon}^*}.
\end{align}
For instance, it suffices to choose
\begin{equation}
	\varepsilon=\frac{2}{3}\left(2\,\log\frac{1}{\bar{\varepsilon}^*}-(1-\epsilon')\left(C(\underline{d}, \underline{i}, h)~D(P_1^{d_h}(\cdot|i_h)\|P_*^{d_h}(\cdot|i_h))+\sum\limits_{a\neq h, h'}~C(\underline{d}, \underline{i}, a)~D(P_2^{d_a}(\cdot|i_a)\|P_*^{d_a}(\cdot|i_a))\right)\right).
	\label{eq:equation_defining_varepsilon}
\end{equation}

With $\varepsilon$ as chosen above, it follows that inside the probability term in \eqref{eq:exp_bound_9}, the left hand side is strictly positive whereas the right hand side goes to $0$ as $n\to \infty$. Therefore, for all sufficiently large values of $n$, the probability term in \eqref{eq:exp_bound_9} is equal to zero.

We now turn attention to the probability term in \eqref{eq:exp_bound_8}. Using the union bound, this term may be upper bounded by
\begin{equation}
	\sum\limits_{a=1}^{K}~P^\pi\left(\frac{N(n, \underline{d}, \underline{i}, a)}{n} < C(\underline{d}, \underline{i}, a)(1-\epsilon')~\bigg|~C\right)=\sum\limits_{a=1}^{K}~P^\pi\left(N(n, \underline{d}, \underline{i}, a) < n\,C(\underline{d}, \underline{i}, a)(1-\epsilon')~\bigg|~C\right).
	\label{eq:exp_bound_11}
\end{equation}
In order to complete the proof, it suffices to show that each term inside the summation in \eqref{eq:exp_bound_11} is $O(1/n^3)$ for a suitable choice of $\epsilon'>0$.

Fix an arbitrary $a\in \mathcal{A}$. From the exposition in Appendix \ref{appndx:proof_of_prop_convergence_of_ML_estimates_of_TPMs}, we know that the process $\{(\underline{d}(t), \underline{i}(t)): t\geq K\}$ has the property that for all $T_0\geq K$, the relation
\begin{equation}
	P^{\pi}(\underline{d}(T_0+M+K+1)=\underline{d}, \underline{i}(T_0+M+K+1)=\underline{i}\mid \underline{d}(T_0)=\underline{d}'', \underline{i}(T_0)=\underline{i}'', C) \geq \rho\quad \forall~(\underline{d}'', \underline{i}'')\in \mathbb{S},
	\label{eq:exp_bound_14}
\end{equation}
where $\rho \geq \left(\frac{\eta}{K}\right)^{2K}\cdot (\bar{\varepsilon}^*)^K>0$,
and $M$ is a positive integer that satisfies
\begin{equation}
	P_1^M(j|i)>0, \quad P_2^M(j|i)>0\quad \forall ~i, j\in \mathcal{S}.
\end{equation}
The existence of such an integer $M$ is guaranteed by \cite[Proposition 1.7]{levin2017markov}. From \eqref{eq:exp_bound_14}, it follows that
\begin{equation}
		P^\pi(\underline{d}(t)=\underline{d}, ~\underline{i}(t)=\underline{i}|C)\geq \left(\frac{\eta}{K}\right)^{2K} (\bar{\varepsilon}^*)^K \quad \forall ~t\geq M+3K+1.
		\label{eq:lower_bound_on_P(d(t)=d, i(t)=i)}
\end{equation}
Additionally, we know that under the policy $\pi=\pi^\star(L,\delta)$,
\begin{equation}
	P^\pi(A_t=a|\underline{d}(t)=\underline{d}, ~\underline{i}(t)=\underline{i}, ~B^{t-1}, ~A^{t-1}, ~\bar{X}^{t-1}, ~C)\geq \frac{\eta}{K} \quad \forall~t\geq K.
	\label{eq:lower_bound_on_P(A_t=a|F_{t-1}}
\end{equation}
As a consequence of \eqref{eq:lower_bound_on_P(A_t=a|F_{t-1}}, it follows that $P^\pi(A_t=a|\underline{d}(t)=\underline{d}, ~\underline{i}(t)=\underline{i}. ~C)\geq \frac{\eta}{K}$ for all $t\geq K$.
Combining this with \eqref{eq:lower_bound_on_P(d(t)=d, i(t)=i)}, we see that for all $t\geq M+3K+1$,
\begin{align}
	P^\pi(\underline{d}(t)=\underline{d}, ~\underline{i}(t)=\underline{i}, ~A_t=a|C)\geq \left(\frac{\eta}{K}\right)^{2K+1} (\bar{\varepsilon}^*)^K.
	\label{eq:lower_bound_on_P(d(t)=d, i(t)=i, A_t=a|C)}
\end{align}
Clearly, for all $n\geq M+3K+1$,
\begin{align}
	N(n, \underline{d}, \underline{i}, a)
	&= \sum\limits_{t=K}^{M+2K}\mathbb{I}_{\{\underline{d}(t)=\underline{d}, ~ \underline{i}(t)=\underline{i},~ A_t=a \}}
	+ \sum\limits_{t=M+2K+1}^n\mathbb{I}_{\{\underline{d}(t)=\underline{d}, ~ \underline{i}(t)=\underline{i},~ A_t=a \}}\nonumber\\
	&\geq \sum\limits_{t=M+2K+1}^n\mathbb{I}_{\{\underline{d}(t)=\underline{d}, ~ \underline{i}(t)=\underline{i},~ A_t=a \}}
	\label{eq:exp_bound_12}
\end{align}
almost surely.
Denoting the right hand side of \eqref{eq:exp_bound_12} as $N'(n, \underline{d}, \underline{i}, a)$, it follows from \eqref{eq:lower_bound_on_P(d(t)=d, i(t)=i, A_t=a|C)} that
\begin{align}
	E^\pi[N'(n, \underline{d}, \underline{i}, a)|C]\geq (n-M-2K)\,\left(\frac{\eta}{K}\right)^{2K+1}(\bar{\varepsilon}^*)^K \quad \forall~n\geq M+3K+1.
	\label{eq:lower_bound_on_E[N'(n,d,i,a)|C]}
\end{align}
Therefore, for all $n\geq M+3K+1$, we have
\begin{align}
	&P^\pi\left(N(n, \underline{d}, \underline{i}, a) < n\,C(\underline{d}, \underline{i}, a)(1-\epsilon')~\bigg|~C\right)\nonumber\\
	&\leq P^\pi\left(N'(n, \underline{d}, \underline{i}, a) < n\,C(\underline{d}, \underline{i}, a)(1-\epsilon')~\bigg|~C\right)\nonumber\\
	&=P^\pi\bigg(N'(n, \underline{d}, \underline{i}, a) -E^\pi[N'(n, \underline{d}, \underline{i}, a)|C] < n\,C(\underline{d}, \underline{i}, a)(1-\epsilon') - E^\pi[N'(n, \underline{d}, \underline{i}, a)|C]~\bigg|~C\bigg)\nonumber\\
	&\leq P^\pi\bigg(N'(n, \underline{d}, \underline{i}, a) -E^\pi[N'(n, \underline{d}, \underline{i}, a)|C] < n\,C(\underline{d}, \underline{i}, a)(1-\epsilon') - (n-M-2K)\,\left(\frac{\eta}{K}\right)^{2K+1} (\bar{\varepsilon}^*)^K~\bigg|~C\bigg)\nonumber\\
	&\leq P^\pi\bigg(N'(n, \underline{d}, \underline{i}, a) -E^\pi[N'(n, \underline{d}, \underline{i}, a)|C] < n\left(C(\underline{d}, \underline{i}, a)(1-\epsilon') - \left(\frac{n-M-2K}{n}\right)\left(\frac{\eta}{K}\right)^{2K+1}(\bar{\varepsilon}^*)^K\right)~\bigg|~C\bigg)\nonumber\\
	&\leq P^\pi\bigg(N'(n, \underline{d}, \underline{i}, a) -E^\pi[N'(n, \underline{d}, \underline{i}, a)|C] < n\left(\frac{1-\epsilon'}{1+\delta} - \left(\frac{n-M-2K}{n}\right)\left(\frac{\eta}{K}\right)^{2K+1}(\bar{\varepsilon}^*)^K\right)~\bigg|~C\bigg),
	\label{eq:exp_bound_13}
\end{align}
where the last line follows by noting that
\begin{align*}
	C(\underline{d}, \underline{i}, a)
	&=\frac{\nu^{\lambda_{h, P_1, P_2, \delta}}(\underline{d}, \underline{i}, a)}{1+\delta}\\
	&<\frac{1}{1+\delta}.
\end{align*}
We now show that for a suitable choice of $\epsilon'$, the right hand side of the probability term in \eqref{eq:exp_bound_13} can be made negative. Because $(n-M-2K)/n \longrightarrow 1$ as $n\to\infty$, it follows that there exists $N_2=N_2(\delta)$ such that for all $n\geq N_2$, $$ \frac{n-M-2K}{n} > \frac{1}{1+\delta}, $$
 as a consequence of which $$ \left(\frac{n-M-2K}{n}\right)\left(\frac{\eta}{K}\right)^{2K+1}(\bar{\varepsilon}^*)^K > \frac{\left(\frac{\eta}{K}\right)^{2K+1}(\bar{\varepsilon}^*)^K}{1+\delta} $$ for all $n\geq N_2$.
Let $\epsilon'$ be chosen such that
\begin{equation}
	1-\epsilon' < \left(\frac{\eta}{K}\right)^{2K+1}(\bar{\varepsilon}^*)^K.
	\label{eq:equation_defining_epsilon'}
\end{equation}
Such a choice of $\epsilon'$ is possible since $$ 0~<~\left(\frac{\eta}{K}\right)^{2K+1}(\bar{\varepsilon}^*)^K~<~1. $$
For instance, it suffices to choose
\begin{equation}
	\epsilon'=1-\frac{1}{3}\cdot \left(\frac{\eta}{K}\right)^{2K+1}(\bar{\varepsilon}^*)^K.
	\label{eq:value_of_epsilon'}
\end{equation}
With this choice of $\epsilon'$, it follows that for all $n\geq \max\{N_2, M+3K+1\}$,
\begin{align}
	P^\pi\left(N(n, \underline{d}, \underline{i}, a) < n\,C(\underline{d}, \underline{i}, a)(1-\epsilon')~\bigg|~C\right) & \leq P^\pi\bigg(N'(n, \underline{d}, \underline{i}, a) -E^\pi[N'(n, \underline{d}, \underline{i}, a)|C] < -n \Delta ~\bigg|~C\bigg),
	\label{eq:exp_bound_17}
\end{align}
where $\Delta$ is given by $$ \Delta = \frac{2}{3}\cdot \frac{\left(\frac{\eta}{K}\right)^{2K+1}(\bar{\varepsilon}^*)^K}{1+\delta}. $$

We now demonstrate that $N'(n, \underline{d}, \underline{i}, a)$ is sub-gaussian. Subsequently, we use sub-gaussian concentration bounds to show that the right hand side of \eqref{eq:exp_bound_17} is bounded above exponentially. Recall that a random variable $Z$ is said to be sub-gaussian with variance factor $v$ \cite[Section 2.3]{boucheron2013concentration} if
\begin{equation}
	\log E\bigg[\exp\bigg(\lambda(Z-E[Z])\bigg)\bigg]\leq \frac{\lambda^2\,v}{2}\quad \forall~\lambda\in \mathbb{R}.
\end{equation}
It is well known that if $Z$ is Bernoulli distributed, then $Z$ is sub-gaussian with variance factor $v=1/4$.
The below lemma demonstrates that given sub-gaussian random variables $X$ and $Y$ (not necessarily independent), their sum is also sub-gaussian.
\begin{lemma}
\label{lem:sum_of_2_subgaussian_rvs_is_subgaussian}
	Suppose $X$ is sub-gaussian with variance factor $v_1$ and $Y$ (not necessarily independent of $X$) is sub-gaussian with variance factor $v_2$. Then, $X+Y$ is sub-gaussian with variance factor $v_1+v_2$.
\end{lemma}
\begin{IEEEproof}[Proof of Lemma \ref{lem:sum_of_2_subgaussian_rvs_is_subgaussian}]
Using Holder's inequality, for all $p>1$, we have
\begin{align}
	E\bigg[\exp\bigg(\lambda(X+Y - (E[X]+E[Y]))\bigg)\bigg] &\leq \left(E\bigg[\exp\bigg(p\,\lambda (X-E[X])\bigg)\bigg]\right)^{\frac{1}{p}}\cdot \left(E\bigg[\exp\bigg((1-p)\,\lambda (Y-E[Y])\bigg)\bigg]\right)^{\frac{1}{1-p}} \nonumber\\
	&\leq \left(\exp\left(\frac{\lambda^2\,p^2\,v_1}{2}\right)\right)^{\frac{1}{p}}\cdot \left(\exp\left(\frac{\lambda^2\,(1-p)^2\,v_2}{2}\right)\right)^{\frac{1}{1-p}}\nonumber\\
	&=\exp\left(\frac{\lambda^2(v_2+p(v_1-v_2)}{2}\right).
	\label{eq:exp_bound_18}
\end{align}
In particular, for $p=1+v_2/v_1$, we have
\begin{align}
	E\bigg[\exp\bigg(\lambda(X+Y - (E[X]+E[Y]))\bigg)\bigg] &\leq \exp\left(\frac{\lambda^2}{2}\left(v_2+\frac{v_1^2-v_2^2}{v_1}\right)\right)\nonumber\\
	&\leq \exp\left(\frac{\lambda^2(v_1+v_2)}{2}\right).
\end{align}
This establishes the desired result.
\end{IEEEproof}
Using Lemma \ref{lem:sum_of_2_subgaussian_rvs_is_subgaussian} and the sub-gaussian property of a Bernoulli random variable mentioned earlier, it follows that $N'(n, \underline{d}, \underline{i}, a)$ is sub-gaussian with variance factor $$ v^*=\frac{n-M-2K}{4}. $$ Using concentration bounds for sub-gaussian random variables \cite[p. 25]{boucheron2013concentration}, we get
\begin{equation}
	P^\pi\bigg(N'(n, \underline{d}, \underline{i}, a)- E^\pi[N'(n, \underline{d}, \underline{i}, a)|C]<-n\,\Delta~\bigg|~C\bigg)\leq \exp\left(-\frac{n^2\,\Delta^2}{2\,v^*}\right)\leq \exp(-2\,n\,\Delta^2).
\end{equation}
We know that there exists $N_3$ such that for all $n\geq N_3$, $$\exp(-2\,n\,\Delta^2)\leq 1/n^3.$$
Therefore, it follows that $$ P^\pi\bigg(N'(n, \underline{d}, \underline{i}, a)- E^\pi[N'(n, \underline{d}, \underline{i}, a)|C]<-n\,\Delta~\bigg|~C\bigg)\leq \frac{1}{n^3}\quad \forall~n\geq \max\{N_2, N_3, M+3K+1\}. $$
{\color{black} This completes the handling of $U_4(n)$.}

{\color{black} Combining \eqref{eq:exp_bound_5} and the above result, and choosing $A'$ in \eqref{eq:exp_bound_5} large if needed, we arrive at \eqref{eq:1/n^4_bound}. This completes the proof of Lemma \ref{lemma:exp_upper_bound_for_a_certain_probability_term}.}
\end{IEEEproof}

\subsection{Completing the Proof of Proposition \ref{prop:upper_bound}}
We first use the result established in Lemma \ref{lemma:exp_upper_bound_for_a_certain_probability_term} to show \eqref{eq:exp_moment_finite}, and later use \eqref{eq:exp_moment_finite} to complete the proof of Proposition \ref{prop:upper_bound}.
\begin{IEEEproof}[Proof of \eqref{eq:exp_moment_finite}]
Let
\begin{align}
	u(L)\coloneqq \frac{\log ((K-1)L)}{(\log L)\cdot	 \bigg[2\,\log\frac{1}{\bar{\varepsilon}^*}-(1-\epsilon')\left(C(\underline{d}, \underline{i}, h)~D(P_1^{d_h}(\cdot|i_h)\|P_*^{d_h}(\cdot|i_h))+\sum\limits_{a\neq h, h'}~C(\underline{d}, \underline{i}, a)~D(P_2^{d_a}(\cdot|i_a)\|P_*^{d_a}(\cdot|i_a))\right)\bigg]},
	\label{eq:u(L)}
\end{align}
where $\epsilon'$ is as given in \eqref{eq:value_of_epsilon'}. Let $\psi(L)=\max\{u(L), N_1, N_2, N_3, M+3K+1\}$, where the constants $N_1, N_2, N_3$ are as determined in the proof of Lemma \ref{lemma:exp_upper_bound_for_a_certain_probability_term}. Then, we have
\begin{align}
	& E^\pi\left[\left(\frac{\tau(\pi)}{\log L}\right)^2~\bigg|~C\right]\nonumber\\
	&=\int\limits_{0}^{\infty}~P^\pi\bigg(\left(\frac{\tau(\pi)}{\log L}\right)^2>x~\bigg|~C\bigg)~dx\nonumber\\
	&=\int\limits_{0}^{\infty}~P^\pi\bigg(\frac{\tau(\pi)}{\log L}>\sqrt{x}~\bigg|~C\bigg)~dx\nonumber\\
	&=\int\limits_{0}^{\infty}~P^\pi\bigg(\tau(\pi)> \lfloor\sqrt{x}\,\log L\rfloor~\bigg|~C\bigg)~dx\nonumber\\
	&=\int\limits_{0}^{\psi(L)}~P^\pi\bigg(\tau(\pi)> \lfloor\sqrt{x}\,\log L\rfloor~\bigg|~C\bigg)~dx + \int\limits_{\psi(L)}^{\infty}~P^\pi\bigg(\tau(\pi)> \lfloor\sqrt{x}\,\log L\rfloor~\bigg|~C\bigg)~dx\nonumber\\
	&= \frac{1}{(\log L)^2}\int\limits_{0}^{\log L\,\sqrt{\psi(L)}}~P^\pi\bigg(\tau(\pi)>\lfloor u \rfloor~\bigg|~C\bigg)\,2\,u\,du~+~\int\limits_{\psi(L)}^{\infty}~P^\pi\bigg(\tau(\pi)> \lfloor\sqrt{x}\,\log L\rfloor~\bigg|~C\bigg)~dx\nonumber\\
	&\leq  \frac{1}{(\log L)^2}\int\limits_{0}^{\log L\,\sqrt{\psi(L)}}~2\,u\,du ~+~\sum\limits_{n\geq \lfloor \log L\,\sqrt{\psi(L)}\rfloor} \left[\left(\frac{n+1}{\log L}\right)^2-\left(\frac{n}{\log L}\right)^2\right]~P^\pi\bigg(\tau(\pi)>n~\bigg|~C\bigg)\nonumber\\
	&\stackrel{(a)}{\leq} \psi(L)~+~\sum\limits_{n\geq \lfloor \log L\,\sqrt{\psi(L)}\rfloor} \frac{2n+1}{(\log L)^2}~P^\pi\bigg(M_h(n) < \log((K-1)L)~\bigg|~C\bigg)\nonumber\\
	&\stackrel{(b)}{\leq}\psi(L) +\frac{1}{(\log L)^2}\sum\limits_{n\geq \lfloor \log L\,\sqrt{\psi(L)}\rfloor} (2n+1)~\frac{B}{n^3}\nonumber\\
	&\leq \psi(L) +\frac{1}{(\log L)^2}~\sum\limits_{n\geq 1} ~(2n+1)~\frac{B}{n^3},
	\label{eq:exp_bound_19}
\end{align}
where $(a)$ above follows by upper bounding noting that $\{\tau(\pi)>n\}\subset \{M_h(n) < \log((K-1)L)\}$, and $(b)$ is due to Lemma \ref{lemma:exp_upper_bound_for_a_certain_probability_term}. Noting that the summation \eqref{eq:exp_bound_19} is finite, we get
\begin{align}
	\limsup\limits_{L\to\infty} E^\pi\left[\left(\frac{\tau(\pi)}{\log L}\right)^2~\bigg|~C\right] &\leq \limsup\limits_{L\to\infty} \psi(L)\nonumber\\
		&<\infty.
\end{align}
This establishes \eqref{eq:exp_moment_finite}.
\end{IEEEproof} 

{\color{black} Combining the almost sure upper bound in \eqref{eq:almost_sure_upper_bound_for_policy} and the uniform integrability result of \eqref{eq:exp_moment_finite}, we arrive at the upper bound for the expected stopping time of the policy $\pi^\star(L, \delta)$ in \eqref{eq:upper_bound_in_terms_of_delta}. This completes the proof Proposition \ref{prop:upper_bound}.}

\bibliographystyle{IEEEtran}
\bibliography{IEEEabrv, oai_restless_arms_with_learning.bib}

\begin{thebibliography}{10}
\providecommand{\url}[1]{#1}
\csname url@samestyle\endcsname
\providecommand{\newblock}{\relax}
\providecommand{\bibinfo}[2]{#2}
\providecommand{\BIBentrySTDinterwordspacing}{\spaceskip=0pt\relax}
\providecommand{\BIBentryALTinterwordstretchfactor}{4}
\providecommand{\BIBentryALTinterwordspacing}{\spaceskip=\fontdimen2\font plus
\BIBentryALTinterwordstretchfactor\fontdimen3\font minus
  \fontdimen4\font\relax}
\providecommand{\BIBforeignlanguage}[2]{{%
\expandafter\ifx\csname l@#1\endcsname\relax
\typeout{** WARNING: IEEEtran.bst: No hyphenation pattern has been}%
\typeout{** loaded for the language `#1'. Using the pattern for}%
\typeout{** the default language instead.}%
\else
\language=\csname l@#1\endcsname
\fi
#2}}
\providecommand{\BIBdecl}{\relax}
\BIBdecl

\bibitem{karthik2021detecting}
P.~N. Karthik and R.~Sundaresan, ``Detecting an {O}dd {R}estless {M}arkov {A}rm
  with a {T}rembling {H}and,'' \emph{IEEE Transactions on Information Theory},
  2021.

\bibitem{naghshvar2013two}
M.~Naghshvar and T.~Javidi, ``Two-{d}imensional {V}isual {S}earch,'' in
  \emph{2013 IEEE International Symposium on Information Theory}.\hskip 1em
  plus 0.5em minus 0.4em\relax IEEE, 2013, pp. 1262--1266.

\bibitem{Whittle1988}
P.~Whittle, ``Restless {B}andits: {A}ctivity {A}llocation in a {C}hanging
  {W}orld,'' \emph{Journal of applied probability}, vol.~25, no.~A, pp.
  287--298, 1988.

\bibitem{mandl1974estimation}
P.~Mandl, ``Estimation and {C}ontrol in {M}arkov {C}hains,'' \emph{Advances in
  Applied Probability}, pp. 40--60, 1974.

\bibitem{borkar1982identification}
V.~Borkar and P.~Varaiya, ``Identification and {A}daptive {C}ontrol of {M}arkov
  {C}hains,'' \emph{SIAM Journal on Control and Optimization}, vol.~20, no.~4,
  pp. 470--489, 1982.

\bibitem{Vaidhiyan2017}
N.~K. Vaidhiyan, S.~Arun, and R.~Sundaresan, ``Neural {D}issimilarity {I}ndices
  that {P}redict {O}ddball {D}etection in {B}ehaviour,'' \emph{IEEE
  Transactions on Information Theory}, vol.~63, no.~8, pp. 4778--4796, 2017.

\bibitem{vaidhiyan2017learning}
N.~K. Vaidhiyan and R.~Sundaresan, ``Learning to {D}etect an {O}ddball
  {T}arget,'' \emph{IEEE Transactions on Information Theory}, vol.~64, no.~2,
  pp. 831--852, 2017.

\bibitem{prabhu2017optimal}
G.~R. Prabhu, S.~Bhashyam, A.~Gopalan, and R.~Sundaresan, ``Optimal {O}dd {A}rm
  {I}dentification with {F}ixed {C}onfidence,'' \emph{arXiv preprint
  arXiv:1712.03682}, 2017.

\bibitem{pnkarthik2019learning}
P.~N. Karthik and R.~Sundaresan, ``Learning to {D}etect an {O}dd {M}arkov
  {A}rm,'' \emph{IEEE Transactions on Information Theory}, vol.~66, no.~7, pp.
  4324--4348, July 2020.

\bibitem{deshmukh2018controlled}
A.~Deshmukh, S.~Bhashyam, and V.~V. Veeravalli, ``Controlled {S}ensing for
  {C}omposite {M}ultihypothesis {T}esting with {A}pplication to {A}nomaly
  {D}etection,'' in \emph{2018 52nd Asilomar Conference on Signals, Systems,
  and Computers}.\hskip 1em plus 0.5em minus 0.4em\relax IEEE, 2018, pp.
  2109--2113.

\bibitem{deshmukh2021sequential}
A.~Deshmukh, V.~V. Veeravalli, and S.~Bhashyam, ``Sequential {C}ontrolled
  {S}ensing for {C}omposite {M}ultihypothesis {T}esting,'' \emph{Sequential
  Analysis}, pp. 1--38, 2021.

\bibitem{prabhu2020sequential}
G.~R. Prabhu, S.~Bhashyam, A.~Gopalan, and R.~Sundaresan, ``Sequential
  {M}ulti-hypothesis {T}esting in {M}ulti-armed {B}andit {P}roblems: {A}n
  {A}pproach for {A}symptotic {O}ptimality,'' \emph{arXiv preprint
  arXiv:2007.12961}, 2020.

\bibitem{Kaufmann2016}
E.~Kaufmann, O.~Capp{\'e}, and A.~Garivier, ``On the {C}omplexity of
  {B}est-{A}rm {I}dentification in {M}ulti-{A}rmed {B}andit {M}odels,''
  \emph{The Journal of Machine Learning Research}, vol.~17, no.~1, pp. 1--42,
  2016.

\bibitem{moulos2019optimal}
V.~Moulos, ``Optimal {B}est {M}arkovian {A}rm {I}dentification with {F}ixed
  {C}onfidence,'' in \emph{Advances in Neural Information Processing Systems},
  2019, pp. 5606--5615.

\bibitem{ortner2012regret}
R.~Ortner, D.~Ryabko, P.~Auer, and R.~Munos, ``Regret {B}ounds for {R}estless
  {M}arkov {B}andits,'' in \emph{International Conference on Algorithmic
  Learning Theory}.\hskip 1em plus 0.5em minus 0.4em\relax Springer, 2012, pp.
  214--228.

\bibitem{grunewalder2019approximations}
S.~Gr{\"u}new{\"a}lder and A.~Khaleghi, ``Approximations of the {R}estless
  {B}andit {P}roblem,'' \emph{The Journal of Machine Learning Research},
  vol.~20, no.~1, pp. 514--550, 2019.

\bibitem{liu2012learning}
H.~Liu, K.~Liu, and Q.~Zhao, ``Learning in a {C}hanging {W}orld: {R}estless
  {M}ultiarmed {B}andit with {U}nknown {D}ynamics,'' \emph{IEEE Transactions on
  Information Theory}, vol.~59, no.~3, pp. 1902--1916, 2012.

\bibitem{aastrom1973self}
K.~J. {\AA}str{\"o}m and B.~Wittenmark, ``On {S}elf {T}uning {R}egulators,''
  \emph{Automatica}, vol.~9, no.~2, pp. 185--199, 1973.

\bibitem{doshi1980strong}
B.~Doshi and S.~E. Shreve, ``Strong {C}onsistency of a {M}odified {M}aximum
  {L}ikelihood {E}stimator for {C}ontrolled {M}arkov {C}hains,'' \emph{Journal
  of Applied Probability}, pp. 726--734, 1980.

\bibitem{borkar1979adaptive}
V.~Borkar and P.~Varaiya, ``Adaptive {C}ontrol of {M}arkov {C}hains {I}:
  {F}inite {P}arameter {S}et,'' \emph{IEEE Transactions on Automatic Control},
  vol.~24, no.~6, pp. 953--957, 1979.

\bibitem{kumar1985survey}
P.~R. Kumar, ``A {S}urvey of {S}ome {R}esults in {S}tochastic {A}daptive
  {C}ontrol,'' \emph{SIAM Journal on Control and Optimization}, vol.~23, no.~3,
  pp. 329--380, 1985.

\bibitem{puterman2014markov}
M.~L. Puterman, \emph{Markov {D}ecision {P}rocesses: {D}iscrete {S}tochastic
  {D}ynamic {P}rogramming}.\hskip 1em plus 0.5em minus 0.4em\relax John Wiley
  \& Sons, 2014.

\bibitem{ausubel1993generalized}
L.~M. Ausubel and R.~J. Deneckere, ``A {G}eneralized {T}heorem of the
  {M}aximum,'' \emph{Economic Theory}, vol.~3, no.~1, pp. 99--107, 1993.

\bibitem{levin2017markov}
D.~A. Levin and Y.~Peres, \emph{Markov {C}hains and {M}ixing {T}imes}.\hskip
  1em plus 0.5em minus 0.4em\relax American Mathematical Soc., 2017, vol. 107.

\bibitem{federgruen1978note}
A.~Federgruen, A.~Hordijk, and H.~C. Tijms, ``A {N}ote on {S}imultaneous
  {R}ecurrence {C}onditions on a {S}et of {D}enumerable {S}tochastic
  {M}atrices,'' \emph{Journal of Applied Probability}, pp. 842--847, 1978.

\bibitem{milgrom2002envelope}
P.~Milgrom and I.~Segal, ``Envelope {T}heorems for {A}rbitrary {C}hoice
  {S}ets,'' \emph{Econometrica}, vol.~70, no.~2, pp. 583--601, 2002.

\bibitem{munkres2000topology}
J.~R. Munkres, \emph{Topology ({A} {F}irst {C}ourse)}.\hskip 1em plus 0.5em
  minus 0.4em\relax Englewood Cliffs, New Jersey. MIT. p, 2000.

\bibitem{chow2012probability}
Y.~S. Chow and H.~Teicher, \emph{Probability {T}heory: {I}ndependence,
  {I}nterchangeability, {M}artingales}.\hskip 1em plus 0.5em minus 0.4em\relax
  Springer Science \& Business Media, 2012.

\bibitem{boucheron2013concentration}
S.~Boucheron, G.~Lugosi, and P.~Massart, \emph{Concentration {I}nequalities:
  {A} {N}onasymptotic {T}heory of {I}ndependence}.\hskip 1em plus 0.5em minus
  0.4em\relax Oxford {U}niversity {P}ress, 2013.

\end{thebibliography}

\end{document}